\documentclass[twocolumn]{aastex631}

\usepackage{float}
\usepackage{xcolor}
\usepackage{graphicx}
\usepackage{natbib}
\usepackage{wrapfig}
\usepackage{amsmath,amssymb,amstext}
\usepackage{xspace}
\usepackage[linguistics]{forest}
\usepackage{upgreek}
\usepackage{mhchem}
\makeatletter
\let\splitbox\relax
\makeatother
\usepackage[splitbox=false]{adjustbox}
\usepackage{makecell}
\usepackage{booktabs}
\usepackage{threeparttablex}

\newcommand{\eureka}{\texttt{Eureka!}\xspace}
\newcommand{\exotic}{\texttt{ExoTiC-MIRI}\xspace}

\newcommand{\planet}{HD\,209458b}
\begin{document}

\title{Magnesium Silicate Clouds in the Atmosphere of HD\,209458b from a Rule-Based Tree-Structured Data Reduction}

\author[0000-0002-4552-4559]{Katy L. Chubb}
\affiliation{University of Bristol, HH Wills Physics Laboratory, Tyndall Avenue, Bristol, UK}

\author[0000-0001-5878-618X]{David Grant}
\affiliation{University of Bristol, HH Wills Physics Laboratory, Tyndall Avenue, Bristol, UK}

\author[0000-0003-4328-3867]{Hannah R. Wakeford}
\affiliation{University of Bristol, HH Wills Physics Laboratory, Tyndall Avenue, Bristol, UK}

\author[0000-0002-6721-3284]{Sarah E. Moran}
\altaffiliation{NHFP Sagan Fellow}
\affiliation{Space Telescope Science Institute, 3700 San Martin Drive, Baltimore, MD 21218, USA}

\author[0000-0003-1240-6844]{Natasha E. Batalha}
\affiliation{NASA Ames Research Center, Moffett Field, CA, 94035, USA}

\author[0000-0002-4701-8916]{Arika Egan}
\affiliation{Johns Hopkins APL, 11100 Johns Hopkins Rd, Laurel, MD 20723, USA}

\author[0000-0001-9665-5260]{Charlotte Fairman}
\affiliation{University of Bristol, HH Wills Physics Laboratory, Tyndall Avenue, Bristol, UK}

\author[0000-0002-4250-0957]{Diana Powell}
\affiliation{Department of Astronomy \& Astrophysics, University of Chicago, 5640 S. Ellis Avenue, Chicago, IL 60637, USA}

\author[0000-0002-7352-7941]{Kevin B. Stevenson}
\affiliation{Johns Hopkins APL, 11100 Johns Hopkins Rd, Laurel, MD 20723, USA}

\author[0000-0001-8703-7751]{Lili Alderson}
\affiliation{University of Bristol, HH Wills Physics Laboratory, Tyndall Avenue, Bristol, UK}

\author[0000-0002-8518-9601]{Peter Gao}
\affiliation{Earth and Planets Laboratory, Carnegie Institute of Washington, Washington, DC, USA}

\author[0000-0003-3759-9080]{Tiffany Kataria}
\affiliation{Jet Propulsion Laboratory, California Institute of Technology, 4800 Oak Grove Drive, Pasadena, CA 91001, USA}

\author[0000-0002-8507-1304]{Nikole K. Lewis}
\affiliation{Department of Astronomy and Carl Sagan Institute, Cornell University, 122 Sciences Drive, Ithaca, NY 14853, USA}

\author[0000-0003-4816-3469]{Ryan J. MacDonald}
\affiliation{School of Physics and Astronomy, University of St Andrews, North Haugh, St Andrews, KY16 9SS, UK}

\author[0000-0002-5251-2943]{Mark Marley}
\affiliation{Lunar and Planetary Laboratory, The University of Arizona, Tucson, AZ 85721, USA}

\author[0000-0003-0814-7923]{Elijah Mullens}
\affiliation{Department of Astronomy and Carl Sagan Institute, Cornell University, 122 Sciences Drive, Ithaca, NY 14853, USA}

\author[0000-0001-6050-7645]{David K. Sing}
\affiliation{Department of Physics \& Astronomy, Johns Hopkins University, Baltimore, MD 21218, USA}
\affiliation{Department of Earth \& Planetary Sciences, Johns Hopkins University, Baltimore, MD 21218, USA}

\author[0000-0003-3305-6281]{Jeff A. Valenti}
\affiliation{Space Telescope Science Institute, 3700 San Martin Drive, Baltimore, MD 21218, USA}



\begin{abstract}

\noindent HD\,209458b is the canonical hot Jupiter: the first to have its atmosphere measured and the first to hint at the role of aerosols in exoplanet atmospheres through the muting of Na absorption signatures in the optical. 
Here we present \textit{JWST} MIRI/LRS transmission observations of HD\,209458b from 5--12\,$\upmu$m, directly measuring the absorption signatures of its clouds for the first time. The observations indicate the presence of magnesium silicates, most likely \ce{Mg2SiO4} or a mixture of \ce{Mg2SiO4} and \ce{MgSiO3}. 
We also present a new methodology to reduce observational data, whereby the analysis is formulated as a rule-based model with a tree structure, enabling key decisions to be identified and uncertain decisions to be incorporated into subsequent modeling. With this data reduction, and using a combination of ARCiS free retrievals and \texttt{PICASO+Virga} self consistent forward models, we are able to show that amorphous \ce{Mg2SiO4} clouds explain the LRS data to high significance over either a clear ($\Delta$$\ln(Z)$=16.63) or gray cloud atmosphere ($\Delta$$\ln(Z)$=22.26). 
By combining the LRS dataset with archival \textit{JWST} NIRCam and \textit{HST} optical and near-infrared observations, we are able to more robustly constrain the properties of the magnesium silicate condensates, finding particle sizes of approximately 0.1\,$\upmu$m and atmospheric pressures of the clouds of roughly 1--10~millibar.
 Our results add to the growing detections of silicate clouds as a dominant atmospheric component of hot Jupiters, with the exact silicate species contextualizing the atmospheric chemistry and potentially formation conditions of these planets.  

\end{abstract}



\section{Introduction} \label{sec:intro}

Aerosols alter the way an atmosphere filters the light from the planet’s host star, obscuring or muting absorption signatures of gas phase atoms and molecules in the atmosphere. The Hubble Space Telescope (\textit{HST}) has played a definitive role in the characterization of exoplanet atmospheres and how aerosols imprint on exoplanet spectra, from the resolved but muted molecular features of H$_2$O in the atmosphere of WASP-39b \citep{wakeford2018_w39}, to the extreme scattering slope of hot Jupiter HD\,189733b \citep{sing2011b}, and the featureless cloud deck of the mini-Neptune GJ\,1214b \citep{kreidberg2014a}. In fact, aerosols have been notably present in exoplanet atmospheric spectra since the first atmospheric characterization of an exoplanet: HD\,209458b observed with \textit{HST} STIS \citep{charbonneau2002}. In particular, {\planet}'s atmosphere showed muted \ce{Na} I absorption, with a visible line core but a lack of line wings. It was hypothesized that this muting was due to the presence of optically thick clouds in the atmosphere of the planet obscuring the absorption signatures \citep[e.g.,][]{Atreya2003}. Further observations from \cite{Sing2016} showed that HD\,209458b has evidence for enhanced scattering in the UV-optical with a distinct but muted H$_2$O feature in the near-IR, both indicative of aerosols. 

Looking across a statistical sample of exoplanet transmission spectra \citep[e.g.,][]{iyer2016,Fu2017,Wakeford2019RNAAS}, the amplitude of the 1.4 $\mu$m water band in the near-infrared (NIR) is reduced compared to theoretical models assuming clear atmospheres. A theoretical study by \citet{Gao2020} suggested that the muting of H$_2$O in the NIR is correlated predominantly with the expected presence of magnesium silicate clouds in the atmosphere at temperatures above 950\,K, and soot/haze in atmospheres below 950\,K.

Observationally, a number of studies have presented evidence of silicate clouds in the atmospheres of L3-L5 brown dwarfs using the broad silicate absorption feature at 10\,$\upmu$m \citep{cushing2006,Looper2008,suarez2022}, with some constraining the size of silicate particles to $<$1 $\upmu$m \citep{Hiranaka2016}.
The low nucleation energy required for the formation of magnesium silicate grains and the high abundances of their gaseous precursors -- Mg, Si, and oxygen -- available at solar abundance means they likely dominate the cloud formation process in H/He atmospheres \citep[e.g.,][]{Gao2020,lee2016}. The stretching vibrational-mode of the major Si-O bond in the molecule results in the observed 10\,$\upmu$m absorption feature, with its strength and shape corresponding to the specific silicate composition, particle size, and abundance of condensate \citep{wakeford2015}.

The launch of \textit{JWST} in 2021 and subsequent observations from Summer 2022 presented the first opportunities for directly measuring the composition of transiting exoplanet aerosols in the mid-IR. The MIRI (Mid-InfraRed Instrument) on \textit{JWST} has detected the tell-tale absorption of aerosols in a handful of transiting exoplanets with the Low Resolution Spectrograph (LRS). The most definitive detections have suggested that silica (\ce{SiO2}, in the form of quartz) instead of magnesium silicates is dominant for small particles at high altitude; as measured in the limb of WASP-17b \citep{Grant2023quartz} and dayside of HD\,189733b \citep{Inglis2024ApJ}. 

Evidence for silica has also been found in brown dwarf atmospheres \citep{Burningham2021} and it has been theorized to be a seed particle for further cloud nucleation \citep{Helling2006}. In directly imaged planets at similar temperatures, \citet{hoch2025} and \citet{miles2023ERS} observed magnesium silicate clouds in YSES-1c and VHS-1256~b, and \citet{Molliere2025SiO} observed a potential nucleation phase of silicates in the form of SiO in PSO J318, using MIRI/LRS, MIRI/MRS, and MIRI/MRS, respectively. Further study of brown dwarfs with MIRI has revealed significant time variability of silicate cloud features, suggesting on-going cloud evolution and weather on these worlds \citep{Biller2024WeatherReport,Chen2025WeatherReport}.

Intriguingly, the cooler and smaller planets WASP-69b and WASP-107b show broader absorption features indicative of magnesium silicates or mixed aerosol compositions in their emission \citep{Schlawin2024AJ} and transmission \citep{Dyrek2024Nature}, but lack the expected $\sim$6\,$\upmu$m absorption feature \citep[e.g.,][]{he2020, corrales2023} of hydrocarbon-rich hazes that are expected to dominate at cooler temperatures below 1000\,K \citep[e.g.,][]{Gao2020}.
This suggests that atmospheric dynamics such as strong vertical winds are dominant over settling timescales, with the ability to loft small sub-micron silicate particles from hotter deeper layers to the top of the atmosphere \citep[e.g.,][]{sing2024}. 
Furthermore, recent \textit{JWST} observations have revealed inhomogeneous aerosol distributions across the limbs of several hot Jupiters, including WASP-107~b, suggesting that clouds either evaporate or gravitationally settle on the hot dayside~\citep{Murphy2025,Espinoza2024,Fu2025,Mukherjee2025}. These limb asymmetries provide interesting constraints on cloud transport, ultimately highlighting the critical interplay between vertical lofting and gravitational settling timescales in determining the observable aerosol properties of hot Jupiter atmospheres.

\begin{figure*}
\centering
\includegraphics[width=1.0\textwidth]{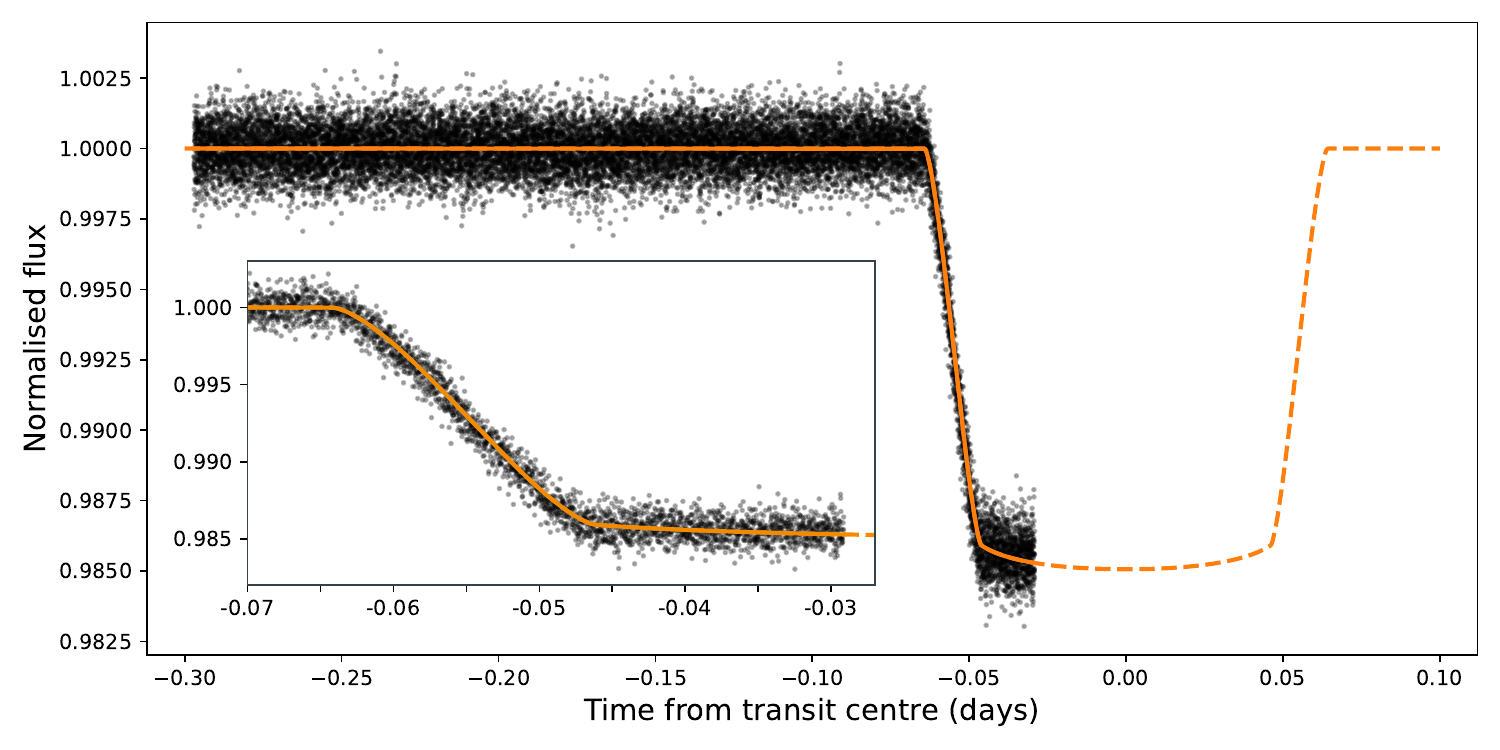}
\caption{Broadband light curve obtained by binning our 5--12 $\mu$m MIRI/LRS data. The start of the data are trimmed leaving a long and stable pre-transit baseline with minimal systematics. The best-fit transit model is shown in orange, where the dashed line represents an extrapolation of this model, and depicts the missed transit information due to erroneous observation timing due to an incorrect orbital ephemeris. Inset panel: a zoom in of the ingress data with the same axis labels as the main figure.}
\label{fig:white_light_curve}
\end{figure*}

In this paper we detail the observations obtained with \textit{JWST} MIRI/LRS of a single transit of the T$_{eq}\approx$1500\,K hot Jupiter HD\,209458b. 
HD\,209458b was the first transiting exoplanet discovered \citep{charbonneau2000,henry2000}, has a mass of 0.73\,M$_J$ and radius of 1.39\,R$_J$, and orbits a bright 6091\,K late F-type star on a 3.5\,day orbit \citep{Stassun2017}. 
\cite{Xue2024ApJ} obtained a \textit{JWST} NIRCam transmission spectrum of HD\,209458b from 2.3--5.1\,$\upmu$m (F322W2 and F444W) that suggests the
presence of an optically thick aerosol deck, a very low C/O$\sim$0.11, and a metallicity of 3\,$\times$\,solar. These point to an environment where oxygen-rich silicate clouds could readily form. Here we test this hypothesis and assess direct aerosol detections in the wavelength range where the vibrational modes of these materials should become apparent for small sub-micron particles. 

In Section \ref{sec:obs} we detail the measurements made and observational set-up. In Section \ref{sec:dt} we introduce the use of a rule-based tree-structured method to assess the optimal set of reduction steps to perform in our analysis of the data. In Section \ref{sec:modelling} we outline our atmospheric modeling frameworks and present our results. We discuss our results in the context of this planet's full spectral characterization and in terms of our current theoretical understanding of cloud formation in Section \ref{sec:discussion}. We present our conclusions in Section \ref{sec:conclusions}. 

\section{Observations} \label{sec:obs}

We observed one transit of HD\,209458b in program GO 2667 (PI: H. R. Wakeford), using \textit{JWST}'s Mid Infrared Instrument (MIRI) Low Resolution Spectrometer (LRS) in slitless mode~\citep{kendrew2015}.
The observations lasted 7.14 hr (transit duration 3.072 hr), using the FASTR1 readout pattern, and is comprised of 20,191 integrations each containing 7 groups.

The observations were carried out on 2023 July 14 yielding a time series of 5--12 $\mu$m spectra. However, errors in the transit ephemeris meant the observations started too early. As a result, these observations only captured a partial transit: ingress plus 19\% of full transit (Figure \ref{fig:white_light_curve}). However, due to the brightness of HD\,209458 (K mag\,=\,6.3) and long pre-transit baseline, the expected light-curve ramp \citep[e.g., see][]{bouwman2023} had significantly decayed prior to ingress, leaving behind data with minimal systematics across the captured portion of the transit\footnote{Due to the partial transit we reach an average spectral precision of 92\,ppm while an expected full transit spectral precision would range from 35-39\,ppm based on extrapolation of the observed lightcurve and Pandexo simulations \citep{BatalhaN2017b}, respectively.}. 

\begin{figure}
\centering
\begin{forest}
for tree={
    calign=fixed edge angles,
    calign angle=60,
    s sep=14mm,
    inner sep=0.1mm,
    l=15mm,
    align=center,
    grow=south,
}
[\hspace{-10mm}\texttt{\_uncal} data, for children={l=10mm}
    [\hspace{-10mm}Groups to mask
        [\hspace{-7mm}..., edge label={node[midway,gray,fill=white,font=\footnotesize]{None}}, edge={draw=black}]
        [\hspace{-7mm}..., edge label={node[midway,gray,fill=white,font=\footnotesize]{Last}}, edge={draw=black}]
        [\hspace{-10mm}Linearity, edge label={node[midway,gray,fill=white,font=\footnotesize]{First \& last}}, edge={draw=black}
            [\hspace{-10mm}Dark, edge label={node[midway,gray,fill=white,font=\footnotesize]{Default}}, edge={draw=black}
                [\hspace{-10mm}Jump, edge label={node[midway,gray,fill=white,font=\footnotesize]{True}}, edge={draw=black}
                    [\hspace{-7mm}..., edge label={node[midway,gray,fill=white,font=\footnotesize]{4$\sigma$}}, edge={draw=black}]
                    [\hspace{-10mm}Flat, edge label={node[midway,gray,fill=white,font=\footnotesize]{15$\sigma$}}, edge={draw=black}
                        [\hspace{-10mm}Background, edge label={node[midway,gray,fill=white,font=\footnotesize]{True}}, edge={draw=black}
                            [\hspace{-7mm}..., edge label={node[midway,gray,fill=white,font=\footnotesize]{Constant}}, edge={draw=black}]
                            [\hspace{-10mm}Aperture, edge label={node[midway,gray,fill=white,font=\footnotesize]{Row-wise}}, edge={draw=black}
                                [\hspace{-7mm}..., edge label={node[midway,gray,fill=white,font=\footnotesize]{Box 9pix}}, edge={draw=black}]
                                [\hspace{-10mm}Baseline, edge label={node[midway,gray,fill=white,font=\footnotesize]{Box 13pix}}, edge={draw=black}
                                    [\hspace{-7mm}..., edge label={node[midway,gray,fill=white,font=\footnotesize]{Long}}, edge={draw=black}]
                                    [\hspace{-10mm}Systematic model, edge label={node[midway,gray,fill=white,font=\footnotesize]{Short}}, edge={draw=black}
                                        [\hspace{-7mm}..., edge label={node[midway,gray,fill=white,font=\footnotesize]{Constant}}, edge={draw=black}]
                                        [\hspace{-10mm}Limb darkening, edge label={node[midway,gray,fill=white,font=\footnotesize]{Linear}}, edge={draw=black}
                                            [\hspace{-10mm}\texttt{\_spec$_i$}, edge label={node[midway,gray,fill=white,font=\footnotesize]{Quad}}, edge={draw=black}]
                                            [\hspace{-10mm}\texttt{\_spec$_j$}, edge label={node[midway,gray,fill=white,font=\footnotesize]{4-param}}, edge={draw=black}]
                                            [\hspace{-10mm}\texttt{\_spec$_k$}, edge label={node[midway,gray,fill=white,font=\footnotesize]{Fitted}}, edge={draw=black}]
                                        ]
                                        [\hspace{-7mm}..., edge label={node[midway,gray,fill=white,font=\footnotesize]{Linear+exp}}, edge={draw=black}]
                                    ]
                                ]
                                [\hspace{-7mm}..., edge label={node[midway,gray,fill=white,font=\footnotesize]{Optimal}}, edge={draw=black}]
                            ]
                            [\hspace{-7mm}..., edge label={node[midway,gray,fill=white,font=\footnotesize]{Narrow}}, edge={draw=black}]
                        ]
                        [\hspace{-7mm}..., edge label={node[midway,gray,fill=white,font=\footnotesize]{False}}, edge={draw=black}]
                    ]
                ]
                [\hspace{-7mm}..., edge label={node[midway,gray,fill=white,font=\footnotesize]{False}}, edge={draw=black}]
            ]
            [\hspace{-7mm}..., edge label={node[midway,gray,fill=white,font=\footnotesize]{Self-cal}}, edge={draw=black}]
        ]
    ]
]
\end{forest}
\caption{Diagram of the key steps considered in the MIRI/LRS Rule-based tree structure model for this dataset going from the \texttt{$\_$uncal} data to the final \texttt{$\_$spec}$_{i,j,k,...}$. At the nodes of the tree, branches are expanded for different reduction and analysis choices; with details provided in the text. Triple dots indicate the continuation of the tree structure. The tree can be described as homogeneous or symmetric in that it can be thought of as a complete set of decisions with no dangling branches.}
\label{fig:total_decision_tree_diagram}
\end{figure}
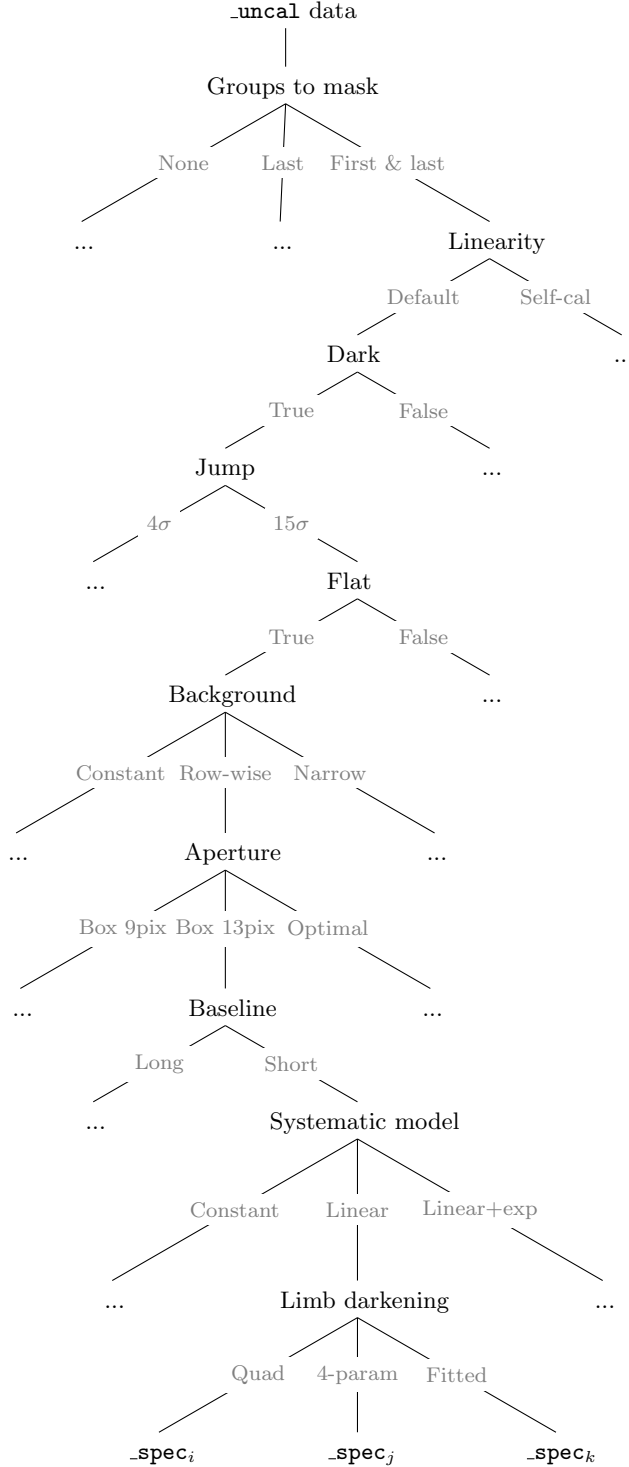

\section{Tree-Structured Data Reduction} \label{sec:dt}
In this work, we introduce a new framework to reduce and analyse any datasets for which the best data reduction practices have yet to be fully established\footnote{In this case for MIRI/LRS data in general, and with all \textit{JWST} time series analysis in mind.}. Traditionally multiple data reduction pipelines are used to reduce and analyze transmission spectra (see Appendix\,\ref{sec:independent_pipeline_check}) with one or more reduction taken forward for interpretation. The core idea we present here is to methodically explore the sensitivity of results to specific data reduction decisions within a single pipeline which can cover selected decisions using a tree-based structure, and generate resulting spectra which include uncertainties from these decisions for the subsequent inferences. This process can be summarized as follows:
\begin{enumerate}
  \item Create a rule-based tree structure which includes all possible data reduction decisions from the uncalibrated (\texttt{\_uncal}) data to the planetary spectrum. Run all the reasonable paths (`branches') through the tree. All of the resulting spectra (`leaves' of the tree) correspond to the end point of a unique branch of reduction decisions.
  \item Quantify the differences between the resulting spectra. The extent to which a given decision drives differences is ascertained by looking at pairs of spectra which differ by only one decision in their branches. These data reduction decisions are then ranked (see Section\,\ref{sec:dt_rank} for details) and consensus on the best reduction paths is reached.
  \item The spectra (`leaves') from the best reduction paths are combined into a mixture model. For the case of transmission spectra, the transit depth distributions are combined as a Gaussian Mixture Model (GMM) and subsequent modeling can involve this in the likelihood computation. In the case of there being a single best path, this results in one spectral data product and the typical normal likelihood.
\end{enumerate}
This process therefore enables the discovery of the decisions which drive differences between data reductions, and provides a framework for marginalizing over uncertainties in the data reduction process when interpreting the spectra. This approach extends previous work marginalizing over potential light-curve systematic models \citep{wakeford2016}, now including possible data reduction choices. These above steps can be applied to any dataset with decisions specific to the observatory, instrument, or planetary configuration (i.e., transit, eclipse, phase curve) used to construct the rule-based tree structure. 

In the following subsections, we walk through the process in greater detail applied specifically to our MIRI/LRS HD\,209458b data. We applied the \exotic \citep{grant_david_2023_8211207} data reduction software, which at the time of this reduction made use of the \texttt{jwst} pipeline v1.8.2 with CRDS map 1077 along with updated bias corrections. 

\subsection{Exploring decisions within a rule-based tree structure} \label{sec:dt_explore}
First, a large rule-based tree structure is created, which contains new branches for all the data-reduction decisions that might reasonably be expected to affect the results. This tree starts with the \texttt{\_uncal} data and branches out to the final spectral data products. For our MIRI/LRS observations we identify ten decisions worth exploring which are shown in Figure \ref{fig:total_decision_tree_diagram}. These ten decisions were selected based on unexplored questions in the MIRI/LRS pipeline (e.g., the necessity of darks and flats) and those with known or uncertain impacts to the final planetary spectrum (e.g., systematic model and limb-darkening prescriptions). In this vein we do not include decisions that would be considered unrecommended such as performing no background removal or, for transit observations, no corrections for limb-darkening.
From top to bottom of the tree, the necessary pipeline steps and hence decisions and choices are as follows:

\begin{itemize}
\item The choice on the number of `\textbf{groups to mask}' in the ramp fitting step of the \texttt{jwst} pipeline can change the correction to the sampling up the ramp method used across \textit{JWST} instruments. 
This is especially important here as MIRI/LRS may contain biased values, often at the start and end groups of each ramp \citep{ressler2015mid, wright2023mid, Bergeron2021jwst.rept.7761B}. Our data, having 7 groups, results in three reasonable choices: mask none of the groups, mask the last group, or mask the first and last groups. 

\item A \textbf{linearity} correction to account for the non-linear response of the pixels as they become progressively full must be applied. However, the default linearity model may not be optimal depending on the brightness of the target and occurrence of the debiasing-induced brighter-fatter effect (BFE) on the pixels \citep{argyriou2023brighter}. As such, the choices here are to use the default linearity correction or create a custom self-calibrated correction \citep[e.g.,][]{Grant2023quartz}. 

\item The \textbf{dark} current step may be chosen to be on or off depending on how well calibration files perform. 

\item For the \textbf{jump} step, the default algorithm can be over zealous in flagging outliers, and so various cleaning thresholds, for example 4$\sigma$ (the \texttt{jwst} default) and 15$\sigma$, can be trialed \citep[e.g., see][]{bell2023first}.

\item Post ramp fitting, the \textbf{flat field} step can be turned on or off, again depending on how well the calibration files perform. 

\item The \textbf{background subtraction} step for MIRI/LRS data has a number of options owing to the background varying along the direction of the dispersion axis, and many viable pixels either side of the spectral trace. Three reasonable choices used here are a constant value per rateimage, a row-dependent value, and a row-dependent value using a narrower set of pixels \citep{Grant2023quartz}, with the aim of excluding columns too close to the trace or subarray edges. In future tree structures it would also be worth incorporating a decision associated with the inner edge of the background window as this may have an additional effect on the data \citep{Stevenson2025}. 

\item The method for extracting and summing the pixel values to generate the stellar spectra may make use of various \textbf{apertures}. Here, we include options for standard box apertures, at full widths of either 9 or 13 pixels (as shown in \citet{bell2023first} to represent a reasonable range of apertures to use for MIRI/LRS measurements), as well as an optimally-weighted aperture following \citet[][]{Horne1986}.
\end{itemize}

At the light curve fitting stage, there are a few further key options to include. 
\begin{itemize}
\item The duration of the data \textbf{baseline} pre-transit \footnote{Or post-transit data for correctly-timed datasets.} can affect results when using MIRI/LRS due to the known charge-trapping ramps \citep{bouwman2023} and the potential degeneracies that can occur from including these types of features in mid-infrared light curves \citep{stevenson2012, may2020}. Here, due to the incredibly long baseline caused by timing errors, we include options for masking the first 2,000 (42-minutes) versus 10,000 (3.5-hours) integrations. These timings will be specific to each dataset and to reduce the number of branches created we choose to test two extremes which represent the removal of the onset of the ramp and at the extreme the removal of all exposures which show a visible ramp structure. These values can be found through simple pre-transit tests on the ramp structure/duration prior to setting the full rule-based tree structure to reduce final branch numbers.

\item The transit light curve model includes a \textbf{systematic model} for the aforementioned ramps which can take the functional form,
$$S(t_s,t_c) = (c_0 + c_1 t_c) \times r_0 \exp(r_1 t_s),$$ with $t_s$ and $t_c$ as the observation times minus the light curve start time and center-of-transit time, respectively. $c_0$, $c_1$, $r_0$ and $r_1$ are constants to be fit or set to zero when not considered as part of the systematic model applied to the data. Here, the systematic model options considered are a constant offset ($c_0$), a linear ramp ($(c_0 + c_1 t_c)$), and a linear plus exponential ramp (the full function $S(t_s,t_c)$ above). 

\item The treatment of \textbf{limb-darkening} presents another decision, whether to fit or fix the limb-darkening parameters, and which law to use. The options presented here were to fix the coefficient for the quadratic or 4-parameter laws \citep[computed using \texttt{ExoTiC-LD;}][]{Grant2024}, as well as fit for the coefficients using the parametrized quadratic law.
\end{itemize}

\begin{table*}[t]
    \centering
    \caption{Data reduction metrics for our rule-based tree structure. Large numbers represent large difference in the final spectrum in this case clearly showing that a constant background or not masking any groups at the reduction stage are unwise choices. Small numbers mean that either decision considered is ``correct'' and only one need be taken to ensure an accurate result. In this case negligible signifies differences where only one or two data points showed differences within 1$\sigma$ and small signifies less than half of the spectrum show differences within 1$\sigma$.}
    \label{tab:decision_tree_metrics}
    \begin{tabular}{lcll} 
    \hline
    \rule{0pt}{3ex}Decision & Mean $l^2_w$-norm & Prune? & Reason \\
    \hline
    \rule{0pt}{3ex} Background: {\color{gray} constant} vs {\color{gray}narrow}  & 8.18 & Prune & Constant background model shows large residuals. \\
    \rule{0pt}{3ex} Background: {\color{gray} constant} vs {\color{gray} row-wise}  & 7.75 & Prune & Constant background model shows large residuals. \\
    \rule{0pt}{3ex} Groups to mask: {\color{gray} none} vs {\color{gray} first \& last}  & 4.95 & Prune & The MIRI last-frame effect is apparent.  \\
    \rule{0pt}{3ex} Groups to mask: {\color{gray} none} vs {\color{gray} last}  & 4.32 & Prune & The MIRI last-frame effect is apparent.  \\
    \rule{0pt}{3ex} Baseline: {\color{gray} long} vs {\color{gray} short}  & 4.20 & Keep &   \\
    \rule{0pt}{3ex} Systematic model: {\color{gray} Linear+exp} vs {\color{gray} Linear}  & 3.45 & Keep &   \\
    \rule{0pt}{3ex} Groups to mask: {\color{gray} last} vs {\color{gray} first \& last}  & 2.90 & Keep &   \\
    \rule{0pt}{3ex} Aperture: {\color{gray} Box 9pix} vs {\color{gray} optimal}  & 2.10 & Prune & Small differences between spectra. \\
    \rule{0pt}{3ex} Aperture: {\color{gray} Box 9pix} vs {\color{gray} Box 12pix}  & 2.07 & Prune & Small differences between spectra. \\
    \rule{0pt}{3ex} Background: {\color{gray} row-wise} vs {\color{gray} narrow}  & 2.03 & Prune & Small differences between spectra. \\
    \rule{0pt}{3ex} Dark: {\color{gray} true} vs {\color{gray} false}  & 1.71 & Prune & Small differences between spectra. \\
    \rule{0pt}{3ex} Limb darkening: {\color{gray} quadratic} vs {\color{gray} 4-param}  & 0.35 & Prune & Negligible differences between spectra. \\
    \rule{0pt}{3ex} Aperture: {\color{gray} Box 12pix} vs {\color{gray} optimal}  & 0.30 & Prune & Negligible differences between spectra. \\
\hline
    \end{tabular}
\end{table*}

With the rule-based tree structure complete, all of the possible data reduction branches can be run. Specific to our HD\,209458b data, initial testing showed that we could omit several branch decisions that either are ill advised based on the specific observational set-up we employed or in general for MIRI observations. These resulted in removing branches that: used the custom self-calibrated linearity correction (meaning we default to the \texttt{jwst} pipeline linearity correction), used the 4$\sigma$ jump detection, did not apply the flat field, used a constant offset systematic model, and fit for the limb-darkening coefficients using the parameterized quadratic model. 
The remaining branches shown in Figure\,\ref{fig:total_decision_tree_diagram} were run, resulting in 432 `leaves' or transmission spectra. We note that based on these decisions the data reduction pipeline does not need to be run 432 times, as the data products at each node in the tree can be used for all lower branches. Summing up the number of decisions made at Stage 1, which wraps and modifies the \texttt{jwst} pipeline, results in just six runs of the data reduction pipeline. This ensures that the computation time does not become unnecessarily long as this is the most time intensive step. The subsequent branches then encompass 72 decisions for each of these six runs to total our 432 leaves. 

\begin{figure*}
\centering
\begin{forest}
for tree={
    calign=fixed edge angles,
    calign angle=60,
    s sep=14mm,
    inner sep=0.1mm,
    l=15mm,
    align=center,
    grow=south,
}
[\hspace{-10mm}\texttt{\_uncal} data, for children={l=10mm}
    [\hspace{-10mm}Groups to mask
        [\hspace{-10mm}Baseline, edge label={node[midway,gray,fill=white,font=\footnotesize]{Last}}, edge={draw=black}
            [\hspace{-10mm}Systematic model, edge label={node[midway,gray,fill=white,font=\footnotesize]{Long}}, edge={draw=black}
                [\hspace{-10mm}\texttt{\_spec$_1$}, edge label={node[midway,gray,fill=white,font=\footnotesize]{Linear+exp}}, edge={draw=black}]
            ]
            [\hspace{-10mm}Systematic model, edge label={node[midway,gray,fill=white,font=\footnotesize]{Short}}, edge={draw=black}
                [\hspace{-10mm}\texttt{\_spec$_2$}, edge label={node[midway,gray,fill=white,font=\footnotesize]{Linear}}, edge={draw=black}]
            ]
        ]
        [\hspace{-10mm}Baseline, edge label={node[midway,gray,fill=white,font=\footnotesize]{First \& last}}, edge={draw=black}
            [\hspace{-10mm}Systematic model, edge label={node[midway,gray,fill=white,font=\footnotesize]{Long}}, edge={draw=black}
                [\hspace{-10mm}\texttt{\_spec$_3$}, edge label={node[midway,gray,fill=white,font=\footnotesize]{Linear+exp}}, edge={draw=black}]
            ]
            [\hspace{-10mm}Systematic model, edge label={node[midway,gray,fill=white,font=\footnotesize]{Short}}, edge={draw=black}
                [\hspace{-10mm}\texttt{\_spec$_4$}, edge label={node[midway,gray,fill=white,font=\footnotesize]{Linear}}, edge={draw=black}]
            ]
        ]
    ]
]
\end{forest}
\caption{The decisions that were kept for our 4-leaf spectrum. Our 1-leaf spectrum is formed from spec3.}
\label{fig:final_decision_tree_diagram}
\end{figure*}
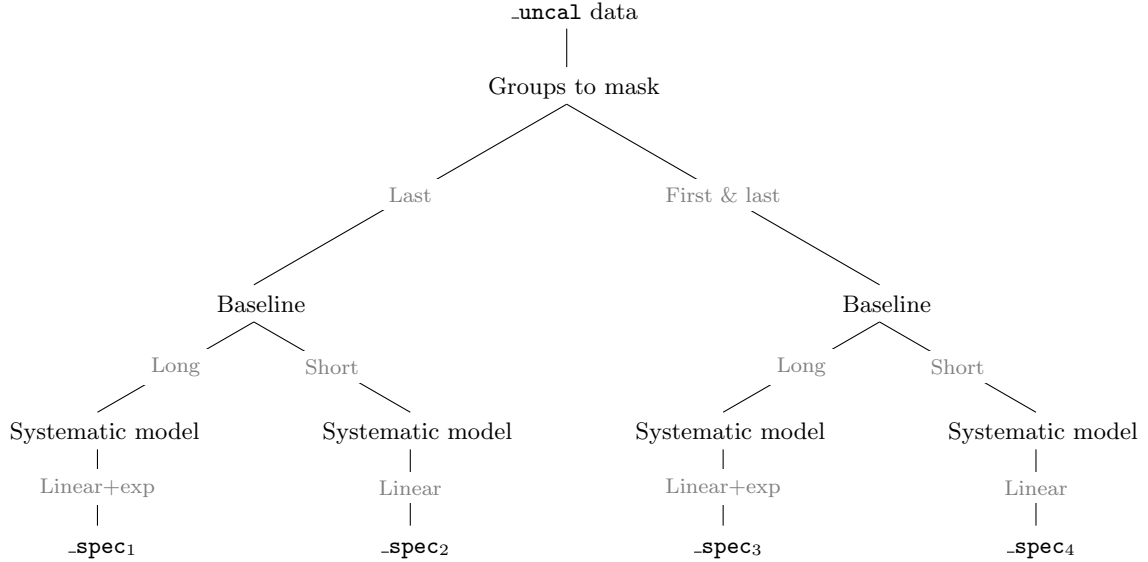

\subsection{Ranking decisions and building consensus} \label{sec:dt_rank}
The next step is to assess what decisions within the tree structure are important, what decisions in the tree make no difference, and what decisions in the tree lead to the best quality spectra. To quantify this, we measure the similarity between pairs of spectra where only one decision is different. In other words, these spectra follow the same branches through the tree except for one decision node. The similarity measure was chosen to be the $l^2_w$-norm (the weighted L2-norm), and this value is averaged over all paths that compare each decision. 
In this case, our $l^2_w$-norm takes the form 
\begin{equation}
     \sqrt{\sum_i \frac{(x_{1,i} - x_{2,i})^2}{\sigma_{1,i}^2 + \sigma_{2,i}^2}} \ ,
\end{equation}\label{l2-norm}
where $\sigma_{1,i}$ is the uncertainty on the transmission spectral point $x_{1,i}$. This is a standard metric in statistics and machine learning applications used to compare two vectors where the importance of considering each vector is balanced by the weighting.

The decisions for our HD\,209458b data are ranked and shown in Table \ref{tab:decision_tree_metrics}. The rankings show that some decisions significantly impact the results (large values), while many make no difference (small values)\footnote{Note, the $l^2_w$-norm does not have a built in cut-off value and the numbers are considered arbitrary depending on the dimensionality of the vectors compared.}. As we have assembled a rule-based tree structure we can use those rules and evidence from other factors such as the standard deviation of the light curve residuals or the red noise properties of the residuals (calculated using a time-averaging method on the residuals to show the binned standard deviation relative to that expected from photon noise \citep[e.g.,][]{pont2006,Alderson2023_ERS}) used to create a beta scaling factor\footnote{see example calculation using the \texttt{ExoTiC-JEDI/extra\_functions.py/noise\_calculator} function (https://github.com/Exo-TiC/ExoTiC-JEDI/)}, to rule in or out any branch decisions in the tree. 

At the bottom of the rankings, we find the decisions that do not impact the resulting spectra meaning that any decision made can be considered ``correct'' or accurate to achieve the final result. 
These include the aperture type, the pixels used for the row-wise background subtractions, the implementation of dark correction, and the limb-darkening law. These rule-based decisions are then deemed unnecessary to explore and pruned from the tree structure, meaning while in many cases a decision will be made it will not change the resulting spectrum (see Section\,\ref{sec:final_analysis} for these decisions).

At the other end of the rankings exist the decisions that greatly impact the results. 
Once again, some of these branches can be pruned as these decisions lead to obviously worse quality data when evaluating the light curves themselves or from an understanding of the expected MIRI reduction systematics. 
These include using a constant background rather than a row-wise subtraction, which produces very large residuals in the light curves, and not masking the final group prior to ramp fitting, which displays the known last-frame effect in MIRI data \citep{Bergeron2021jwst.rept.7761B}. 

The remaining decisions are those in the middle of the rankings. These decisions are the ones that impact the results meaningfully (often more than half the data points are different but still within 1$\sigma$ of the uncertainties), but cannot be pruned as it is not certain which branches lead to the most accurate spectrum. For other datasets in future work, it could also be worth investigating the decisions as a function of wavelength, computing the $l^2_w$-norm over a sliding window of wavelength bins. This may help in understanding what features in the spectra each reduction decision affects more specifically. We note that additional steps can be added to the tree structure and any rule-based decisions for a particular reduction or with different instruments will necessitate different rule-based decisions being applied. For example for near-IR JWST data the correction for 1/f noise is an important step \citep[e.g.,][]{Schlawin2020AJ,Rustamkulov2022,Birkmann2022SPIE,Alderson2023_ERS} which can be incorporated into the tree structure, or for light curves for active or young stars the incorporation of branches which consider the inclusion of single or multiple spot models or stellar activity at the light curve level \citep[e.g.,][]{Barat2025AJ_V1298Taub} may need to be considered. This rule-based tree structure aims to incorporate expert knowledge of the observations from the observer and community while accounting for decision in which we may be ignorant or are inherently dataset or target dependent.

\subsection{Our final data reduction steps}\label{sec:final_analysis}
The remaining tree (shown in Figure \ref{fig:final_decision_tree_diagram}) is comprised of a minimal set of decision paths that all lead to high-quality spectra, but from which it is not clear what is the most correct approach. This tree's remaining decisions involve which groups to mask (last or first and last), how much pre-transit baseline to include (long cutting 2,000 integrations or short cutting 10,000 integrations), and what systematic model to fit (linear or linear + exponential). Noting that the baseline and systematic model are correlated such that we only include the more complex systematics model with the longer baseline, and vice versa.

For all other decisions in the data reduction and light curve analysis we default to the following procedures. 
We use the default \texttt{jwst} pipeline procedures for the linearity correction, dark subtraction, jump step (setting the threshold to 15$\sigma$), ramp fitting, gain, and flat fielding steps. 
For background subtraction we use a row wise subtraction over a narrow set of columns from [12:22] and [50:68] left and right of the spectral trace, respectively.
We extract the stellar spectrum centered around column 36 using a box aperture of 9 pixels (extracting the central column and $\pm$4 pixels either side). 

At the light curve fitting stage our HD\,209458b system parameters are fixed for a circular orbit with a period of 3.52474859\,days, an orbital inclination of 86.71$^\circ$, and a/R$_*$ of 8.78. We fix the center of transit times in our spectroscopic fits to 2460140.48408 (JD). 

We compute our limb-darkening coefficients using the \texttt{ExoTiC-LD} package \citep{Grant2024} for [M/H]\,=\,0, T$_\mathrm{eff}$\,=\,6026\,K, and log(g)\,=\,4.3. We use the Kurucz 1D stellar grid \citep{kurucz1993atlas9}, locating the nearest grid point in the grid and computing the coefficient for the quadratic limb-darkening law across the default MIRI/LRS throughput curve. 
We compute all of our transmission spectra in 0.25\,$\upmu$m bins starting from 5.06\,$\upmu$m on the detector and fit them using an MCMC \citep{emc32019JOSS....4.1864F} with 6,000 samples, half of which is burn in, and 16 walkers. Our MCMC set-up ensures that the minimum number of walkers and samples are needed to provide robust and converged results for this dataset thus minimizing the computational burden across our full tree structure. However, we note that this is a high signal-to-noise light curve and additional tests may be required to ensure the MCMC set-up for other datasets is robust to the data.

\subsubsection{Comparison to \eureka reduction}
For completeness and to better assess this proposed rule-based tree structure, we also performed a standard additional reduction with the well-vetted \texttt{Eureka!} \citep{bell2022} pipeline, described in Appendix \ref{sec:independent_pipeline_check}. Using the \eureka pipeline we identify the same steps as shown in Figure~\ref{fig:final_decision_tree_diagram} as the most important decisions relevant to the final spectrum. With this comparison we note a number of steps that are not incorporated into the tree structure but should be considered for data reduction purposes as tests of the data quality and information content. For example, \citet{Stevenson2025} showed that the starting point and width of bins can be important for the shape of MIRI transmission spectra and the resulting inferences. Here we tested how the binwidth of 0.125, 0.25, or 0.5\,$\upmu$m impacted the spectrum outside of our tree structure and found that the shape was consistent while the 0.25\,$\upmu$m bins provided a balance between resolution and precision. However, we do suggest that future implementation of the rule-based tree structure also incorporate shifts in the starting point of the spectral bins (see Appendix\,\ref{sec:independent_pipeline_check} for direct comparisons in this case) but that this would need to be done with care to avoid wavelength correlated noise between bins and may be best incorporated as an external weighting to supplement decisions made using the $l^2_w$-norm.

\subsection{Finalizing the spectrum as a mixture model} \label{sec:dt_finalise}
It may be the case that only one leaf remains, meaning an optimal set of data reduction decisions has been found, and the final transit spectrum can therefore be modeled in the typical manner using a normal likelihood function. However, if more than one leaf is present, as is the case for our HD\,209458b data, we propose the spectra are combined into a Gaussian mixture model (GMM). For these data, this can be thought of as having $K$ independently and identically distributed draws from a mixture of $D$-dimensional multivariate normal distributions, where $K$ is the number of spectra and $D$ is the number of wavelength bins. A GMM is defined as
\begin{equation}\label{eq:gmm_1}
    p(\boldsymbol{x} | \boldsymbol{\theta}) = \sum_{k = 1}^{K} \pi_k \mathcal{N}(\boldsymbol{x} | \boldsymbol{\mu}_k, \boldsymbol{\Sigma}_k),
\end{equation}
where $\boldsymbol{x}$ is a vector of transit depths to evaluate the GMM, $\pi_k$ is the weight of each mixture component (the probability of selecting each component, often initialized as a uniform weighting and is updated based on the likelihood calculated from the spectral uncertainties), and
\begin{align}
    \label{eq:gmm_2}
    \mathcal{N}(\boldsymbol{x} | \boldsymbol{\mu}_k, \boldsymbol{\Sigma}_k) = &(2 \pi)^{-D/2} |\Sigma|^{-1/2} \\ &\times \exp\left[-\frac{1}{2}(\boldsymbol{x} - \boldsymbol{\mu}_k)^T \boldsymbol{\Sigma}_k^{-1} (\boldsymbol{x} - \boldsymbol{\mu}_k) \right],
\end{align}
is the multivariate normal distribution, having vectors $\boldsymbol{\mu}_k$ and matrices $\boldsymbol{\Sigma}_k$ for the mean and covariance of the $k$th component. In this way, any subsequent analysis will be able to marginalize over any uncertainties in the data reduction decisions, and inferences will be robust to any uncertainty in data reduction practices. Hereafter, we refer to the typical single transit spectrum as a 1-leaf spectrum, and mixture models as an n-leaf spectrum. 

The total log-likelihood, $\ln \mathcal{L}$ can be computed from an array of log-likelihoods $x$ of length $n$, where $n$ is the number of leaves, and evaluated using:
\begin{equation}\label{log_lik}
\ln \mathcal{L} = \log \left( \sum_{i=1}^{n} e^{x_i} \right).
\end{equation}
A GMM is weighted such that each component is not considered an independent event which means models averaging clusters (underlying sub-populations of results) should be less likely than models which fall directly on a specific cluster, and for this we need to treat each spectrum as a multivariate normal distribution (each dataset has a shared origin and is not considered an independent observation).
We note that if datasets produce spectra that are wildly non-Gaussian in their transit depths, then the GMM may need to be replaced with a kernel density estimator, although this may then add significant additional computation time at the retrieval stage. 

\begin{figure*}
\centering
\includegraphics[width=1.0\textwidth]{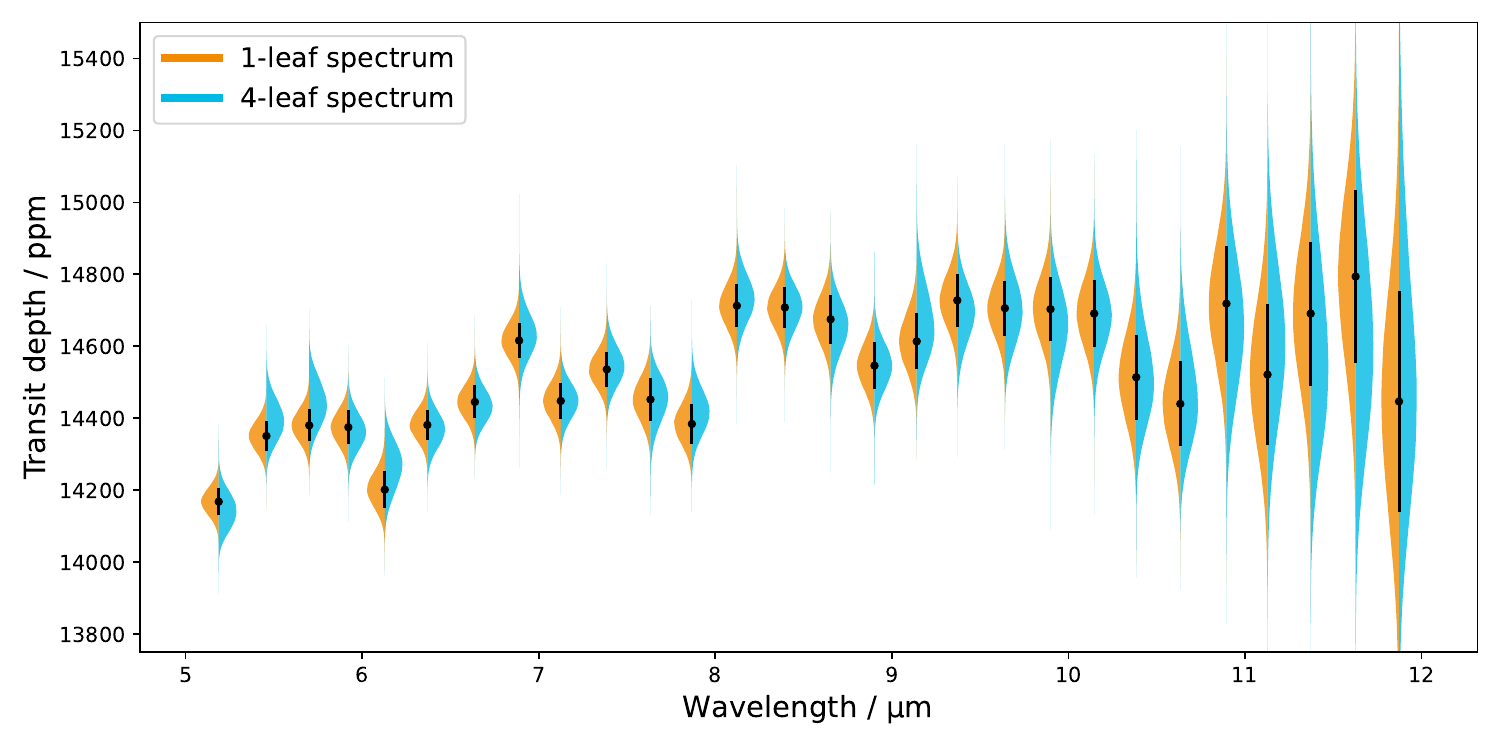}
\caption{Transmission spectra from the 1-leaf (orange) and 4-leaf (light blue) reductions as a violin plot. For each reduction option the transit depth distributions are shown as a function of wavelength. For the classical 1-leaf spectrum, the black data points corresponding to the usual 1$\sigma$ error bars are also shown.}
\label{fig:transmission_spectrum}
\end{figure*}

In Figure \ref{fig:transmission_spectrum}, we display the resulting spectrum from our rule-based tree structure data reduction. We show the typical 1-leaf spectrum, corresponding to \texttt{\_spec3} (or sometimes referred to as \texttt{L3}) in Figure \ref{fig:final_decision_tree_diagram}, as well as the 4-leaf spectrum as a GMM. In our interpretive analysis we primarily use the 1-leaf spectrum to explore and discuss our results as this allows for direct combination with other datasets of this planet. We detail the impact of the 4-leaf spectrum on the MIRI/LRS results in Section\,\ref{sec:4_leaf_retrieval}.

\section{Atmospheric Modelling} \label{sec:modelling}

With the reduced data in hand, we employ a variety of forward model grids and retrievals to interpret the atmospheric composition of HD\,209458b from our MIRI/LRS spectra. We follow a strategy of first performing free retrievals using the ARCiS code~\citep{18OrMi,Min2020}, then examining whether the best-fits drawn from these retrievals can be reproduced by more physically motivated radiative-convective equilibrium and cloud forward models from the \texttt{PICASO} \citep{batalhaNE2019a,Mukherjee2023} and \texttt{Virga} \citep{Batalha2025} codes. 

First, we find the combination of cloud species and morphologies that best explain the LRS data alone, using the 1-leaf spectrum which corresponds to \texttt{\_spec3} in Figure \ref{fig:final_decision_tree_diagram}. Then, we combine together transmission spectra of HD\,209458b from \textit{HST} (STIS+WFC3) \citep{Sing2016}, \textit{JWST} NIRCam~\citep{Xue2024ApJ}, and \textit{JWST} MIRI/LRS (this work) to conduct the most comprehensive analysis of HD\,209458b's atmospheric transmission spectra to date, spanning 0.3~--~12~$\mu$m. The importance of optical wavelength data in atmospheric retrievals has been highlighted by various studies -- in particular \cite{24FaWaMa} showed that wavelengths below 0.6 $\mu$m are important to constrain cloud scattering slopes and alkali species Na and K, with {\planet} used as a case study. The inclusion of spectra in this optical wavelength region thus allows us to probe the particle size via the scattering slope, as well as spectral features of Na and K. Meanwhile, including the NIRCam observations allows us to place tighter constraints on the H$_2$O and CO$_2$ abundances. With the subset of best-fitting cloud species found from the LRS data, we perform further analysis on the full 0.3~--~12~$\mu$m spectrum to constrain the particle size distributions of these cloud species using both the ARCiS and \texttt{PICASO+Virga} approaches. In these approaches we assume that the same aerosols causing the spectral features in the mid-IR are also fully responsible for the optical slope, i.e. we do not include multiple cloud or haze layers of different compositions.

We then examine the physical plausiblity of our best fits in the context of global climate models and microphysical cloud formation models, though we leave full exploration of these more detailed analyses to future work.

\subsection{ARCiS retrievals}\label{sec:ARCiS}

For our retrievals, we use the atmospheric modeling and Bayesian retrieval code ARCiS (ARtful modelling Code for exoplanet Science)~\citep{18OrMi,Min2020}, which uses the Multinest~\citep{Feroz2009,Feroz2019} Monte Carlo nested sampling algorithm. The set of molecules and atoms we consider for our free retievals, with k-tables computed at R~=~$\frac{\lambda}{\Delta\lambda}$~=~1000 from the ExoMolOP database~\citep{20ChRoAl.exo}, are as follows:
H$_2$O~\citep{ExoMol_H2O}, CO$_2$~\citep{ExoMol_CO2}, CO~\citep{15LiGoRo.CO}, NH$_3$~\citep{ExoMol_NH3}, C$_2$H$_2$~\citep{ExoMol_C2H2}, H$_2$S~\citep{ExoMol_H2S}, HCN~\citep{ExoMol_HCN}, CH$_4$~\citep{ExoMol_CH4}, 
Na~\citep{19AlSpLe.broad,KURonline}, K~\citep{16AlSpKi.broad,KURonline}, SO$_2$~\citep{ExoMol_SO2}, SiO~\citep{21YuTeSy}, TiO~\citep{19McMaHo}, and VO~\citep{16McYuTe}. Our model atmosphere consists of 100 layers between 100~bar and 10$^{-9}$~bar, with a H$_2$:He number density ratio of 0.85:0.15 for the background atmosphere. For the MIRI/LRS retrievals we assume an isothermal pressure-temperature profile, but when considering multiple datasets across a wider wavelength we use a parameterised Guillot profile~\citep{10Guillot.exo}, as a wider range of pressures and temperatures contribute to the spectrum. We retrieve a planetary radius at a reference pressure of 10~bar in all instances. We use 1000 live points and a sampling efficiency of 0.3 in Multinest for all retrievals. 
The free parameters and their priors used across our suite of retrieval setups can be found in Table~\ref{tab:arcis_retrievals}.

\begin{table*}[t]
    \centering
    \caption{Summary of free parameters and their priors used in the ARCiS retrievals.}
    \label{tab:arcis_retrievals}
    \begin{tabular}{lccc}
    \hline 
\rule{0pt}{3ex}Parameter & Prior range & Prior type &Description\\
    \hline 
\rule{0pt}{3ex}\textbf{All retrievals:} & & & \\
\rule{0pt}{3ex}$R_p$ & 1.104, 1.656 & Linear & Planet radius ($R_J$)  \\
\rule{0pt}{3ex}VMR(mol) & -12, -1 & Logarithmic & Molecule/atom volume mixing ratio (VMR) \\
\rule{0pt}{3ex}\textbf{Isothermal P-T profile:} & & & \\
\rule{0pt}{3ex}$T_p$  & 400, 2300 & Linear & Temperature (K)  \\
\rule{0pt}{3ex}\textbf{Guillot P-T profile:} & & & \\
\rule{0pt}{3ex}$\gamma_T$  & -2, +2 & Logarithmic & Ratio of mean optical to infrared opacities  \\
\rule{0pt}{3ex}$\beta_T$  & 0, 1 & Linear &  Measure of atmospheric re-circulation efficiency \\
\rule{0pt}{3ex}$\kappa$  & -4, +4 & Logarithmic & Optical depth at infrared wavelengths  \\
\rule{0pt}{3ex}$T_{\rm int}$  & 1, 3 & Logarithmic & Internal planet temperature (K) \\
\rule{0pt}{3ex}\textbf{Clouds:} & & & \\
\rule{0pt}{3ex}$P$  & -7, +2 & Logarithmic &  Central cloud pressure (bar) \\
\rule{0pt}{3ex}$\sigma_P$  & -1, 1 & Logarithmic &  Cloud gaussian width \\
\rule{0pt}{3ex}$\tau$  & -4, 3 & Logarithmic &  Optical thickness at $P$ and $\lambda$~=~9$\mu$m \\
\rule{0pt}{3ex}$C_{\rm cov}$  & 0, 1 & Linear &  Cloud coverage fraction \\
\rule{0pt}{3ex}$r_{\rm cloud}$  & -2, 1 & Logarithmic &  Cloud particle radius ($\mu$m) \\
\rule{0pt}{3ex}\textbf{Mixed material clouds:} & & & \\
\rule{0pt}{3ex}abun  & 0, 1 & Linear &  Relative abundance of one cloud type to another \\
\hline
    \end{tabular}
\end{table*}

The cloud setup for the ARCiS retrievals consists of a Gaussian cloud layer which uses the following parameterisation for $f_{\rm cloud}$, the mass fraction of the cloud particles in the local atmosphere (i.e., $\frac{\rho_{\rm cloud}}{\rho_{\rm gas}}$) as a function of pressure $P$: 
	\begin{equation}\label{eq:cloud_layer_arcis}
		f_{\rm cloud} = \frac{g_p \tau_{\rm cloud}}{\kappa_{\rm cloud} P \sigma_P \sqrt{2\pi}} \exp\left({\frac{-1}{2 \sigma_P^2} \left[log{\frac{P}{P_0}}\right]^2}\right)
	\end{equation}
	Here, $g_p$ is the gravitational acceleration of the planet (cm/s$^2$) (where we use $\log{g_p}$=2.9708 (cgs)~\citep{kataria2016}),
and $\kappa_{\rm cloud}$ is the mass extinction coefficient of the cloud particles (cm$^2/g$) at 9~$\mu$m. Central cloud pressure $P$, the width of the gaussian cloud layer distribution $\sigma_P$, and optical depth $\tau$ (at a reference wavelength of 9~$\mu$m and reference pressure of $P_0$=10~bar), are all retrieval parameters.
We also include a cloud coverage parameter $C_{\rm cov}$ to allow for patchy clouds, implemented in the code as a linear combination of clear and cloudy model atmospheres. For the cloud opacity, we use the species and corresponding refractive indices given in Table~\ref{tab:arcis_clouds}; where more than one cloud species is included in a retrieval, we mix the refractive indices together using effective medium theory, using the standard Bruggeman mixing rule~\citep{35Bruggeman} (see Eq. 26 of \cite{23ChStHe}), and retrieve an abundance ratio of one species relative to the other (we do not mix more than two species). We retrieve for a single particle size across our cloud layer with no size distribution, and consider spherical particles only in these retrievals. 
Further details of the cloud setup used here can be found in the ARCiS GitHub documentation\footnote{\url{https://github.com/michielmin/ARCiS.git}}, with a similar setup for the Gaussian cloud layer described in the supplementary information of \cite{24DyMiDe}. For the retrievals on the full observational dataset where we use a Guillot profile for the pressure-temperature structure, we set an internal temperature of 280~K based on the calculations of \cite{24WiChMa}. We do not find our results to be sensitive to the internal temperature.

\begin{table*}[t]
\raggedright
\begin{ThreePartTable}
    \caption{Summary of cloud species and their source for refractive indices that we consider in our ARCiS retrievals, along with $\ln(Z)$ and BPICS~\citep{thorngren2025bayesian} for each retrieval on our LRS 1-leaf spectra, and on the LRS+HST+NIRCam (full dataset) for the most preferred LRS fits. The letter in brackets after the material refers to amorphous (A), crystalline (C), or metallic (M). A model is more preferred for larger values of $\ln(Z)$ and for smaller (more negative) values of BPICS. $\Delta$$\ln(Z)$ and $\Delta$BPICS are measured against the forsterite model highlighted in bold, with a higher $\Delta$$\ln(Z)$ (more negative $\Delta$BPICS) meaning the forsterite model is more preferred over the given model. We follow \cite{thorngren2025bayesian}: $\Delta$$\ln(Z)$$\geq$5.91 ($\Delta$BPICS$\leq$-11.82) can roughly be thought of as a ``detection'', and $\Delta$$\ln(Z)$$\geq$14.37 ($\Delta$BPICS$\leq$-28.74) a stronger detection or ``discovery''.  
    The combinations which include amorphous MgSiO$_3$ and Mg$_2$SiO$_4$ are using the refractive indices from \cite{dorschner1995} and \cite{Scott1996}, respectively. In bold are the two cloudy models we present results for in this work. 
    }
    \label{tab:arcis_clouds}
	\begin{tabular}{lllllllll} 
		\hline
\rule{0pt}{3ex} Material &	Name &	$\ln(Z)$ 	&	$\Delta$$\ln(Z)$&	 $\ln(Z)$ 	&	BPICS	&	$\Delta$BPICS	&	BPICS	&	 Reference  \\

\rule{0pt}{3ex} 	&	  	&	(LRS) 	&	(LRS)	&	 (Full) 	&	(LRS)	&	(LRS)	&	(full)	&	   \\

\hline																
  \rule{0pt}{3ex}Al$_2$O$_3$  (A)	&	 Corundum 	&	197.08	&	3.43	&	 	&	-408.16	&	-6.56	&		&	\cite{95KoKaYa}\tnote{a} \\
    \rule{0pt}{3ex}Al$_2$O$_3$  (C)	&	 Corundum 	&	198.28	&	2.23	&	 	&	-410.08	&	-4.64	&		&	\cite{13ZePoMu}\tnote{a,b}  \\
    \rule{0pt}{3ex}Fe (M)	&	 Iron  	&	181.04	&	19.47	&	 	&	-382.14	&	-32.58	&		&	 \cite{12Palik}\tnote{a} \\
           \rule{0pt}{3ex}MgSiO$_3$ (A)	&	 Enstatite  	&	198.92	&	1.59	&	 	&	-413.83	&	-0.89	&		&	 \cite{Scott1996}\tnote{c} \\
    \rule{0pt}{3ex}MgSiO$_3$ (A)	&	 Enstatite  	&	199.1	&	1.41	&	1740.72	&	-414.71	&	-0.01	&	-3553.74	&	 \cite{dorschner1995}\tnote{a,d} \\
    \rule{0pt}{3ex}MgSiO$_3$ (C) 	&	 Enstatite  	&	196.26	&	4.25	&	  	&	-413.36	&	-1.36	&		&	 \cite{jaeger1998}\tnote{a} \\ 
   \rule{0pt}{3ex}Mg$_2$SiO$_4$ (A)	&	 Forsterite 	&	197.94	&	2.57	&	 	&	-408.97	&	-5.75	&		&	 \cite{Scott1996}\tnote{c} \\
    \rule{0pt}{3ex}\textbf{Mg$_2$SiO$_4$ (A)} 	&	 \textbf{Forsterite} 	&	 \textbf{200.51} 	&	\textbf{0}	&	 \textbf{1759.46} 	&	\textbf{-414.72}	&	\textbf{0}	&	\textbf{-3558.82}	&	 \cite{03JaDoMu}\tnote{a,e} \\
      \rule{0pt}{3ex}Mg$_2$SiO$_4$ (C) 	&	 Forsterite 	&	194.7	&	5.81	&	 	&	-402.59	&	-12.13	&		&	 \cite{Eckes2013}\tnote{a,f} \\ 
   \rule{0pt}{3ex}Mg$_2$SiO$_4$ (C) 	&	 Forsterite 	&	198.92	&	1.59	&	 	&	-414.7	&	-0.02	&		&	 \cite{Eckes2013}\tnote{a,g} \\ 
  \rule{0pt}{3ex}SiO$_2$ (A)	&	 Quartz  	&	196.38	&	4.13	&	 	&	-385.07	&	-29.65	&		&	 \cite{97HeMu,12Palik}\tnote{g} \\
    \rule{0pt}{3ex}SiO$_2$ (C) 	&	 $\beta$-Quartz  	&	195.03	&	5.48	&	 	&	-376.6	&	-38.12	&		&	 \cite{13ZePoMu}\tnote{b,g} \\
  \rule{0pt}{3ex}SiO (A)	&	 Silicon oxide 	&	197.76	&	2.75	&	 	&	-406.96	&	-7.76	&		&	\cite{12Palik,wetzel2013}\tnote{h}\\
    \rule{0pt}{3ex}Clear 	&	  	&	183.88	&	16.63	&	1670.26	&	-382.57	&	-32.15	&	-3367.12	&	\\
\rule{0pt}{3ex}Gray cloud 	&	  	&	178.25	&	22.26	&	1725.94	&	-386.2	&	-28.52	&	-3485.61	&	\\						
  \hline															
  \hline															
  \rule{0pt}{3ex}Combinations & & & & & & & \\
  \hline															
 \multicolumn{2}{l}{\rule{0pt}{3ex}\textbf{MgSiO$_3$+Mg$_2$SiO$_4$ (A)}}	 	&	 \textbf{202.54}  	&	\textbf{-2.03}	&	\textbf{1758.47}	&	\textbf{-418.5}	&	\textbf{3.78}	&	\textbf{-3550.7} & 	\\
  \multicolumn{2}{l}{\rule{0pt}{3ex}MgSiO$_3$+SiO$_2$ (A)}	  	&	197.65	&	2.86	&	1742.32	&	-388.53	&	-26.19	&	-3506.04 &	\\
\hline
\end{tabular}
\begin{tablenotes}
\footnotesize
\item[a] Downloaded from POSEIDON aerosol database~\citep{Mullens2024} \url{https://poseidon-retrievals.readthedocs.io/en/latest/content/opacity_database.html}
\item[b] Measured at 928~K
\item[c] Taken directly from Table 1 of the original paper 
\item[d] Used a melting and quenching technique
\item[e] Laboratory data used the sol-gel method
\item[f] Measured at 295~K, refractive indices averaged over each crystallographic index
\item[g] Measured at 1000~K, refractive indices averaged over each crystallographic index
\item[h] Downloaded from LX-Mie \url{https://github.com/NewStrangeWorlds/LX-MIE/tree/master/compilation}
\end{tablenotes}
\end{ThreePartTable}
\end{table*}

\noindent
\begin{table*}[ht!]
\caption{Retrieved atmospheric parameters from this work compared with previous studies for our free chemistry retrievals (upper table) and equilibrium chemistry retrievals (lower table). ``Combo'' refers to our combined \ce{Mg2SiO4} and \ce{MgSiO3} retrievals, with \ce{MgSiO3} fraction defined as \ce{MgSiO3} / (\ce{MgSiO3} + \ce{Mg2SiO4}). $P_{\rm midcloud}$ is the central cloud pressure and $P_{\rm cloudtop}$ the cloud top pressure. We assume a Jupiter radius, $R_{\rm Jup}$, of 69,911~km. Where \textit{HST} is labelled this includes WFC3+STIS, and N stands for NIRCam. 
The equilibrium chemistry results of \cite{25VeGoAv} given here are from their Table 5, with O kept fixed and C changed when computing C/O, as is done in our retrievals. 
The results of \cite{25BaKrMo} are using a new HST data reduction and are for their (a) cloudy+CO prior and (b) cloudy with H$_2$O and CO$_2$ only retrievals. 
}
\label{tab:ret_pars}
\raggedright
\footnotesize
\begin{tabular}{l|cccccccc}
\toprule 
Free chem & \ce{Mg2SiO4}  & Combo  & \ce{Mg2SiO4}  & Combo & \ce{Mg2SiO4} & \cite{25VeGoAv} & \multicolumn{2}{c}{\cite{25BaKrMo}}  \\
& LRS & LRS & full & full & N+HST & N+HST & N+HST(a) & N+HST(b)\\

\midrule
$R_p$ ($R_{\rm Jup}$)                 & $1.32^{+0.02}_{-0.04}$ & $1.31^{+0.02}_{-0.02}$ & $1.20^{+0.02}_{-0.02}$ &$1.22^{+0.02}_{-0.02}$  & $1.31^{+0.02}_{-0.02}$ & $1.40^{+0.00}_{-0.00}$ & $1.36^{+0.01}_{-0.01}$ & $1.36^{+0.01}_{-0.01}$ \\
$T$ at $10^{-3}$ bar (K) & $1424^{+354}_{-507}$ & $1373^{+347}_{-270}$ & $1961^{+278}_{-204}$ & $1614^{+124}_{-127}$ & $1353^{+130}_{-129}$   & $\sim1300$ & $1094^{+112}_{-119}$ & $1177^{+121}_{-119}$  \\
$\log\,P_{\rm midcloud}$ (bar)   & $-3.30^{+2.20}_{-1.70}$& $-4.08^{+2.17}_{-1.53}$& $-2.21^{+0.67}_{-0.79}$ & $-2.95^{+0.46}_{-0.59}$ &  $-1.62^{+0.81}_{-0.89}$ & - & - & -\\
$\log\,P_{\rm cloudtop}$ (bar)    & - & - & - & - & - & $-2.61^{+0.53}_{-0.63}$ & $-2.84^{+0.40}_{-0.40}$ & $-1.47^{+0.25}_{-1.00}$\\
$r_{\mathrm{eff}}$ ($\mu$m)         & $0.015^{+0.062}_{-0.012}$ & $0.009^{+0.031}_{-0.007}$ & $0.16^{+0.02}_{-0.02}$ & $0.11^{+0.01}_{-0.01}$  & $0.19^{+0.14}_{-0.03}$  & -  & - & - \\
Cloud fraction                       & $0.58^{+0.18}_{-0.19}$ & $0.50^{+0.20}_{-0.16}$ & $0.73^{+0.03}_{-0.03}$ & $0.73^{+0.03}_{-0.03}$ & $0.61^{+0.06}_{-0.10}$  & $0.53^{+0.09}_{-0.09}$ & - & -\\
\ce{MgSiO3} fraction             & - & $0.14^{+0.14}_{-0.07}$ & - &  $0.90^{+0.07}_{-0.12}$ & -  & - & - & - \\
\hline\noalign{\vskip 2pt} 
$\log\,$VMR(H$_2$O)                  & $-4.54^{+1.15}_{-1.06}$ & $-4.35^{+1.11}_{-0.91}$ & $-4.28^{+0.21}_{-0.17}$ & $-4.54^{+0.13}_{-0.13}$ & $-4.86^{+0.21}_{-0.24}$ & $-4.53^{+0.32}_{-0.29}$ & $-2.80^{+0.42}_{-0.41}$ & $-4.51^{+1.15}_{-0.48}$ \\
$\log\,$VMR(CO$_2$)                  & $-7.87^{+2.30}_{-2.34}$ & $-7.55^{+2.36}_{-2.60}$ & $-6.61^{+0.17}_{-0.18}$ & $-6.52^{+0.23}_{-0.20}$ & $-7.58^{+0.24}_{-0.23}$ & $-7.52^{+0.30}_{-0.26}$ & $-6.16^{+0.38}_{-0.39}$ & $-7.67^{+1.07}_{-0.36}$ \\
$\log\,$VMR(CO)                     & $-8.88^{+2.13}_{-1.93}$ & $-8.50^{+2.24}_{-2.14}$ & $-9.44^{+1.84}_{-1.63}$ & $-9.11^{+1.92}_{-1.90}$ & $-9.40^{+1.76}_{-1.69}$ & $-9.21^{+1.84}_{-1.84}$ & $-4.15^{+0.51}_{-0.30}$ & - \\
$\log\,$VMR(Na)                      & - & - & $-6.10^{+0.27}_{-0.24}$ & $-6.14^{+0.22}_{-0.22}$ & $-6.50^{+0.23}_{-0.24}$ & $-5.67^{+0.51}_{-0.45}$ & $-7.88^{+3.65}_{-3.39}$  & - \\
$\log\,$VMR(K)                     & - & - & $-6.52^{+0.33}_{-0.32}$ & $-6.72^{+0.29}_{-0.29}$ & $-7.44^{+0.42}_{-0.46}$ & $-6.93^{+0.54}_{-0.47}$ & -  & - \\
\end{tabular}
\begin{tabular}{l|c c c c}
\toprule
Eq chem
& \makecell{\ce{Mg2SiO4}\\full}
& \makecell{\cite{25VeGoAv}\\NIRCam+WFC3}
& \makecell{\cite{25VeGoAv}\\NIRCam+HST}
& \makecell{\cite{Xue2024ApJ}\\NIRCam+WFC3} 
\\
\midrule
$R_p$ ($R_{\rm Jup}$)                 & $\;\;1.35^{+0.01}_{-0.01}$ & $1.367^{+0.005}_{-0.005}$ & $1.384^{+0.004}_{-0.006}$ & $\;\;1.34^{+0.01}_{-0.01}$ \\
$T$ at $10^{-3}$ bar (K)             & $894^{+35}_{-38}$        & $1306^{+83}_{-81}$ &$\sim1100$ & $1290^{+83}_{-81}$ \\
$\log\,P_{\rm midcloud}$ (bar)       & $-4.22^{+0.27}_{-0.24}$   & - & - \\
$\log\,P_{\rm cloudtop}$ (bar)       & -                         & $-4.00^{+0.38}_{-0.35}$ & $-0.49^{+1.64}_{-1.69}$ & $-3.68^{+0.45}_{-0.44}$ \\
$r_{\mathrm{eff}}$ ($\mu$m)          & $\;\;0.13^{+0.02}_{-0.01}$ & - & - & -\\
Cloud fraction                       & $\;\;0.60^{+0.04}_{-0.03}$ & $0.80^{+0.09}_{-0.08}$ &$0.56^{+0.05}_{-0.05}$ & $\;\;0.82^{+0.09}_{-0.09}$ \\
\hline\noalign{\vskip 2pt}
C/O                                  & $\;\;0.16^{+0.06}_{-0.04}$ & $0.20^{+0.12}_{-0.09}$ & $0.56^{+0.10}_{-0.12}$  & $\;\;0.23^{+0.12}_{-0.15}$ \\
Z [M/H] (dex)                                & $\;\;0.18^{+0.11}_{-0.11}$ & $0.84^{+0.25}_{-0.26}$ & $0.13^{+0.20}_{-0.16}$ & $\;\;0.69^{+0.34}_{-0.25}$ \\
\bottomrule
\end{tabular}
\end{table*}

\subsection{Identifying cloud species from their vibrational modes between 5--12 $\mu$m} \label{sec:cloud_species}
We run retrievals on the LRS 1-leaf dataset alone to determine whether we can identify the presence of clouds in the spectrum, and if so, whether we can pinpoint the particular species contributing to any cloud features. In order to assess the strength of our models according to how well they fit the data, we use the Nested Sampling Global Log-Evidence from Multinest~\citep{Feroz2019} as $\ln(Z)$. This already penalizes a model with more free parameters compared to one with fewer, so the difference between $\ln(Z)$ of two retrievals gives the log Bayes factor ($\ln{B_{12}}$=$\Delta$$\ln(Z)$), which can be interpreted as a measure of how one model compares to another. \cite{thorngren2025bayesian} recommend $\ln(Z)$ is used alongside the Bayesian Predictive Information Criterion Simplified (BPICS), which is calculated directly from the posterior likelihoods. We therefore also compute BPICS for each model, using P$_{D2}$ (Eq. 26 of \cite{thorngren2025bayesian}) in the BPICS computation.  We use the suggested interpretation of \cite{thorngren2025bayesian} (see their Table 1) to assess the significance of $\Delta$$\ln(Z)$ and $\Delta$BPICS of each of our models compared to our most favoured (highest $\ln(Z)$ model): $\Delta$$\ln(Z)$$\geq$5.91 ($\Delta$BPICS$\leq$-11.82) can roughly be thought of as a ``detection'', and $\Delta$$\ln(Z)$$\geq$14.37 ($\Delta$BPICS$\leq$-28.74) a stronger detection or ``discovery''. 

For the LRS retrievals we use a base clear atmosphere which contains H$_2$O, CO$_2$, and CO. We do not find any evidence for the addition of any of the other molecules listed in Section~\ref{sec:ARCiS} which have absorption features in the LRS wavelength range (i.e. NH$_3$, SO$_2$, SiO, C$_2$H$_2$, HCN, CH$_4$). The inclusion of these species all give $\Delta$$\ln(Z)$$\leq$0 for the cloudy atmosphere compared to the base atmosphere with only H$_2$O, CO$_2$, and CO. For the clear atmosphere the only positive Bayes factors were for the addition of SiO ($\Delta$$\ln{Z}$~=~2.2) and NH$_3$ ($\Delta$$\ln{Z}$~=~1.4). We tested the evidence for a clear atmosphere with SiO and NH$_3$ included against a cloudy (\ce{Mg2SiO4}) atmosphere without them, and find a Bayes factor of 15 in favour of the cloudy atmosphere, classified as strong evidence according to \cite{thorngren2025bayesian}. We therefore do not find significant evidence to support the inclusion of SiO or NH$_3$ in our presented models. 

We consider refractive indices for different structures of the solid cloud species given in Table~\ref{tab:arcis_clouds}, i.e., that of well-ordered crystalline or randomly ordered amorphous glass materials. The species tested include silica (\ce{SiO2}), the forsterite end-member composition of the olivine series (\ce{Mg2SiO4}), the enstatite end-member of the pyroxene series (\ce{MgSiO3}), SiO which has been suggested as a nucleation phase of silicate clouds \citep{Molliere2025SiO}, and both aluminum oxide and iron particles, though these latter two are expected to be sequestered at depth at this planet's equilibrium temperature \citep[e.g.,][]{visscher2010}.

We find (see Table~\ref{tab:arcis_clouds}) that the ARCiS retrievals on the LRS spectra disfavor gray clouds ($\ln(Z)$~=~178.25) compared to a clear atmosphere model ($\ln(Z)$~=~183.88), with $\Delta$$\ln(Z)$~=~5.6 for the cloud-free over gray models. This is not surprising, as without any cloud-specific features, gray clouds give no improvement to the fit and therefore there is no reason to include the extra free parameters required for parameterizing the gray clouds in the retrieval. 
When including wavelength-dependent species-specific cloud refractive indices, however, ARCiS prefers magnesium silicates ($\ln(Z)$~=~200.51 for amorphous Mg$_2$SiO$_4$ and $\ln(Z)$~=~202.54 for amorphous MgSiO$_3$+Mg$_2$SiO$_4$) over a clear atmosphere, with $\Delta$$\ln(Z)$ of 16.63 and 18.66, respectively. This leads to a strong detection of these magnesium silicate clouds over either a clear atmosphere or gray clouds according to Table 1 of \cite{thorngren2025bayesian}. Table~\ref{tab:ret_pars} gives the retrieved atmospheric parameters for our amorphous \ce{Mg2SiO4} and amorphous \ce{Mg2SiO4}+\ce{MgSiO3} (``combo'') LRS retrievals which are highlighted in bold in Table~\ref{tab:arcis_clouds}. We compare these values to other works in Section~\ref{sec:compare}. 

\begin{figure}
\centering
\includegraphics[width=0.5\textwidth]{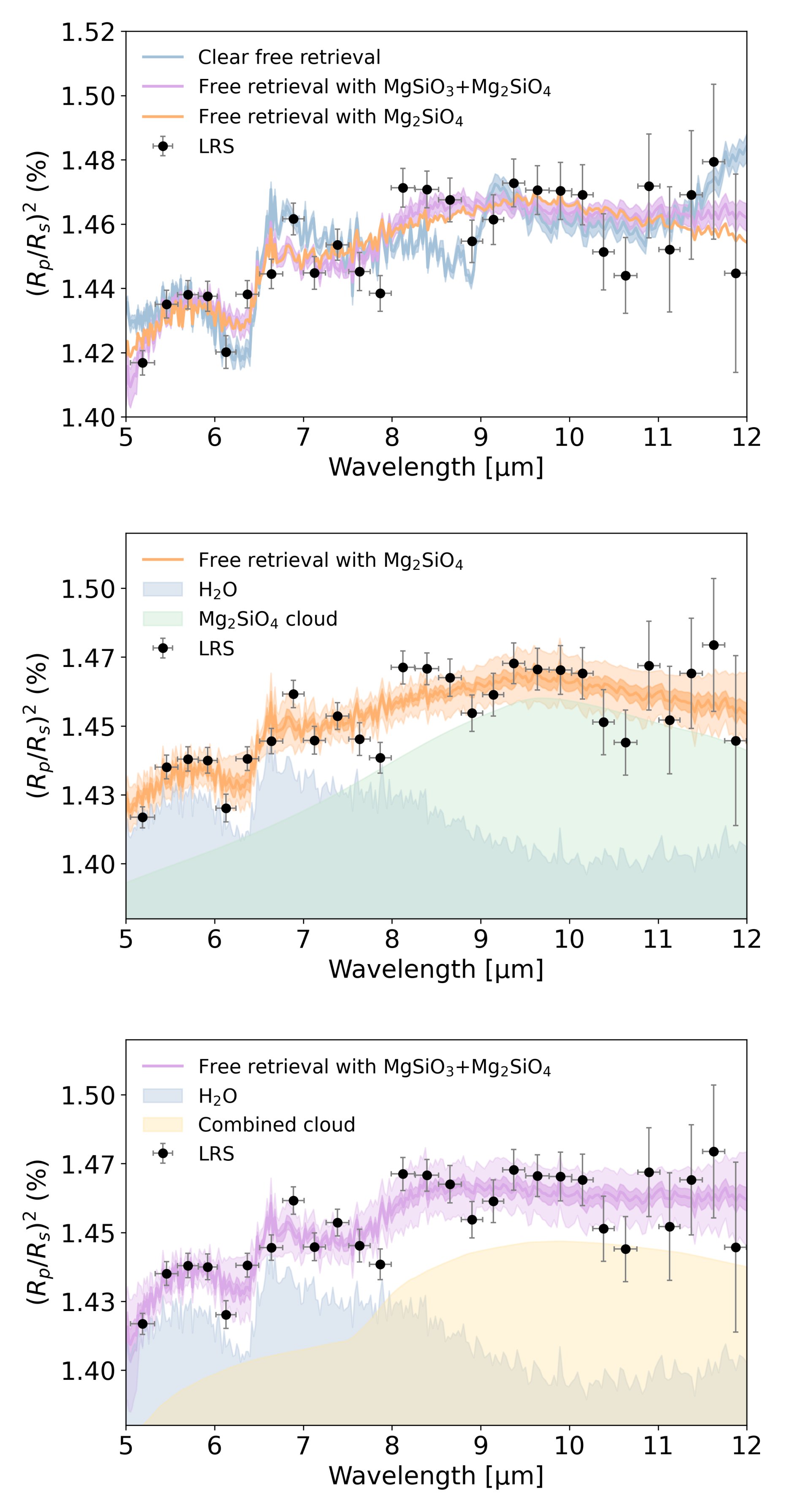}
\caption{Upper: retrieved 1-leaf LRS transmission spectra for: Mg$_2$SiO$_4$ clouds (orange), Mg$_2$SiO$_4$+MgSiO$_3$ clouds (purple), and a cloud-free atmosphere (blue). Middle: opacity contributions for the Mg$_2$SiO$_4$ cloud setup. Lower: opacity contributions for the Mg$_2$SiO$_4$+MgSiO$_3$ cloud setup. }
\label{fig:LRS_all3}
\end{figure}

As shown in Table~\ref{tab:arcis_clouds}, the magnesium silicates are the most preferred of the tested species, although only slightly preferred over \ce{Al2O3} or SiO according to Table~1 of \cite{thorngren2025bayesian}. We discuss the plausability of different species in terms of the full atmospheric context in Section~\ref{sec:plausibility}. Comparing the magnesium silicates to each other, we find that we do not significantly prefer \ce{Mg2SiO4} (i.e., forsterite) over amorphous \ce{MgSiO3} (i.e., enstatite) ($\Delta$$\ln(Z)$=1.4) when considering the LRS spectra only.  
Also as per Table~\ref{tab:arcis_clouds}, we cannot distinguish between particle morphology to high statistical significance given the data signal-to-noise, particularly when using the \ce{Mg2SiO4} crystalline refractive indices of \cite{Eckes2013} which were measured at 1000~K, as the higher temperatures smooth out the sharp crystalline features.
When considering a combination of \ce{Mg2SiO4} and \ce{MgSiO3}, we find a slight but not high confidence preference for the combination (which introduces one extra free parameter for the abundance) over \ce{Mg2SiO4} alone when considering the LRS data in isolation. The peak of \ce{Mg2SiO4}'s Si-O stretching mode is slightly redder than that of \ce{MgSiO3}, which likely leads to our preference for forsterite over enstatite given the absorption feature peaking at $\sim$9.5~$\mu$m in the LRS data.  While the cloud composition inferred from the LRS data alone is most consistent with magnesium silicates, the gas phase abundances of the overall atmospheric chemistry and the particle size distributions of these clouds are relatively unconstrained, as shown in Figure~\ref{fig:cornerplots_all4}. Figure~\ref{fig:LRS_all3} gives a comparison of our retrieved spectra for the LRS data alone from atmospheres with: \ce{Mg2SiO4}, a combination of \ce{Mg2SiO4} and \ce{MgSiO3}, and a clear atmosphere. We note that different refractive indices of \ce{Mg2SiO4} for the \ce{Mg2SiO4}-only and the \ce{Mg2SiO4} and \ce{MgSiO3} combination are used for the middle and lower panels; see Appendix~\ref{sec:cloud_refrinds} for a discussion on the different refractive indices.

\subsection{The constraining power of the full 0.3-12 $\mu$m spectrum} \label{sec:full_power}

After investigating different cloud species using the LRS data alone, we added in the NIRCam data of \citet{Xue2024ApJ} and \textit{HST} (STIS and WFC3) data of \citet{Sing2016} to further put constraints on the overall composition of the atmosphere. We applied our best-fitting single (i.e. amorphous \ce{Mg2SiO4}) and combined (amorphous \ce{Mg2SiO4}+\ce{MgSiO3}) cloud composition retrievals to the full dataset; these are highlighted in bold in Table~\ref{tab:arcis_clouds}. We also test the combination \ce{MgSiO3}+\ce{SiO2}, and \ce{MgSiO3} alone, to interpret our results in the context of \cite{Calamari2024} and \cite{Calamari2026} (see Section~\ref{sec:plausibility}).

We include Na and K opacities as we are now including the HST data, and we allow for wavelength-independent scaling between the observational data (\textit{HST}, NIRCam, and LRS), keeping NIRCam as the anchor. We treat the \textit{HST} data as a single dataset from the original study of \cite{Sing2016}. This study used a consistent set of system parameters and data analysis treatments to analyse the full wavelength range. In previous studies an offset was not included or required when fitting models (see, for example, \cite{Barstow2017,macdonald2017,Pinhas2019}) and thus to reduce the computational complexity we use the dataset as complete and already calibrated to each other. We find, as expected, that the visible-to-NIR infrared \textit{HST} data allow us to place tighter constraints on the particle size of the cloud to $\sim$0.1\,$\upmu$m at $\sim$1--10~millibar pressures. The strong signatures of K and Na in the STIS data further allows us to infer a cloud coverage fraction of 73$\pm$3\%, as some patchiness lets the wings of the alkali gases manifest in the data. With NIRCam and LRS together, we can constrain the log(\ce{H2O}) and log(\ce{CO2}) VMRs tighter than with LRS alone, as can be seen in Table~\ref{tab:ret_pars}. For example, we find $\log\,$VMR(H$_2$O) of $-4.54^{+1.15}_{-1.06}$ with LRS, and $-4.28^{+0.21}_{-0.17}$ with the full dataset, for our \ce{Mg2SiO4} retrieval.
As such, we are able to show that near-IR data over the 2--5 $\mu$m range -- as provided in this context by NIRCam -- is required to fully contextualize the chemistry that gives rise to the clouds observed in the mid-infrared. We also run a retrieval on the full data with \ce{Mg2SiO4} which assumes equilibrium chemistry, using GGchem~\citep{Woitke2018} coupled to ARCiS for computing the equilibrium chemistry abundances. In these equilibrium chemistry retrievals we use the full list of molecules considered for the free retrievals (Section~\ref{sec:ARCiS}), in addition to the following species: AlO~\citep{AlO_ExoMol}, BeH~\citep{BeH_ExoMol}, 
CH~\citep{14MaPlVa.CH}, 
CS~\citep{CS_ExoMol}, H$_2$CO~\citep{H2CO_ExoMol},
HCl~\citep{13LiGoHa.HCl}, 
N$_2$O~\citep{N2O_ExoMol},
NO~\citep{NO_ExoMol}, NS~\citep{NS_ExoMol}, OH~\citep{16BrBeWe.OH}, PS~\citep{PS_ExoMol}, 
SiH~\citep{SiH_ExoMol},
SiS~\citep{SiS_ExoMol}, 
SO$_3$~\citep{SO3_ExoMol}, SO~\citep{SO_ExoMol}.
Under the assumption of equilibrium chemistry, we find a metallicity (M/H) of 0.18$^{+0.11}_{-0.01}$ dex and C/O of 0.16$^{+0.06}_{-0.04}$, which agrees within uncertainties to the value of M/H found by \cite{25VeGoAv} when they consider NIRCam and HST/WFC3+STIS observations, and to the values of C/O for \cite{25VeGoAv} and \cite{Xue2024ApJ} when they consider NIRCam and HST/WFC3 observations only. We find that our free retrieval on the full data with \ce{Mg2SiO4} is strongly preferred ($\Delta$$\ln{Z}$~=~42.96, $\Delta$BPICS~=~-99.06) over the constrained equilibrium chemistry retrieval, indicating that there may be some disequilibrium processes impacting the atmosphere.

We test for the inclusion of other molecules (i.e. CH$_4$, HCN, C$_2$H$_2$, SO$_2$, NH$_3$, H$_2$S, SiO, TiO, VO) in addition to our base atmosphere of H$_2$O, CO$_2$, CO, Na, and K, for our free retrievals on the full dataset including the NIRCam and \textit{HST} data. 
Of these, only C$_2$H$_2$ ($\Delta$$\ln{Z}$~=~3.2) and HCN ($\Delta$$\ln{Z}$~=~4.1) had Bayes factors greater than 0 for the cloudy retrievals, and only C$_2$H$_2$ ($\Delta$$\ln{Z}$~=~2.3), HCN ($\Delta$$\ln{Z}$~=~3.4) and NH$_3$ ($\Delta$$\ln{Z}$~=~13.2) had a Bayes factor greater than 0 for the clear retrievals. It is notable that this high Bayes factor reduces to -2.3 for the cloudy (\ce{Mg2SiO4}) retrievals, and we find a significant Bayes factor of 71.6 for the cloudy over the clear full dataset retrievals when NH$_3$ is included in both. We therefore do not find overall evidence for the inclusion of NH$_3$ in our retrievals.

Finally, we find that we cannot significantly distinguish between \ce{Mg2SiO4} (forsterite) only and a combined magnesium silicate (forsterite and enstatite) cloud when the full 0.3 -- 12 $\mu$m spectrum is considered ($\Delta$ln(Z)~=~2.0; $\Delta$BPICS~=~8.1).  
Thus, when all the available transmission data is taken collectively, our retrievals show evidence for a 
sub-solar C/O atmosphere with small ($\sim$0.1~micron) \ce{Mg2SiO4} or combined \ce{Mg2SiO4} and \ce{MgSiO3} clouds at $\sim$1--10~millibar pressure levels.

\begin{figure*}
\centering
\includegraphics[width=0.9\textwidth]{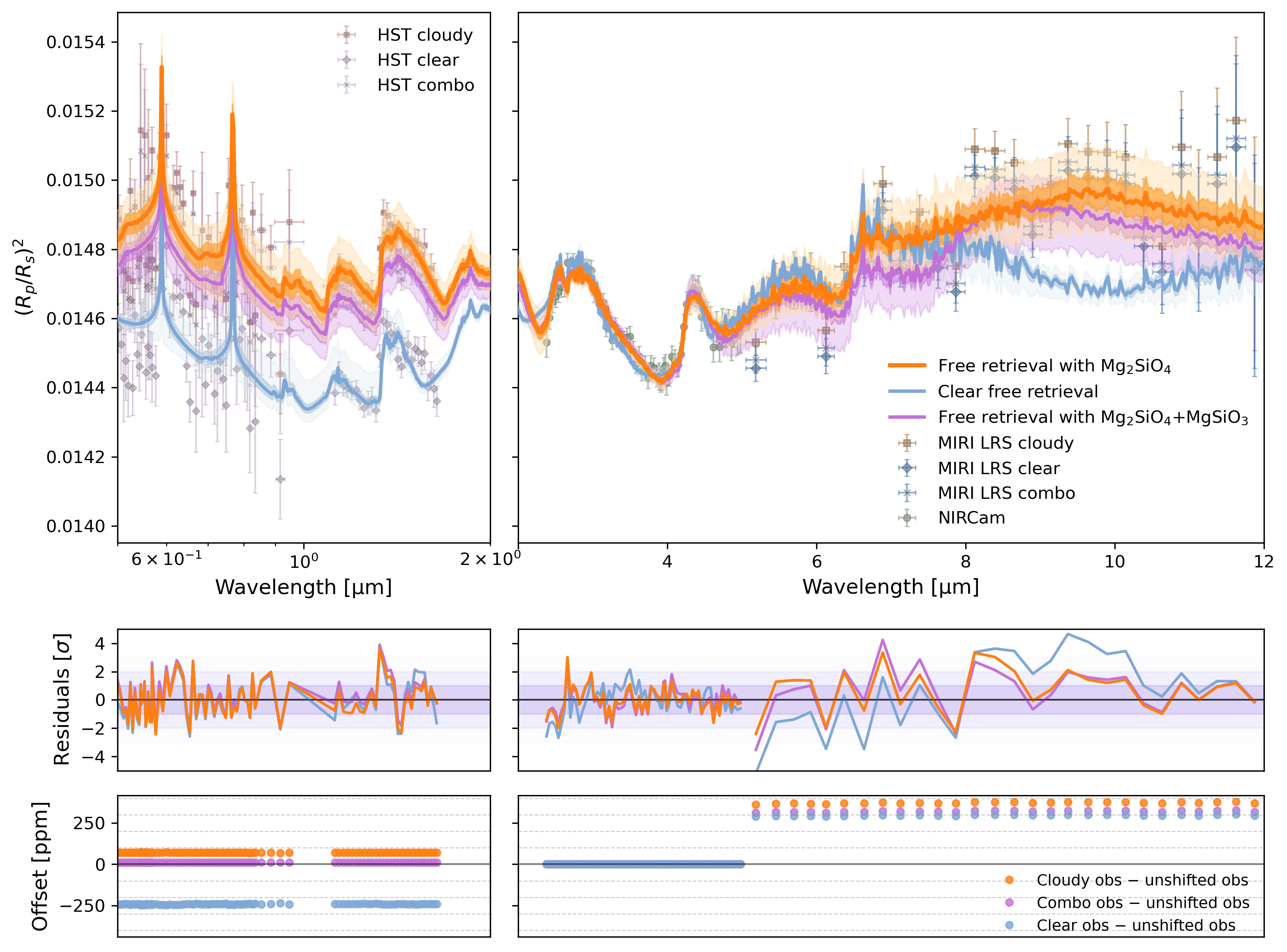}
\includegraphics[width=0.75\textwidth]{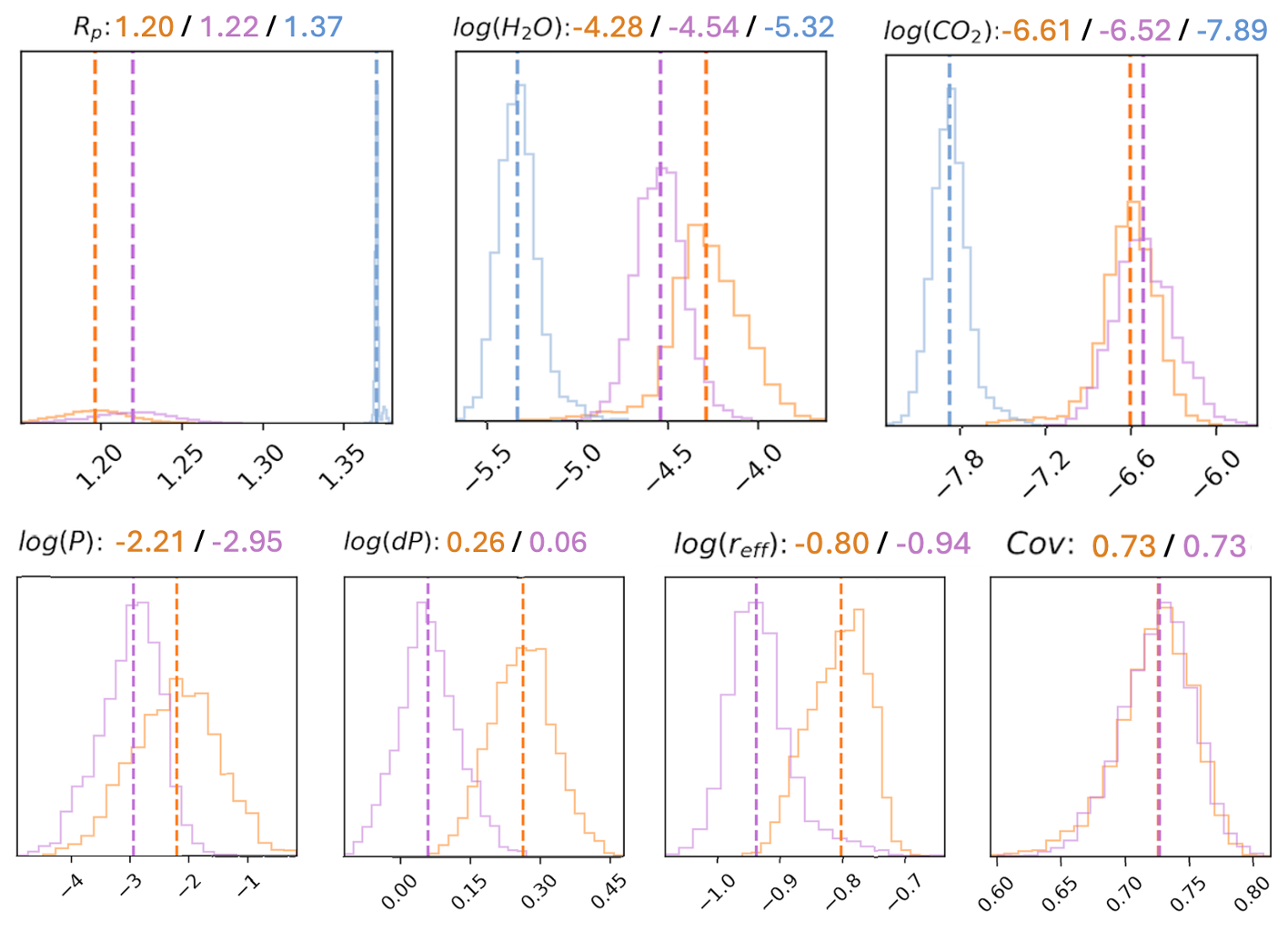}
\caption{Free retrieval of the LRS+NIRCam+HST spectra using: amorphous Mg$_2$SiO$_4$ (``cloudy'', orange), Mg$_2$SiO$_4$+MgSiO$_3$ (``combo'', purple), and a clear atmosphere (blue). The residuals for each model are given in the panel beneath, followed by the offset of the observations used in the retrievals compared to the unscaled observations (where NIRCam is kept anchored).
The posterior panels use the labels described in Table~\ref{tab:arcis_retrievals}, with the median values listed in the following order: Mg$_2$SiO$_4$ (orange), Mg$_2$SiO$_4$+MgSiO$_3$ (purple), and clear (blue), where applicable.}
\label{fig:all_data_residuals}
\end{figure*}

\begin{figure}
\centering
\includegraphics[width=0.5\textwidth]{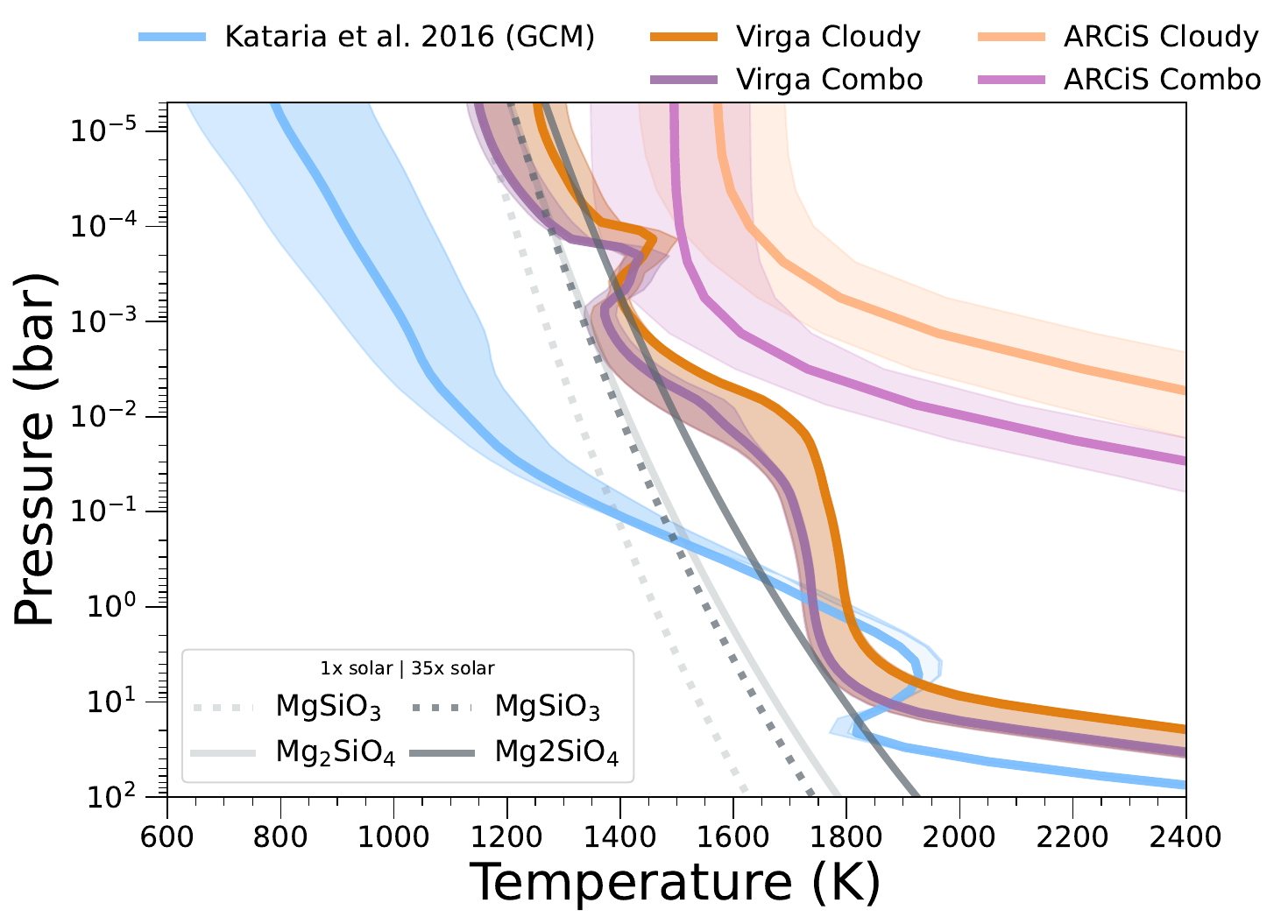}
\caption{Magnesium silicate cloud condensation curves (at 1$\times$ solar metallicity, faint lines; and best-fit Virga metalllicity, dark lines), along with retrieved pressure-temperature profiles from the best-fit, \ce{Mg2SiO4} (oranges) and combo (\ce{Mg2SiO4}/\ce{MgSiO3}, purples) ARCiS and Virga retrievals of the \textbf{LRS+HST+NIRCam} spectra, with shaded regions corresponding to the upper and lower 1$\sigma$ bounds. Also shown is the range of pressure-temperatures from the 
{\planet} GCM of \citet{kataria2016} (blue), with the global average temperature shown in bold and shaded regions corresponding to the day- and night-side temperatures.}
\label{fig:PT_profiles_full_data}
\end{figure}

\begin{figure}
\centering
\includegraphics[width=0.5\textwidth]{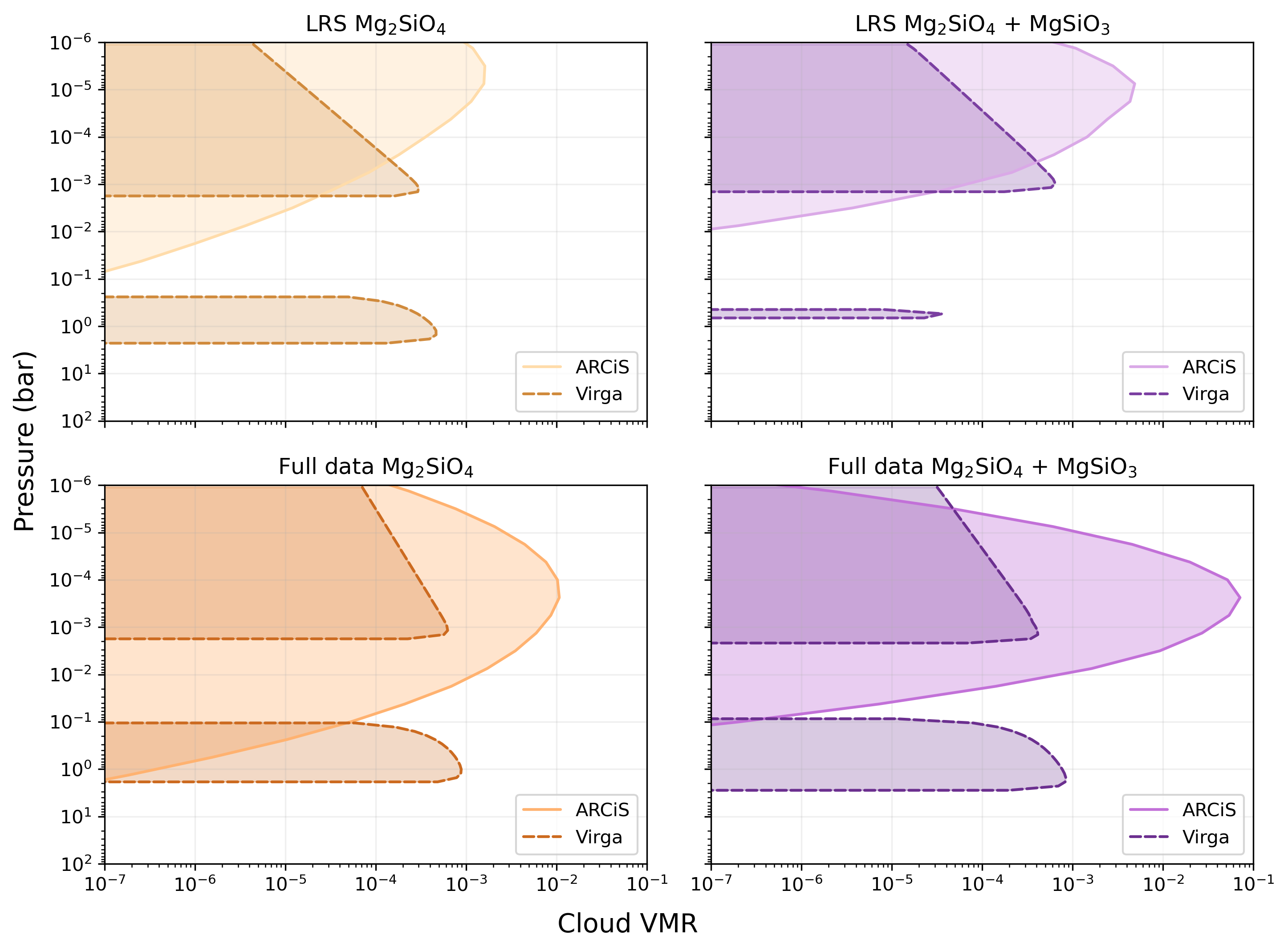}
\caption{The extent of the cloud layer for the best-fit \ce{Mg2SiO4} (oranges) and combination (\ce{Mg2SiO4} and \ce{MgSiO3}, purples) ARCiS and \texttt{Virga} retrievals on the LRS (upper panels) and the LRS+HST+NIRCam (lower panels) data. Here VMR is the volume mixing ratio of the cloud species. }
\label{fig:cloud_extent_combo}
\end{figure}

Our retrieved spectra on the full observed dataset for \ce{Mg2SiO4}, \ce{Mg2SiO4}+\ce{MgSiO3}, and a clear atmosphere, are shown in Figure~\ref{fig:all_data_residuals}, along with residuals, observational offsets, and selected diagonal panels from a combined corner plot. The opacity difference between the clear and cloudy models between 8~--~12~$\mu$m is predominantely due to magnesium silicate absorption. The residuals panel demonstrates, in particular for the LRS spectra, how much better the cloudy models fit the observations than the clear model. The offset panel illustrates that only a small shift is required between the NIRCam and HST observations for the two cloudy models, whereas a larger shift is required for the clear model. All three models require a relatively large shift between the LRS and NIRCam observations, which is likely linked to the LRS transmission observation capturing only a partial transit.
The full corner plot for the two cloudy retrievals, comparing retrieved parameters from using LRS vs the full dataset, is given in Figure~\ref{fig:cornerplots_all4}, and the full corner plot comparing the two cloudy retrievals and the clear retrieval on the full dataset can be found in Figure~\ref{fig:cornerplots_clear_full}. These posterior plots demonstrate how including the \textit{HST} and NIRCam data alongside the LRS significantly improves the constraints on the retrieved parameters in comparison to using the LRS alone. 
We do not include the retrieved parameters used to construct the pressure-temperature profile in these corner plots, but give the retrieved pressure-temperature profiles in Figure~\ref{fig:PT_profiles_full_data}, and the retrieved isothermal temperatures for the LRS retrievals in Table~\ref{tab:ret_pars}. It can be seen that, while the LRS-only retrievals find a temperature lower than the condensation curves of \ce{Mg2SiO4} and \ce{MgSiO3}, the retrieved pressure-temperature profiles using the full data are slightly hotter than expected for \ce{Mg2SiO4} and \ce{MgSiO3} condensation; see Section~\ref{sec:plausibility} for a discussion on this. The extent of the gaussian cloud layer in the atmosphere for each of our four main ARCiS retrievals can be seen in Figure~\ref{fig:cloud_extent_combo}, with the volume mixing ratio of the cloud species on the x-axis indicating the thickness of the cloud at different pressure levels.

\subsection{\texttt{PICASO} and \texttt{Virga} Forward Models and Grid-trievals} 

To further assess the plausibility of our retrieved atmosphere and cloud constraints, we compare our ARCiS retrieval results to those of self-consistent forward models. We generate a grid of clear-sky radiative-convective thermochemical equilibrium (RCTE) \texttt{PICASO} 3.0~\citep{Mukherjee2023} models tailored to {\planet}, spanning a range of metallicities (9 values from 1-100$\times$ solar), C/O ratios (4 values from 0.115 to 0.687), and heat redistribution factors (4 values from 0.5 to 0.8, representing full to nearly dayside only). We set the intrinsic temperature to 300 K. Previous studies have found relatively little sensitivity to this parameter for hot giant planets over 1000 K \citep[e.g.,][]{Grant2023quartz,Alderson2023_ERS,Boehm2025}, provided the intrinsic temperature is at least a few hundred kelvin, which is suggested by interior modeling \citep{thorngren2019b,24WiChMa}.

The climate module relies on the correlated-k opacities of \citet{lupu_2022_6600976}, which includes contributions from \ce{N2}, \ce{C2H2}, \ce{C2H4}, and \ce{C2H6} \citep{hitran2012}; \ce{CH4} \citep{yurchenko13vibrational,yurchenko_2014}, CO \citep{HITEMP2010,HITRAN2016,li15rovibrational}, \ce{CO2} \citep{HUANG2014reliable}, CrH \citep{Burrows02_CrH}, Fe \citep{Ryabchikova2015,oBrian1991Fe,Fuhr1988Fe,Bard1991Fe,Bard1994Fe}, FeH \citep{Dulick2003FeH,Hargreaves2010FeH}, \ce{H2} and OCS \citep{HITRAN2016}, \ce{H3+} \citep{Mizus2017H3p}, \ce{H2O} \citep{Polyansky2018H2O}, \ce{H2S} \citep{azzam16exomol}, HCN \citep{Harris2006hcn,Barber2014HCN,hitran2020}, LiCl and LiF \citep{Bittner2018Lis}, LiH \citep{Coppola2011LiH}, MgH \citep{Yadin2012MgH,GharibNezhad2013MgH,GharibNezhad2021}, \ce{NH3} \citep{yurchenko11vibrationally,Wilzewski16}, \ce{PH3} \citep{sousa14exomol}, SiO \citep{Barton2013SiO}, TiO \citep{McKemmish2019TiO,GharibNezhad2021}, VO \citep{McKemmish16,GharibNezhad2021}; and Li, Na, K, Rb, and Cs \citep{Ryabchikova2015,Allard2007AA,Allard2007EPJD,Allard2016,Allard2019}. The chemical abundances as a function of temperature, pressure, C/O ratio, and metallicity are derived from the models presented in \citet{Marley2021} which follow the thermochemical equilibrium calculations of \citet{visscher2010}. From the atmospheric profiles we generate for {\planet}, we produce simulated transmission spectra using the \texttt{PICASO V3} opacity database, which extends from 0.3 to 15 $\mu$m resampled at R=15,000 from a line-by-line calculation performed at R$\sim$10$^6$, available on Zenodo~\citep{natasha_batalha_2025_14861730}.
This database contains \ce{C2H2}, \ce{CH4}, CO, \ce{CO2}, \ce{H2-H2}, \ce{H2O}, K, Na, \ce{H2S}, OCS, and \ce{SO2}.

For the cloudy models, we follow similar methodology in our \texttt{PICASO} modelling as was performed for WASP-17b \citep{Grant2023quartz}. 
That is, we post-process the \texttt{PICASO} models with \texttt{Virga} \citep{Batalha2025}, which uses the \texttt{EddySed} \citep{ackerman2001} formulation of varying $f_{\rm{sed}}$ (sedimentation efficiency) and $K_{\rm{zz}}$ (the turbulent diffusion, or mixing parameter) to find the mass and particle sizes of clouds self-consistently. 
These \texttt{EddySed}-style retrievals allow us to determine whether the expected condensation curves of the magnesium silicates (given by \citealt{Batalha2025}) can produce cloud features in the LRS region given reasonable mixing and sedimentation in the atmosphere. Here our free parameters are only $f_{\rm{sed}}$, $K_{\rm{zz}}$, and $\sigma$, which is the width of the log-normal particle size distribution. The priors for each of these parameters can be found in Table~\ref{tab:virga_cloud} and are $\mathcal{U}$ [0.1, 10], $\mathcal{U}$ [10$^5$, 10$^{10}$ cm$^{2}$ s$^{-1}$], and $\mathcal{U}$ [0.5, 2.5], respectively. 

We perform our \texttt{PICASO+Virga} grid retrieval fits using only the best fit single composition and combined composition cloud species found by the ARCiS retrievals, i.e., amorphous \ce{Mg2SiO4} and a combined cloud of amorphous \ce{Mg2SiO4} and \ce{MgSiO3}, using the same set of refractive indices as used in ARCiS for each species. 
We perform our full \texttt{PICASO+Virga} model fitting on both the LRS data alone as well as the combined HST+NIRCam+LRS data, for a total of four sets of model grid fits with \texttt{PICASO+Virga}. 

We use Ultranest \citep{buchner2021} to sample the transmission spectra modelled from our \texttt{PICASO+Virga} profiles. In addition to the fixed \texttt{PICASO} grid and free cloud \texttt{Virga} parameters discussed above, we also fit for a vertical offset from the model to the data. When fitting the combined 0.3 -- 12 $\mu$m spectrum, we hold the NIRCam data as fixed and fit for offsets of the model, \textit{HST}, and LRS data. As in ARCiS, we treat the \textit{HST} data from STIS and WFC3 as a single dataset, keeping the offsets fixed from \citet{Sing2016}. Finally, we note that the current \texttt{Virga} configuration does not self-consistently account for the depletion of cloud-forming material by other cloud species \citep{Batalha2025}, and thus treats both \ce{Mg2SiO4} and \ce{MgSiO3} as able to form according to their respective condensation curves, ignoring the fact that as more and more magnesium and silicon are taken up into one cloud species, less material is available for the other. The condensation curves do account for the atmospheric metallicity from the model grid.

\begin{table*}[t]
\centering
\caption{Summary of \texttt{PICASO} cloudy grid retrieval results, with the median and 1$\sigma$ upper and lower limits. ``Full Spectrum'' column corresponds to the combination of MIRI/LRS, NIRCam, and \textit{HST} (STIS+WFC3) combined spectrum. The \ce{Mg2SiO4} runs used the refractive indices for amorphous particles from \citealt{jager2003}, and the ``Combo'' retrievals used the refractive indices of amorphous \ce{Mg2SiO4} and \ce{MgSiO3} from \citealt{Scott1996} and \citealt{dorschner1995} respectively. ln(L) refers to the maximum log-likelihood of the fit.}
\begin{tabular}{@{}lllcccc@{}}
\hline
\hline
\multicolumn{7}{c}{PICASO Grid Fit with Virga Clouds} \\
\hline
Parameter & Type & Prior & \multicolumn{1}{c}
{\begin{tabular}[c]{@{}c@{}}LRS \\ \ce{Mg2SiO4}\end{tabular}} & \multicolumn{1}{c}{\begin{tabular}[c]{@{}c@{}}LRS \\ Combo\end{tabular}} & \multicolumn{1}{c}{\begin{tabular}[c]{@{}c@{}}Full Spectrum\\ \ce{Mg2SiO4}\end{tabular}} & \multicolumn{1}{c}{\begin{tabular}[c]{@{}c@{}}Full Spectrum\\ Combo\end{tabular}} \\
\hline
M/H [dex] & Grid & [1,100] & $1.76^{+0.17}_{-0.30}$ & $1.89^{+0.09}_{-0.13}$ & $1.82^{+0.02}_{-0.18}$ & $1.56^{+0.03}_{-0.04}$ \\
C/O & Grid & [0.115,0.687] & $0.234^{+0.200}_{-0.096}$ &  $0.262^{+0.184}_{-0.106}$ & $0.123^{+0.047}_{-0.006}$ & $0.213^{+0.013}_{-0.019}$ \\
Heat Redis. & Grid & [0.5,0.8] & $0.60^{+0.08}_{-0.08}$ & $0.58^{+0.09}_{-0.06}$ & $0.74^{+0.00}_{-0.02}$ & $0.67^{+0.02}_{-0.01}$ \\
f$_{\mathrm{sed}}$ & Free & $\mathcal{U}$(0.1,10) & $0.401^{+0.345}_{-0.274}$ & $0.262^{+0.157}_{-0.120}$ & $0.336^{+0.058}_{-0.060}$ &  $0.389^{+0.080}_{-0.072}$  \\
log(K$_{\mathrm{zz}}$) [cm$^2$/s]& Free & $\mathcal{U}$(5,10) &  $7.29^{+0.77}_{-0.83}$& $7.47^{+0.53}_{-0.47}$ & $8.88^{+0.12}_{-0.12}$ & $8.76^{+0.15}_{-0.16}$ \\
ln($\sigma$) & Free & $\mathcal{U}$(0.5,2.5) & $0.50^{+0.29}_{-0.34}$ & $0.62^{+0.23}_{-0.48}$ & $0.89^{+0.02}_{-0.07}$ & $0.52^{+0.30}_{-0.22}$ \\
ln(L) & - & - & -27.4 & -37.6 & -189.4 & -204.7
\end{tabular}
\label{tab:virga_cloud}
\end{table*}

\begin{figure*}
\centering
\includegraphics[width=0.99\textwidth]{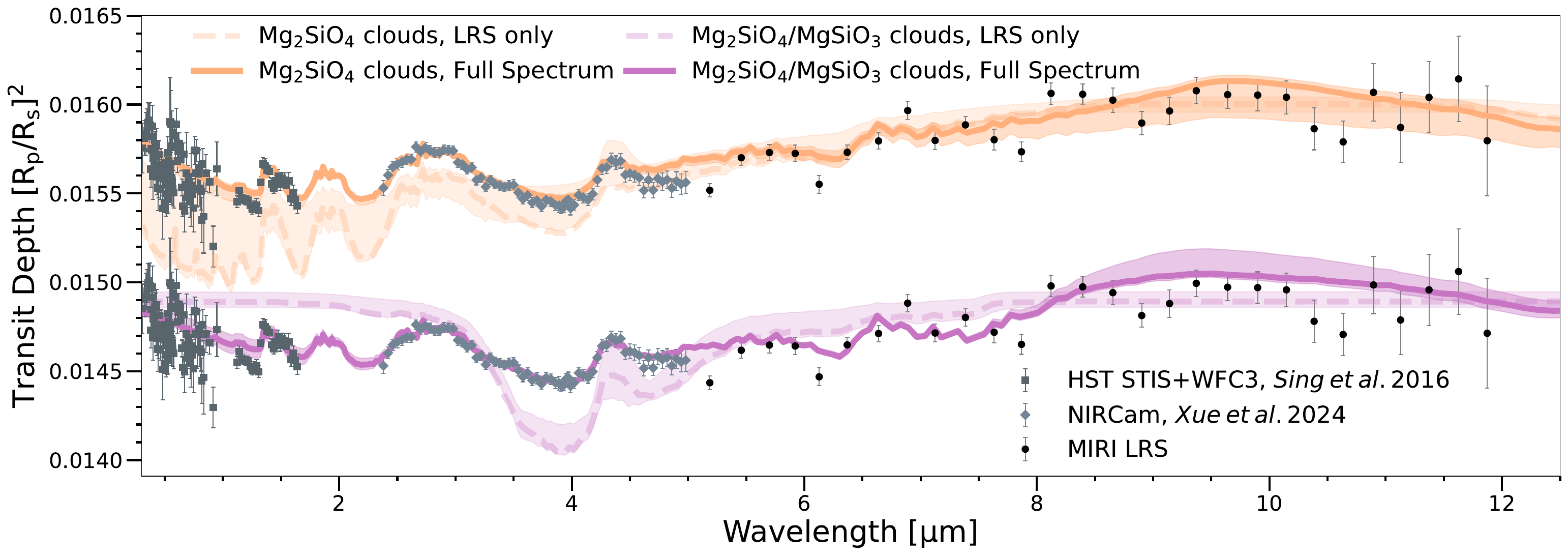}
\caption{Our PICASO/Virga best fit spectra with Mg$_2$SiO$_4$ (oranges) or both Mg$_2$SiO$_4$ and MgSiO$_3$  clouds (purples).The LRS only fit is shown with dashed lines. We apply an offset between the two compositional runs for clarity.}
\label{fig:LRS_clouds_arcis_picaso}
\end{figure*}

In general, our grid retrievals agree broadly with the results of the ARCiS free retrieval runs. For the LRS only fits, we find a strong preference (likelihood ratio of 10) for the the \ce{Mg2SiO4} cloud alone over the combined \ce{Mg2SiO4}-\ce{MgSiO3}, with weak constraints on the metallicity, C/O ratio, and parameters controlling the cloud particle size and extent (i.e., $\sigma$, f$_{\rm{sed}}$, and K$_{\rm{zz}}$). Our preference for the \ce{Mg2SiO4} only cloud likely occurs due to the cooler condensation onset of \ce{MgSiO3}, making it more challenging to achieve particle sizes of similar sizes by both species at once to match the feature. As shown in Figure~\ref{fig:PT_profiles_full_data}, our single and combined \texttt{Virga} retrievals find similar best-fit temperature-pressure profiles, though the combined cloud is slightly colder, again to better allow \ce{MgSiO3} particles to form. 

Using the full spectrum, we find evidence for a super-solar metallicity, significantly sub-solar C/O ratio, and cloud parameters that combine to give mean particle sizes at 1 millibar of $\sim$0.02-0.2 $\mu$m. The \ce{Mg2SiO4} cloud set-up is strongly preferred over the combined cloud (with a likelihood ratio of 15), and there are significant differences between the metallicities, C/O ratios, and width of the particle size distribution between each cloud composition run. All of these sets of results, for both the combined and single composition cloud, are shown in Table \ref{tab:virga_cloud}.  

Both the \ce{Mg2SiO4} and the combined \ce{Mg2SiO4}-\ce{MgSiO3} cloud \texttt{Virga} grid retrievals on the combined \textit{HST}+NIRCam+LRS spectrum (see Figure~\ref{fig:LRS_clouds_arcis_picaso} for the best fit spectra) find significantly higher metallicities ($\sim$65$\times$ solar and $\sim$35$\times$ solar respectively) than either the ARCiS fits ($\sim$13$\times$ solar) or those found by \citet{Xue2024ApJ} ($\sim$5$\times$ solar) or \citet{25VeGoAv} ($\sim$1.5$\times$ solar). The two full spectrum \texttt{Virga} runs find different best-fit C/O ratios, but all are in general agreement with both ARCiS and that found by \citet{Xue2024ApJ}, being highly sub-solar. Most interestingly, the best-fit particle size distribution for the preferred \ce{Mg2SiO4} only \texttt{Virga} fit is quite wide compared to the combined cloud fit. The mean particle size for the \ce{Mg2SiO4} cloud run is $\sim$0.02 $\mu$m, but the width of the size distribution means that particles an order of magnitude larger still strongly contribute to the opacity throughout the spectrum. On the other hand, the combined cloud has an extremely tight particle size distribution, with essentially only particles of $\sim$0.1 $\mu$m contributing opacity throughout the spectrum. Without more detailed microphysical models, it is unclear which scenario is more physical, but such investigations should be the subject of future study. Even in the combined cloud run, we find that the mass mixing ratio and thus opacity contribution from \ce{MgSiO3} is an order of magnitude smaller than that of \ce{Mg2SiO4}, suggesting that \ce{Mg2SiO4} is the primary cloud.

For either cloud composition in the \texttt{Virga} fits, the pressure extent where the cloud contributes the most opacity, as well as the particle sizes that contribute the most opacity, are similar to those found by ARCiS. This is shown in Figure~\ref{fig:cloud_extent_combo}, i.e., at pressures near 1 millibar with particle sizes that, averaging over the full size distribution from \texttt{Virga}, are around 0.1 $\mu$m. These self-consistent forward model results generally support the free retrievals, suggesting our inferred magnesium silicate cloud detection is both robustly supported by the data as well as a physical understanding of the atmospheric processes which give rise to these condensates.  

\subsection{Retrieval of 4-leaf LRS data compared to 1-leaf}\label{sec:4_leaf_retrieval}

We have modified the ARCiS code so we can run a retrieval on a $n$-leaf spectra simultaneously, providing an efficient framework for atmospheric retrieval that accounts for uncertainties introduced during the data reduction stage in the parameter inference. The four 1-leaf spectra are treated as separate transmission spectra within ARCiS, with the total log-likelihood for the 4-leaf spectra, $\ln \mathcal{L}$, computed in ARCiS using Eq.~\ref{log_lik}. 

\begin{figure}
\centering
\includegraphics[width=0.5\textwidth]{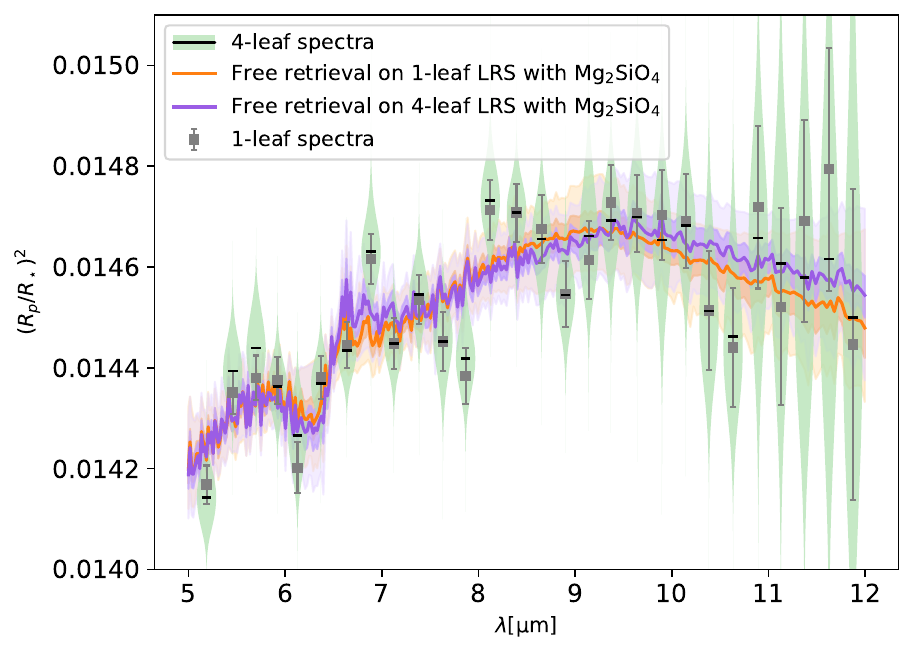}
\caption{Retrieved spectra using amorphous Mg$_2$SiO$_4$ 1-leaf vs 4-leaf LRS spectra.}
\label{fig:spectra_4leaf_vs_1leaf}
\end{figure}

Based on our best model setups for the 1-leaf retrieval, we run a retrieval including amorphous \ce{Mg2SiO4} on the 4-leaf MIRI/LRS spectra described in Section~\ref{sec:dt}, in order to test the robustness of our results with multiple decisions included in the reduced spectra. 
As can be seen by Figures~\ref{fig:spectra_4leaf_vs_1leaf} and \ref{fig:corner_4leaf_vs_1leaf}, even though the constraints are tighter with the 1-leaf retrieval, our retrieved parameters using the 1-leaf vs the 4-leaf combined retrieval remain robust to the decisions in the decision-tree. 
We include an exaggerated example of a simulated 2-leaf spectra in Appendix~\ref{sec:simulated_dtdr}, to demonstrate how decisions which substantially impact the spectra can result in bimodal posteriors.

\section{Discussion}\label{sec:discussion}
\subsection{Comparison to other works}\label{sec:compare}

Table~\ref{tab:ret_pars} compares our retrieved atmospheric parameters with those from other works which analyzed the \textit{HST}+NIRCam spectra. See Appendix \ref{appendix:corner_plots} for the corresponding corner plots of these retrievals.

Our VMRs are consistent with the very low C/O inferred by \cite{Xue2024ApJ}, who also find notable H$_2$O and CO$_2$ features. \cite{25BaKrMo}, who perform a new reduction on the \textit{HST}/WFC3 observations, also find robust detections of H$_2$O and CO$_2$, along with a 3.6$\sigma$ preference for clouds. 
VMRs of Na and K, assuming solar abundance, are predicted to be  $\sim$10$^{-6}$ and $\sim$10$^{-7}$ respectively in the atmosphere of HD\,209458b which we are probing. As can be seen in Table~\ref{tab:ret_pars}, both are within our uncertainties of VMRs of Na and K for both our full data retrievals, and are also consistent with the retrieved abundances from \cite{25VeGoAv}. 
Although we do detect H$_2$O with high significance, we do not significantly detect CH$_4$, or NH$_3$, and only find weak evidence for C$_2$H$_2$ or HCN. All of these species were detected via ground-based high-resolution spectroscopy by \cite{21GiBrGa}, who found the amplitude of their detected molecular lines were dampened by clouds but were nevertheless detectable. The nature of the high-resolution cross-correlation method means that molecules present at smaller abundances can be detected, and with more confidence, due to matching of molecular lines.
However, unlike the methods used here, deducing absolute abundances using cross-correlation methods remains tricky. Ideally the two methods would be combined to maximise the information content~\citep{Brogi_2017}, which remains challenging. \cite{22GaMiCh} predicted that NIRCam should be able to detect VMR of C$_2$H$_2$ and CH$_4$ of above $\sim$~10$^{-6}$ in HD\,209458b, assuming solar abundances of other species such as H$_2$O. However, the presence of clouds does increase the abundance limit required for detection due to muting of spectral features, so it is possible that these species are in the atmosphere but at lower abundances than we are able to detect with the current observations. More data from other \textit{JWST} instruments such as NIRISS or NIRSpec could help further constrain the atmospheric parameters, particularly in the presence of clouds~\citep{22GaMiCh}. These additional data would be particularly valuable for scaling the different observations relative to one another, as they include overlapping wavelength coverage with the current dataset.

As shown in Tables~\ref{tab:ret_pars} and \ref{tab:virga_cloud}, the metallicity is larger for the \texttt{PICASO} grid-retrievals than for the equilibrium chemistry retrievals of ARCiS and other works. This is likely to be mainly due to methodological differences between the frameworks. In particular, the Virga cloud model used in \texttt{PICASO} fixes the gas mean mass mixing ratio to a constant value, which could be driving the metallicity up to compensate for the low cloud opacity.

Our retrieved particle size on the full observational data with amorphous Mg$_2$SiO$_4$ clouds is $\sim$0.1~$\mu$m, which is an order of magnitude larger than the \ce{SiO2} particles found on WASP-17b \citep{Grant2023quartz}. We do not consider a particle size distribution in our retrievals, but only a single particle size which we take to be the size with the strongest contribution to the opacity. This is backed up by our Virga models, which do consider a particle size distribution which is consistent with $\sim$0.1~$\mu$m particles dominating the opacity contribution.
Previous studies such as \cite{Barstow2020} find the \textit{HST} spectra for HD209458b to be consistent with large aerosol particles, due to the relatively weak scattering slope in the optical.

\subsection{Plausibility of the observed cloud species in light of the full atmospheric context}\label{sec:plausibility}
While in general silicate condensates are expected in hot Jupiters \citep[e.g.,][etc.]{visscher2006,wakeford2015,helling2016}, the exact species of silicate can reveal deeper insight into the primordial composition of the planet overall, as well as the details of the cloud formation process. Recently, \citet{Calamari2024} and \cite{Calamari2026} showed the link between bulk Mg/Si and the relative proportions of \ce{Mg2SiO4}, \ce{MgSiO3}, and \ce{SiO2} sequestered at depth. The star HD\,209458 has a sub-solar C/O ratio and Mg/Si$\sim$1 \citep[see Table in Appendix;][]{Polanski2022}. Under the paradigm of \citet{Calamari2024}, we would not expect \ce{SiO2} and \ce{Mg2SiO4} together, while \ce{SiO2} and \ce{MgSiO3} are unfavored given the star's Mg/Si ratio. It is therefore not unexpected that HD\,209458b the planet has highly oxygen-rich clouds made of enstatite and forsterite together, if we assume that the planet has inherited stellar abundances and that the same sequestration at depth is maintained up to higher altitudes.  This sequestration of condensates is supported by our retrievals which do not prefer a combination including \ce{SiO2} (Table~\ref{tab:arcis_clouds}). Furthermore, past observations of HD\,209458b detected Mg I escape with \textit{HST}, indicating plentiful amounts of Mg available to make clouds at high altitudes \citep{vidal2013}. Further work is needed to investigate the link between stellar inheritance, planetary atmospheric composition, and cloud species, but we demonstrate here that at least for HD\,209458b, such chemical connections can be made. Indeed, the two other confirmed cloud detections on hot Jupiters, HD 189733~b \citep{Inglis2024ApJ} and WASP-17~b \citep{Grant2023quartz}, orbit stars with Mg/Si $<$ 0.9 and have \ce{SiO2} clouds, which also fits within this paradigm. 

Such arguments, however, naturally follow from an ``equilibrium'' understanding of clouds. While our self-consistent cloud forward models use \texttt{EddySed}, this is only one method of cloud parametrization that supports the formation of magnesium silicate clouds in this atmosphere. \cite{16HeLeDo} predict the cloud composition of HD\,209458b using a GCM with a kinetic cloud formation model, and also predict the upper atmosphere of the terminator regions (which we are probing with our observations) to be dominated by MgSiO$_3$[s] and Mg$_2$SiO$_4$[s] (with the [s] indicating solid state), adding additional confidence that the cloud species we detect are physically likely in this atmosphere even under different physical assumptions on the cloud formation process. 

We also note that our retrievals only slightly disfavor corundum (\ce{Al2O3}) ($\Delta$ln(Z)~=~3.4; $\Delta$BPICS~=~-6.6) and silicon monoxide (SiO) ($\Delta$ln(Z)~=~2.8; $\Delta$BPICS~=~-7.8) clouds over \ce{Mg2SiO4} clouds. However, both equilibrium and kinetic cloud models suggest {\planet} is either too hot and/or \ce{Al2O3} is too low abundance to expect its opacity to strongly contribute to the planet's spectrum (see, e.g. Figure 4 of \citealt{16HeLeDo}). Meanwhile, SiO is expected to exist over a wide range of pressures and temperatures in the gas phase, with gas-phase SiO required for magnesium silicate formation. The latter occurs preferentially over solid-state SiO formation, provided there is sufficient Mg~\citep{visscher2010}, as appears to be the case with HD~209458b~\citep{vidal2013}. \cite{16HeLeDo} do not predict SiO condensate to be present in the high abundances expected for \ce{Mg2SiO4} and \ce{MgSiO3} in the observable region of HD~209458b. Thus, despite being only slightly disfavored statistically, our current physical understanding of these species' material properties suggests we can discount them in favor of magnesium silicate clouds. 

In each of our cloudy retrievals, we tested both crystalline and amorphous refractive indices for the silicate clouds we examined. The exact formation conditions that would lead to either amorphous or crystalline cloud particles differ, where amorphous particles would suggest that material condenses and then is quickly lofted to cooler atmospheric regions, while crystalline particles would indicate that particles remain at hotter atmospheric layers \citep[e.g.,][]{jaeger1998,cushing2006}. Moreover, the available laboratory data for such particle morphologies uses different methods to generate material for measurement, with only some measured at elevated temperatures (see Appendix \ref{sec:cloud_refrinds} for further discussion). As both magnesium silicate species we prefer have similar condensation profiles (see Figure~\ref{fig:PT_profiles_full_data}), we chose to perform our combined cloud runs using self-similar particle types, as both magnesium silicate species should be either crystalline or amorphous assuming they form under similar conditions. We only weakly disfavor crystalline clouds with our ARCis retrievals on the LRS data, likely due to a combination of the signal-to-noise and the particle size being a free parameter. It is also possible that the temperature-dependence \citep{Moran2024}, fractal shape~\citep{Lodge2024,Moran_2025}, or alignment of silicate grains \citep{Mullens2025} could induce spectral effects that lead to different particle morphology inferences than we find here. Although some indication of temperature-dependence is given by the improved fit of crystalline \ce{Mg2SiO4} measured at 1000~K over 295~K (see Table~\ref{tab:arcis_clouds}), such detailed investigations are beyond the scope of this work, and ideally require further laboratory measurements and theoretical modeling.

As shown in Figure~\ref{fig:PT_profiles_full_data}, the pressure-temperature profiles for the ARCiS retrievals are too hot for magnesium silicate condensation; we do not place any physical constraints between the atmospheric temperature and condensation formation in these retrievals. Furthermore, we do not expect the temperature structure to be well-constrained by free-retrievals of the transmission spectra. However, the pressure-temperature profiles from the Virga models are consistent with magnesium silicate formation, which does place more physical constraints on the models. The fact that we find magnesium silicate absorption features in both our modeling approaches demonstrates that the cloud features are invariant to the pressure-temperature profiles we're considering. This finding supports our approach of using the retrievals to narrow down the parameter space, and then the Virga models to further assess the physical plausibility of our atmosphere. In order to further explore the impact of the pressure-temperature profile on our conclusions, we first ran a free retrieval using a guillot profile instead of isothermal for the MIRI LRS observations only, and find a Bayes factor of 1 over the isothermal case, which indicates that the extra parameters do not significantly improve the fit to observations. We also ran some ARCiS free retrievals on the full dataset where we do not allow the pressure-temperature profile to vary, but instead fix it to the pressure-temperature profile output from \texttt{PICASO}. In this case we find the hotter atmosphere (i.e. with free Guilott profile parameters) is only weakly preferred ($\Delta\ln{Z}$~=~2.7) over the atmosphere with the cooler profile fixed from \texttt{PICASO} for the cloudy (\ce{Mg2SiO4}) retrieval on the full dataset, with the retrieved aerosol particle size remaining unchanged. This indicates that the hotter pressure-temperature profile is not required in the free retrievals for the observations to fit the cloudy models, i.e. the presence and size of the magnesium silicate clouds are not sensitive to the temperature. We thus do not consider the pressure-temperature profiles retrieved from the ARCiS free retrievals to be an accurate representation of the atmospheric thermal structure. We note that the cloudy free retrievals and the grid retrievals all probe relatively high up in the atmosphere, as demonstrated by the transmission contributions of Figure~\ref{fig:trans_cf}.

In our combined retrieval, we mix the refractive indices together using effective medium theory rather than forming two separate cloud layers (see Section~\ref{sec:ARCiS}). Although it offers less flexibility, this approach has the benefit of greatly reducing the number of retrieved parameters compared to considering two separate cloud layers. Physically, it is not unreasonable to expect some mixed-material composition particles (see e.g. \cite{16HeLeDo}).
When we consider the combined \ce{MgSiO3}+\ce{Mg2SiO4} atmosphere, the retrieval finds only 14\% \ce{MgSiO3} for the LRS spectra alone, but this switches to 90\% when considering the full set of observations. We note that this retrieved fraction is dependent on the refractive indices used for a given species, and there are degeneracies with the way the HST observations are shifted compared to NIRCam. This further suggests that we cannot meaningfully distinguish between whether the clouds are \ce{Mg2SiO4} or some combination of \ce{Mg2SiO4}+\ce{MgSiO3}. Further observations with NIRISS or NIRSpec, which overlap the HST and NIRCam wavelength regions, would help break this degeneracy and place tighter constraints on the exact nature of the magnesium silicates, as would additional MIRI/LRS observations capturing the full transit to increase the signal-to-noise. While the Bayes factor, $\Delta\ln{Z}$, indicates that both \ce{Mg2SiO4} or a combination of \ce{Mg2SiO4}+\ce{MgSiO3} are preferred over \ce{MgSiO3} alone, $\Delta$BPICS does not prefer the combination above \ce{MgSiO3} alone. This potentially is due to the BPICS metric offering a stronger penalization for models with more parameters than the Bayes factor.

As discussed above, we have an idea of what species are likely to form in this atmosphere given our knowledge of its atmospheric properties and our understanding of atmospheric chemistry. However, previous microphysical modeling has shown that local effects like the disequilibrium gas abundances~\citep{PowellZhang2024,Kiefer2024}, vertical mixing~\citep{Powell2018}, and material properties~\citep{Gao2020} impact which clouds are able to form in a given environment. While a detailed microphysical study of this object is outside the scope of the current work, we note that this study will be necessary to better understand the cloud formation, cloud properties, and bulk atmospheric properties of this planet.

\section{Conclusions}\label{sec:conclusions}

We observed the canonical hot Jupiter HD\,209458b in transit using \textit{JWST} MIRI/LRS from 5--12\,$\upmu$m. Due to timing errors we captured only a partial transit (ingress plus 19\% full transit) with an extended pre-transit baseline. 
We introduce a rule-based tree structure for data reduction, highlighting a transparent route for tracking methodological choices and propagating associated uncertainties through subsequent modeling. We demonstrate that for MIRI data reduction the treatment of the number of groups to mask, background subtraction, and length of stellar baseline combined with the appropriate systematic model are important decisions that must be included but can result in subtle differences in the final spectrum. We present a method of combining these using a Gaussian mixture model which can then be used to more accurately model the resultant spectrum of the planet. This rule-based tree structure can be easily applied to different instruments and observing set-ups with additions and modifications specific to the observations in question allowing the use to explore all options and marginalize over the results to obtain a more accurate spectrum. We run our retrievals on our 4-leaf spectra obtained from the rule-based tree structure as a comparison against our 1-leaf spectra, and demonstrate our findings are valid when including all 4 main decision branches identified. 

We find evidence for magnesium silicate clouds in the \textit{JWST} MIRI/LRS transmission spectra of HD\,209458b, most likely either composed of \ce{Mg2SiO4} (forsterite), or a combination of \ce{Mg2SiO4} and \ce{MgSiO3} (forsterite and enstatite). We employ Bayesian atmospheric retrievals to identify the most strongly preferred models, and subsequently examine the feasibility of these solutions using physically motivated cloud forward models. By combining our MIRI/LRS reduction with archival NIRCam (F322W2 and F444W) and \textit{HST} observations (STIS G430L and G750L and WFC3-IR G141), our analysis points to $\sim$0.1 $\mu$m grains composed of either \ce{Mg2SiO4} or a mixture of \ce{Mg2SiO4} and \ce{MgSiO3}, at pressures of $\sim$1--10~millibars. These findings add to a growing number of detections of species-specific cloud signatures in the mid-IR which can now be observed with \textit{JWST}'s MIRI instrument, offering a more detailed view of aerosol composition than previously possible. We estimate that with a full transit of HD\,209458b not only would this detection be significantly more precise but we it would also provide the opportunity to map the clouds around the limb to a high precision and accuracy. 

Supplementary data including transmission spectra and light curves are available on Zenodo~\citep{HD209_zenodo}.

\section{Acknowledgments}
K.L.C, D.G, and H.R.W were funded by UK Research and Innovation (UKRI) under the UK government’s Horizon Europe funding guarantee as part of an ERC Starter Grant [grant number EP/Y006313/1]. S.E.M. is supported by NASA through the NASA Hubble Fellowship grant HST-HF2-51563 awarded by the Space Telescope Science Institute, which is operated by the Association of Universities for Research in Astronomy, Inc., for NASA, under contract NAS5-26555. Resources supporting this work were provided by the NASA High-End Computing (HEC) Program through the NASA Advanced Supercomputing (NAS) Division at Ames Research Center as well as the University of Maryland supercomputing resources (https://hpcc.umd.edu). S.E.M. is grateful to M. Weiner Mansfield for enabling UMD supercomputing resource access. K.L.C thanks Michiel Min for discussions on the updates required to ARCiS for the $n$-leaf retrievals. H.R.W. thanks Grant Stevens for discussions on the machine learning.  \\ 

\textit{HD\,209458b cloud babies:} Ariah, Finn, Forest Moon.\\

\textit{Author contributions: }
KLC led the paper writing and conducted the ARCiS analysis. 
DG conducted the rule-based decision tree data reduction.
HRW devised the observations and advised throughout the analysis and interpretation.  
SEM performed the PICASO+Virga analysis working with NEB.
AE performed the Eureka! data reduction with the advice of KBS.
CF performed retrieval analysis and comparative studies to assess the clouds.
DP provided context when considering micro-physical processes. 
All other authors provided comments on the paper, engaged during discussion meetings, or were part of the original telescope proposal. \\
%

\textit{Data Availability:} Data is available through MAST using the DOI: 10.17909/0dj4-mh60 
\facilities{JWST(MIRI)}


\software{ExoTiC-LD \citep[][]{Grant2024}\footnote{https://exotic-ld.readthedocs.io/en/latest/}, ExoTiC-MIRI \citep[][]{grant_david_2023_8211207}\footnote{https://exotic-miri.readthedocs.io/en/latest/}, Eureka! \citep[][]{bell2022}\footnote{https://eurekadocs.readthedocs.io/en/latest/}, batman \citep[][]{Kreidberg2015_batman}, ARCiS\footnote{\url{https://github.com/michielmin/ARCiS.git}}, PICASO \citep[][]{Mukherjee2023}\footnote{https://natashabatalha.github.io/picaso/}, Virga \citep[][]{Batalha2025}\footnote{https://natashabatalha.github.io/virga/}, numpy \citep[][]{harris2020array}, SciPy \citep[][]{2020SciPy-NMeth}, matplotlib \citep[][]{Hunter:2007}, xarray \citep{hoyer2017xarray, hoyer_stephan_2022_6323468}, astropy \citep{2013A&A...558A..33A,2018AJ....156..123A, price2022astropy}.}



\appendix

\section{An independent data analysis check with the Eureka! pipeline}\label{sec:independent_pipeline_check}

\begin{figure*}[h!]
\centering
\includegraphics[width=0.8\textwidth]{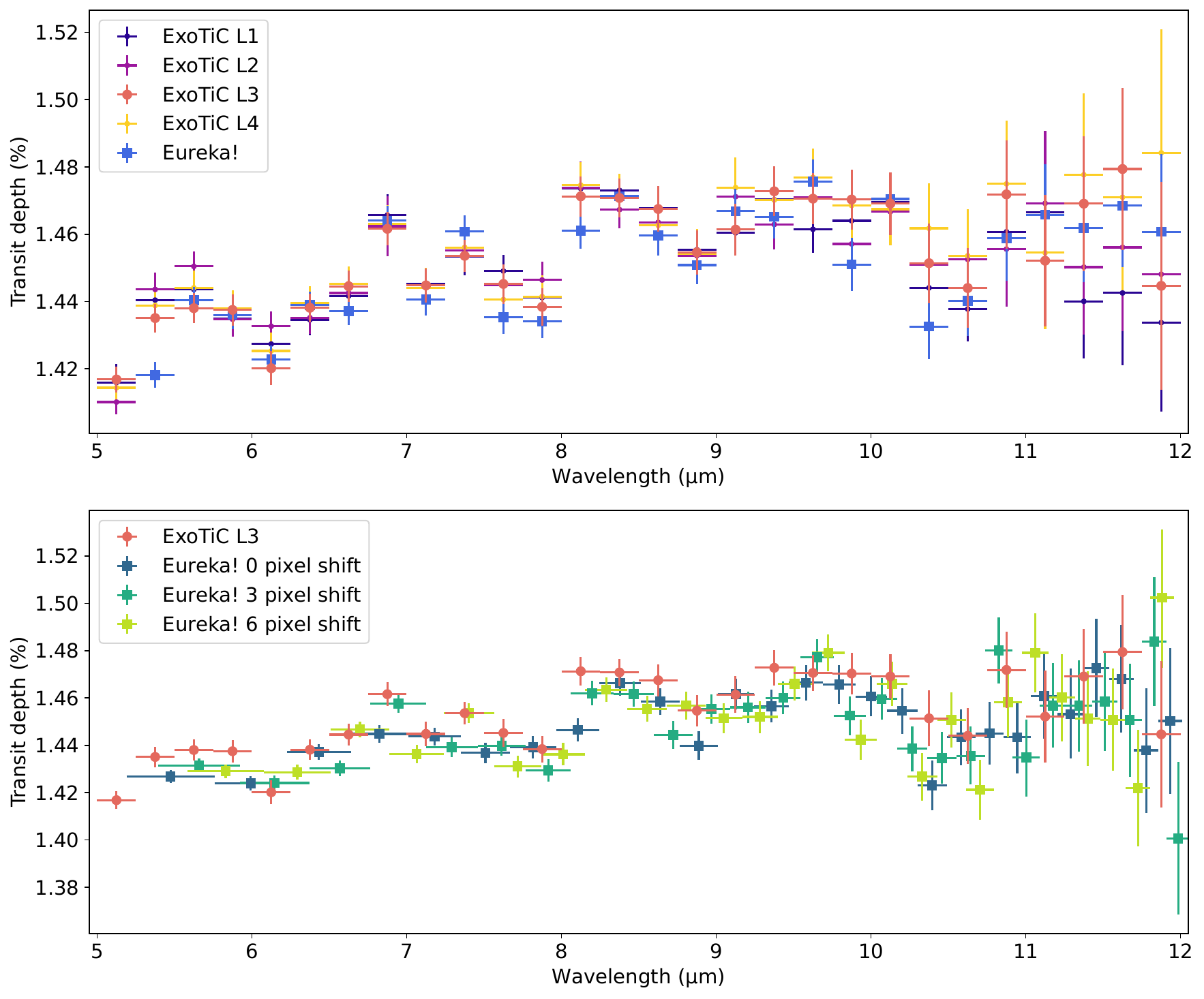}
\caption{Transmission spectra comparison for two independent pipelines: \exotic and \eureka. Top: For \exotic we show all four of the ``leaves'' that make up the final ends of the tree of decisions considered in this paper noting that \texttt{L3} is the main result taken forward to our modeling stage, and the \eureka spectrum decisions most closely matching \texttt{L4} at the light curve stage. Bottom: For \eureka we show the data reduction extracted considering three different starting wavelengths (shown to be important at the light curve fitting stage by \citealt{Stevenson2025}) compared to the \exotic \texttt{L3} spectrum. }
\label{fig:transmission_spectrum_comparison_4leafs}
\end{figure*}

A \texttt{Eureka!} \citep{bell2022} reduction is carried out as a comparison to the \exotic rule-based tree structure reduction to better explore and understand differences in the spectrum. For our \eureka reduction we follow the standard set-up starting at the \texttt{\_uncal} files in Stage 1 as a wrapper of the \texttt{jwst} pipeline (v1.11.4, CRDS 1149). We test for the number of groups to mask finding the best performance when masking the first and last groups in the ramp fitting stage. We use the default linearity correction, do not correct for darks or flat fielding, and use the default jump threshold of 4-$\sigma$ (noting that newer versions of the \texttt{jwst} pipeline are better at treating this step in the reduction). We perform background subtraction in a per-column basis utilizing the detector regions outside an aperture half-width of 12 pixels centered on the spectral trace, testing for the inner edge of the aperture. We then extract our spectrum using a box aperture of 8 pixels centered on the spectral trace. 

At the light curve fitting stage we tested a range of stellar baseline cutoffs (2,000, 5,000, and 10,000 integrations) finding the removal of 10,000 integrations optimal for reducing light curve systematics which were fit with a simple linear trend in time. We calculate and fix the limb-darkening coefficients using \texttt{ExoTiC-LD} for a quadratic model using the 3D Stagger grid \citep{magic2015stagger} for a T$_\mathrm{eff}$\,=\,6000\,K, [M/H]\,=\,0.00, and log(g)\,=\,4.5. For our spectroscopic light curves we test a series of binsizes and starting wavelengths.  

In Figure\,\ref{fig:transmission_spectrum_comparison_4leafs} we show the two spectral reductions. Fist we show the final \eureka spectrum compared to the final four-leaves of the \exotic tree structure where the \eureka data reduction steps are most closely aligned to \texttt{l4}. We then show the comparison of our 1-leaf spectrum (\texttt{L3}) compared to the \eureka reduction computed by considering 9-pixel wide bins with starting wavelengths shifted by 0, 3, and 6 pixels, a process demonstrated to make a difference in MIRI/LRS spectra by \citet{Stevenson2025}. In each case our reductions all sit within the expected uncertainty range and the structure of the transmission spectrum is consistent. As demonstrated the proposed rule-based tree structure can take into account all the decisions deemed necessary in the extraction, reduction, and light curve fitting stages of the analysis. Combining the final ``leaves'' of the tree using a Gaussian mixture model means that all decisions regarded important can then be folded into the retrieval analysis stages where even large differences can be accounted for (see Appendix\,\ref{sec:simulated_dtdr}).

\section{Simulated 2-leaf decision-tree data reduction}\label{sec:simulated_dtdr}

To further demonstrate how an $n$-leaf spectrum can affect retrieval results, here we show a simulated 2-leaf spectrum and the resulting inferences.

\begin{figure}[h]
\centering
\includegraphics[width=0.8\textwidth]{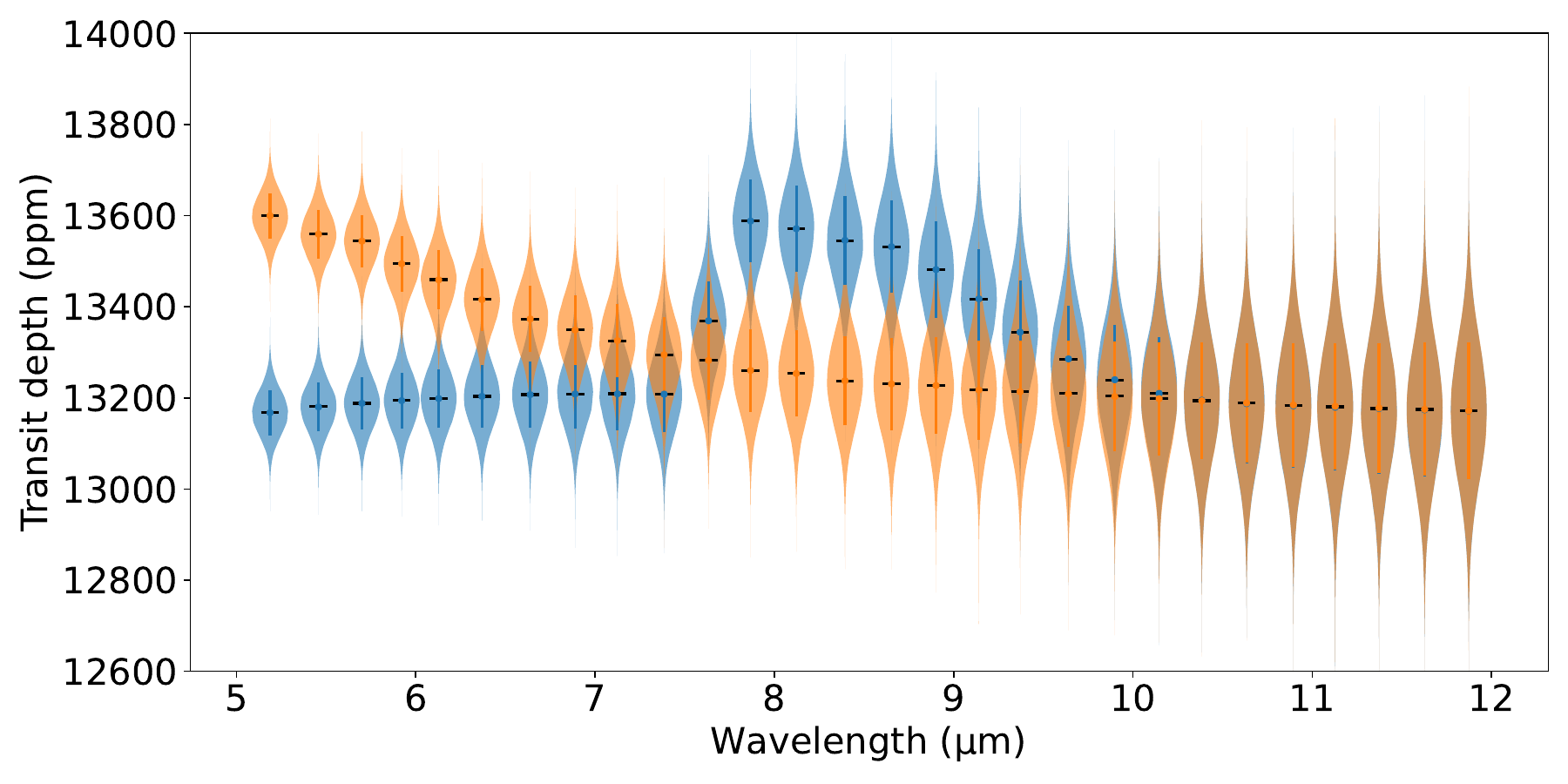}
\caption{Our simulated 2-leaf transmission spectra, here shown as two separate spectra. The spectra with SiO is in blue, and with SiH in orange.}
\label{fig:demo_transmission_spectrum}
\end{figure}

1) We generate two model isothermal spectra (see Figure~\ref{fig:demo_transmission_spectrum}), one with SiH~\citep{SiH_ExoMol} (orange) and one with SiO~\citep{21YuTeSy} (blue), as a proxy for two decision branches in a hypothetical data reduction pipeline which result in very different spectra.
We apply wavelength-dependent uncertainties on the transit depth, linearly increasing from 50~ppm to 150~ppm, to represent typical uncertainties in LRS transit depth measurements. 

2) The transit depth distributions are combined as a Gaussian mixture model; we run one retrieval on these two leaves and include it in ARCiS' likelihood computation (see Eq.~\ref{log_lik}).

\begin{figure*}[h]
\centering
\includegraphics[width=0.49\textwidth]{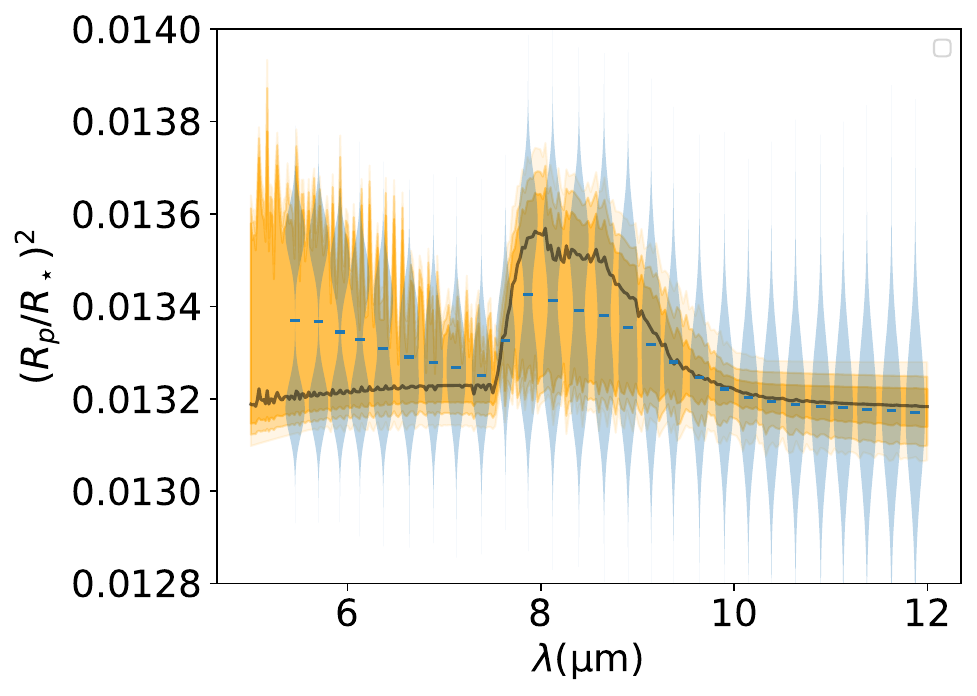}
\includegraphics[width=0.49\textwidth]{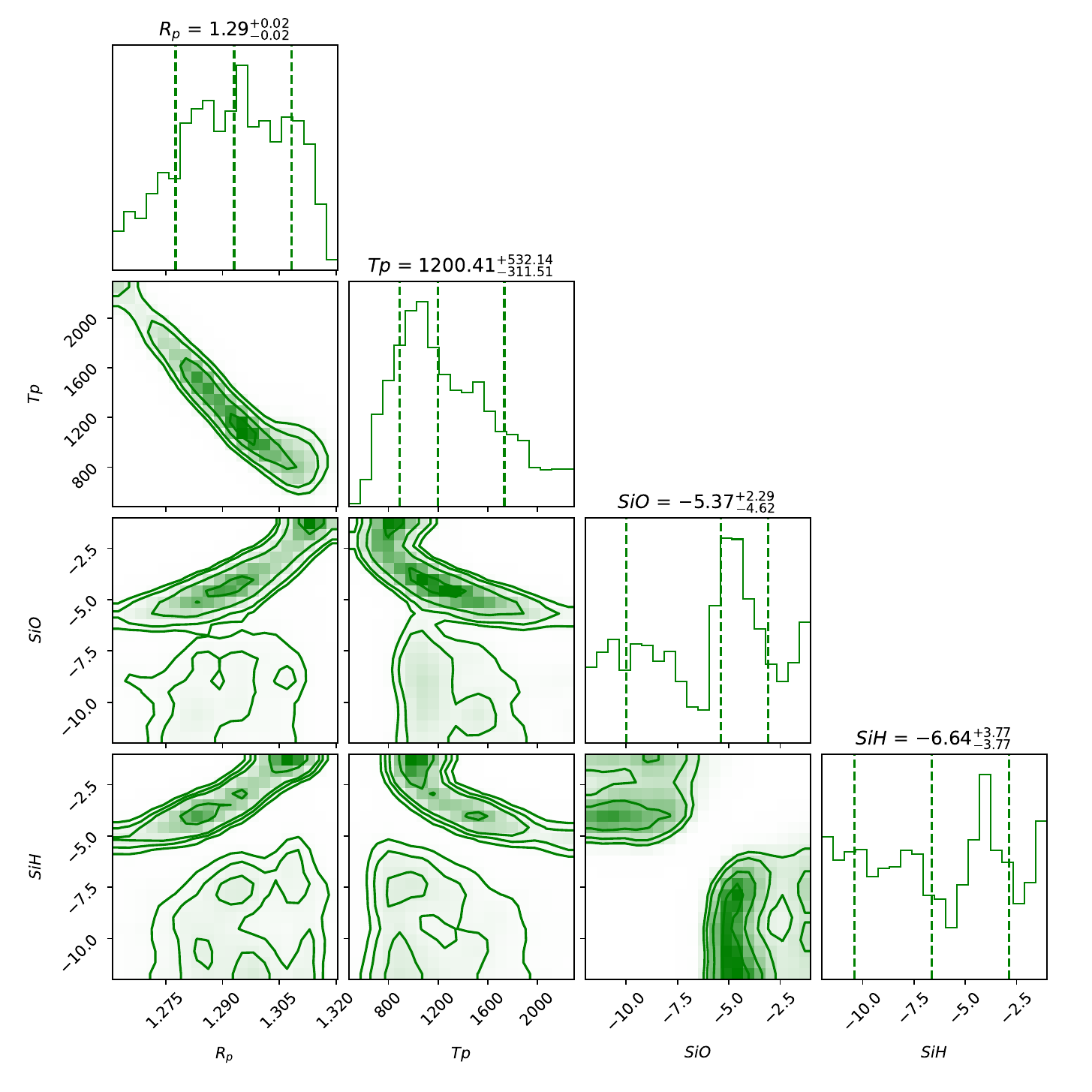}
\caption{Our retrieved transmission spectra (left) and posteriors (right) for the simulated 2-leaf observations. In blue on the left is our combined 2-leaf simulated observations, with our bestfit retrieved spectra and shaded 1, 2, and 3 $\sigma$ regions in orange.}
\label{fig:demo_corner_plot}
\end{figure*}

3) The results of the retrieval on our simulated 2-leaf spectra can be seen in Figure~\ref{fig:demo_corner_plot}. The inferences now take into account the data reduction differences, rather than just picking one leaf and ending up with results biased towards one choice or the other. This can be seen by the bimodal nature of the SiO and SiH posteriors in Figure~\ref{fig:demo_corner_plot} (right panel). If different data reductions produce spectra that can be modelled by two distinct atmospheres, it will show up in the posteriors.  The extent of the orange shading in Figure~\ref{fig:demo_corner_plot} (left) reflects the fact that the retrieval is unable to meaningfully distinguish between the SiO and SiH models, as both are equally weighted in the combined 2-leaf observational spectra.

\section{HD\,209458 Stellar Abundances}

The HD\,209458 stellar properties which we use in this work are given in Table~\ref{tab:placeholder}. They were measured using \texttt{KeckSpec} on Keck/HIRES data and presented in Table 1 of \citet{Polanski2022}.

\begin{table}[h!]
    \centering
    \begin{tabular}{cccccccccc}
\hline
Teff & log (g) & v sin (i) & [C/H] & [N/H] & [O/H] & [Na/H] & [Mg/H] & [Al/H] & [Si/H]\\
\hline
6042$\pm$77 & 4.30$\pm$0.09 & 3.0$\pm$0.9 & 0.00$\pm$0.05 &-0.01$\pm$0.08 &0.13$\pm$0.07 &-0.03$\pm$0.05 &0.03$\pm$0.04 &-0.03$\pm$0.04 & 0.03$\pm$0.03  \\
\hline
& \\
\hline
[Ca/H]  & [Ti/H] & [V/H] & [Cr/H] & [Mn/H] & [Fe/H] & [Ni/H] & [Y/H] & C/O & Mg/Si  \\
\hline
0.08$\pm$0.03 & 0.06$\pm$0.04 & 0.04$\pm$0.06 &0.05$\pm$0.04 &-0.04$\pm$0.05 &0.05$\pm$0.03 &0.01$\pm$0.04 &0.06$\pm$0.08 &0.42$\pm$0.08  &1.05$\pm$0.12  \\
\hline
    \end{tabular}
    \caption{HD\,209458 stellar properties measured using \texttt{KeckSpec} on Keck/HIRES data and presented in Table 1 of \citet{Polanski2022}.}
    \label{tab:placeholder}
\end{table}

\section{Cloud Refractive Indices}\label{sec:cloud_refrinds}

\begin{figure*}
\centering
\includegraphics[width=1.0\textwidth]{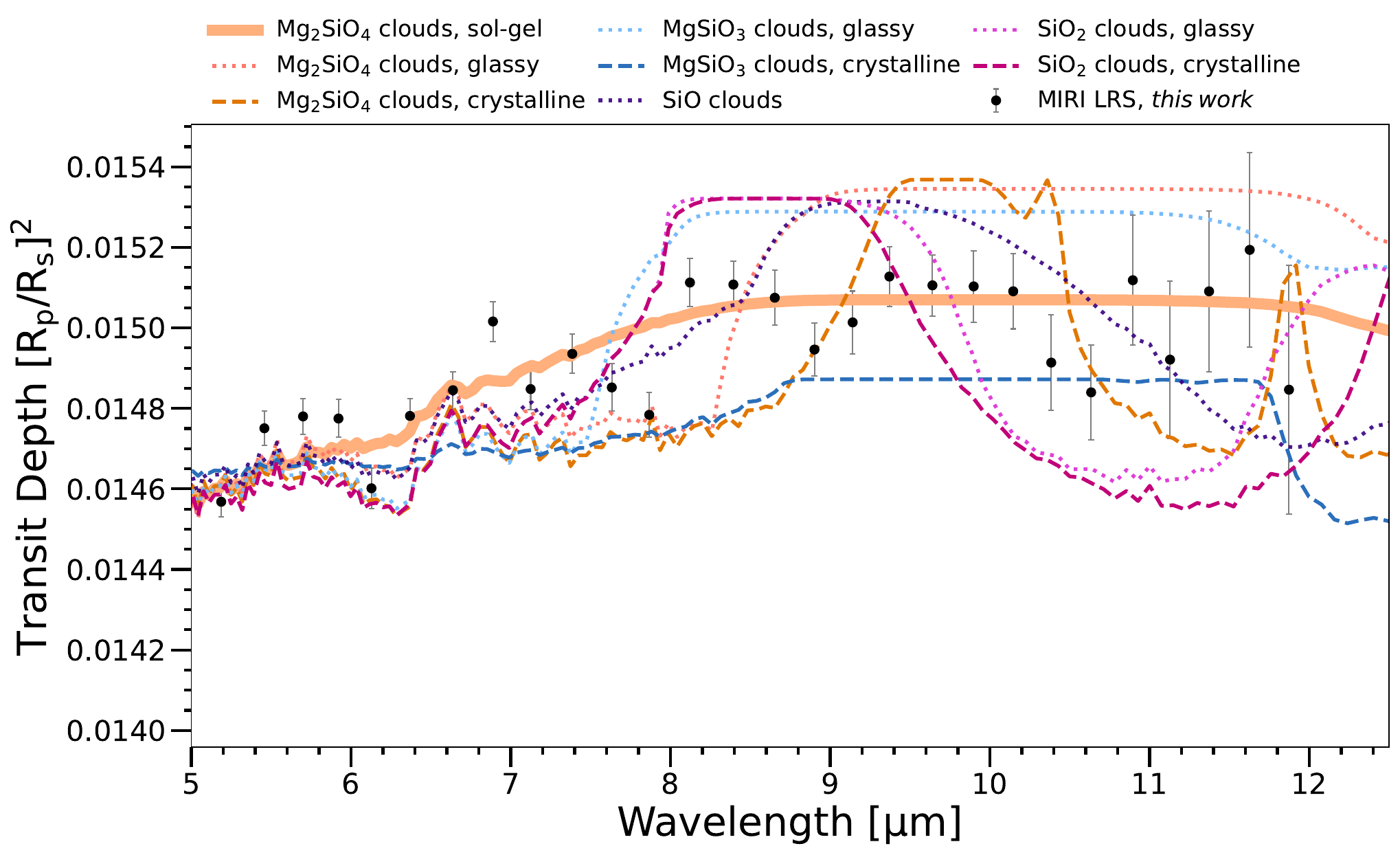}
\caption{Model comparison for the best-fit \texttt{Virga} parameters, changing only the cloud optical properties of the best fit model for the LRS data. Citations for refractive indices used follow those in Table \ref{tab:arcis_clouds}, with \ce{Mg2SiO4} sol-gel from \cite{03JaDoMu}, \ce{Mg2SiO4} glassy from \cite{Scott1996}, and \ce{MgSiO3} glassy from \cite{dorschner1995}.}
\label{fig:optical_properties_comparison}
\end{figure*}

It is non-trivial to decide which refractive indices to use \citep[see, e.g.,][]{Mullens2024}. Here in this Appendix in Figure~\ref{fig:optical_properties_comparison}, we show the differences between the refractive indices of the species and morphologies we tested. This highlights both the difficulty in distinguishing between certain compositions as well as why others are easily disfavored. For the crystalline MgSiO$_3$ and \ce{Mg2SiO4}, we use the refractive indices of \citealt{jaeger1998} and \cite{Eckes2013}, respectively, averaged over each crystallographic index, following the method of both \citealt{Mullens2024} and \citealt{kitzmann2018} which averaged the refractive index rather than the extinction efficiencies as was done in \citealt{lunamorley2021}. All other data used follow those found in Table~\ref{tab:arcis_clouds} in the main text, with amorphous (glassy) \ce{MgSiO3} from \cite{95DoBeHe}. Of the existing laboratory data used to provide such refractive indices, only some are measured at temperatures relevant to the upper atmospheres of hot Jupiters. The method used to synthesize the material also differs. For example, amorphous forsterite as used in our retrievals has two sources. In one, forsterite was melted and then quenched to achieve amorphous \ce{Mg2SiO4} as in \citet{Scott1996}, and in the other it was prepared via the sol-gel technique, where a liquid solution is mixed with solids to form a gel, as in \citet{jager2003}. While the quenching method may be more representative of the formation conditions in a hot Jupiter, the temperature dependence on the morphology is likely not fully captured by either technique (e.g., see \citealt{Moran2024}). 

Figure~\ref{fig:optical_properties_comparison} shows how our median best-fit \texttt{Virga} model on the LRS data alone compares when alternative cloud materials are used. The \ce{Mg2SiO4} model is the result of the grid retrieval, and all other models shown use the same parameters as this best fit, simply substituting alternative refractive indices as shown in the legend. Given the broadness of the cloud feature, this Figure highlights the challenge of isolating a single composition when comparing materials that all share a Si-O bond, particularly if the particle size and abundance is otherwise unconstrained. Nevertheless, we find robustly that magnesium silicates, and in particular \ce{Mg2SiO4} with potentially some added \ce{MgSiO3}, fit the data best when performing both our free and grid retrievals.

\section{Corner plots}\label{appendix:corner_plots}

\begin{figure*}
\centering
\includegraphics[width=1.0\textwidth]{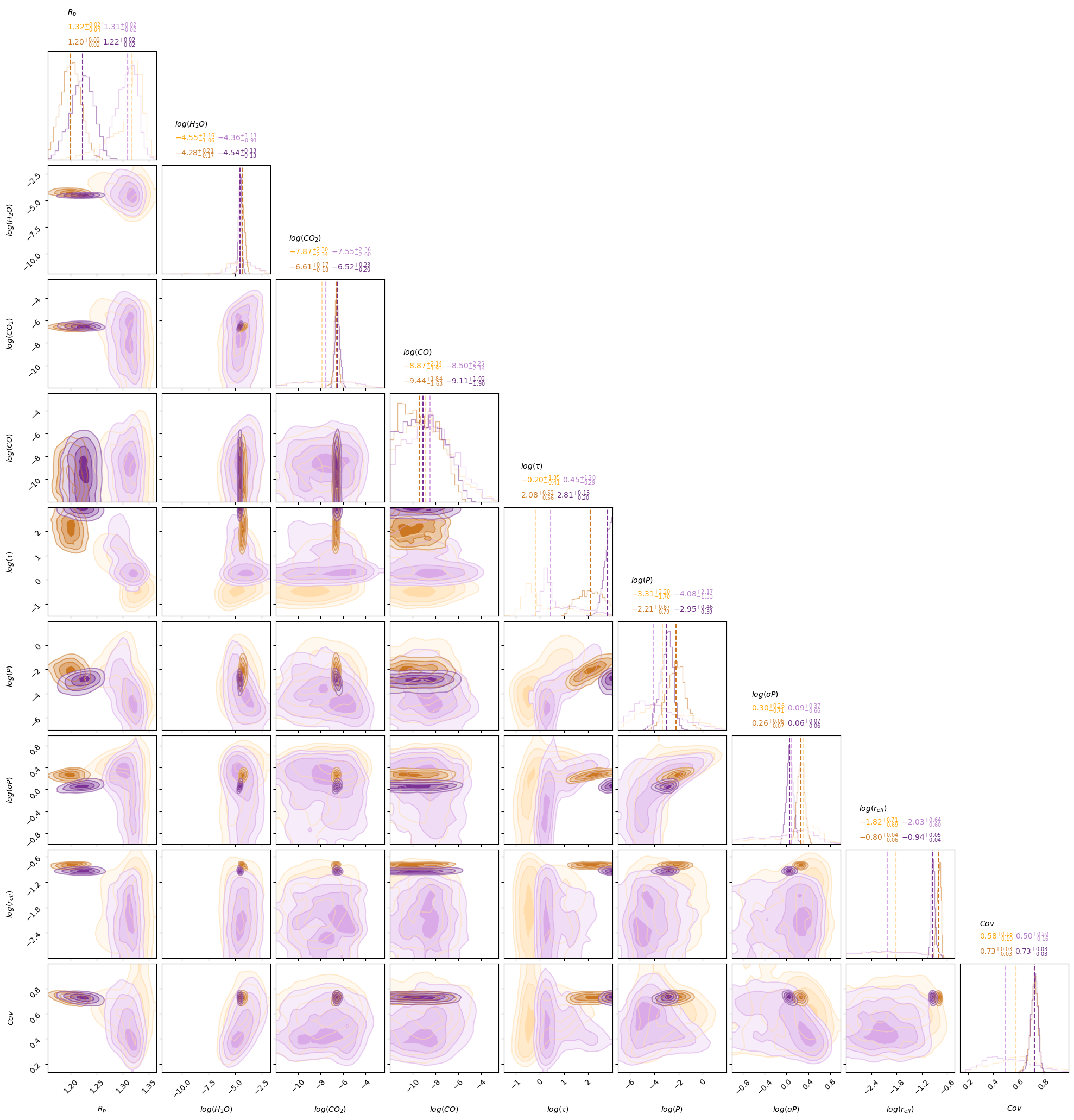}
\caption{Corner plot showing the retrieved parameters for the ARCiS retrieval using the \ce{Mg2SiO4} (light orange) and combined \ce{Mg2SiO4} and \ce{MgSiO3} (light purple) atmosphere on the LRS spectra, and \ce{Mg2SiO4} (dark orange) and combined \ce{Mg2SiO4} and \ce{MgSiO3} (dark purple) on the full HST+NIRCam+LRS dataset. The median values are listed in the order shown in the parameter key.}
\label{fig:cornerplots_all4}
\end{figure*}

\begin{figure*}
\centering
\includegraphics[width=1.0\textwidth]{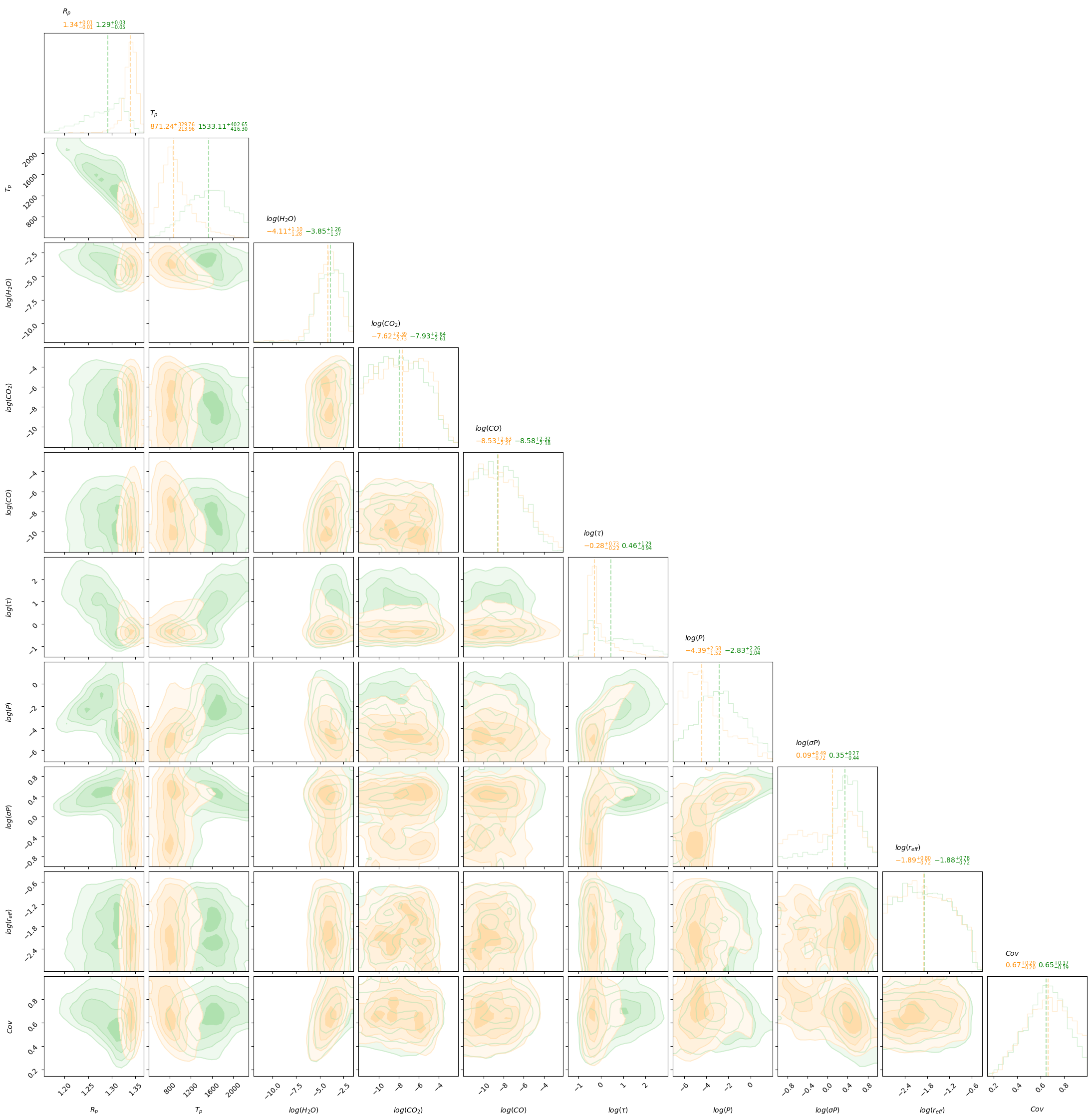}
\caption{Corner plot of the ARCiS retrievals using the Mg$_2$SiO$_4$ atmosphere on the 1-leaf (orange) vs 4-leaf (green) LRS spectra.}
\label{fig:corner_4leaf_vs_1leaf}
\end{figure*}

\begin{figure*}
\centering
\includegraphics[width=1.0\textwidth]{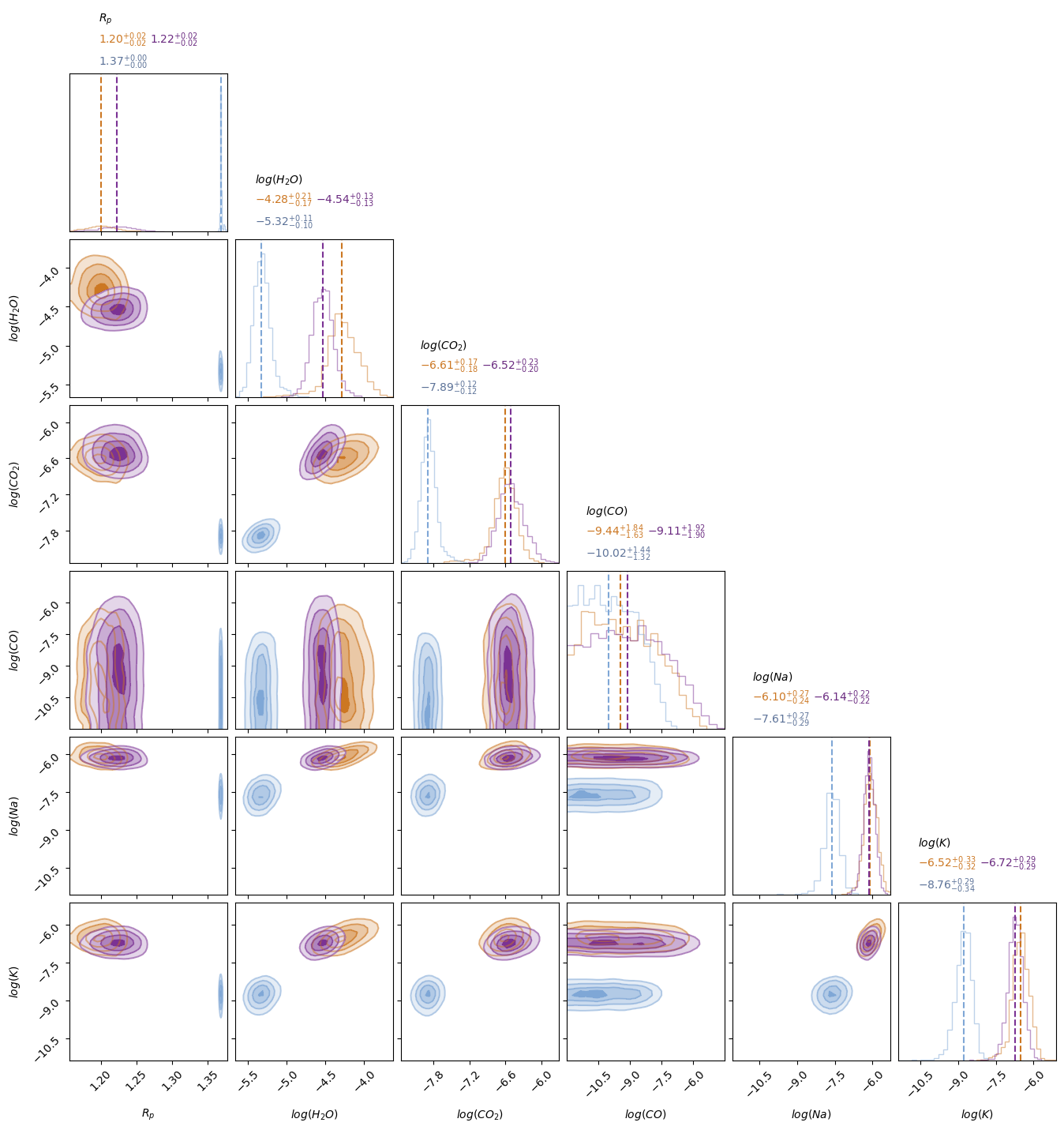}
\caption{Corner plot of ARCiS retrievals using \ce{Mg2SiO4} (orange), combined \ce{Mg2SiO4} and \ce{MgSiO3} (purple) and clear (blue) atmospheres on the full dataset.}
\label{fig:cornerplots_clear_full}
\end{figure*}

\section{Transmission contribution functions}\label{appendix:trans_cf}

\begin{figure*}
\centering
\includegraphics[width=1.0\textwidth]{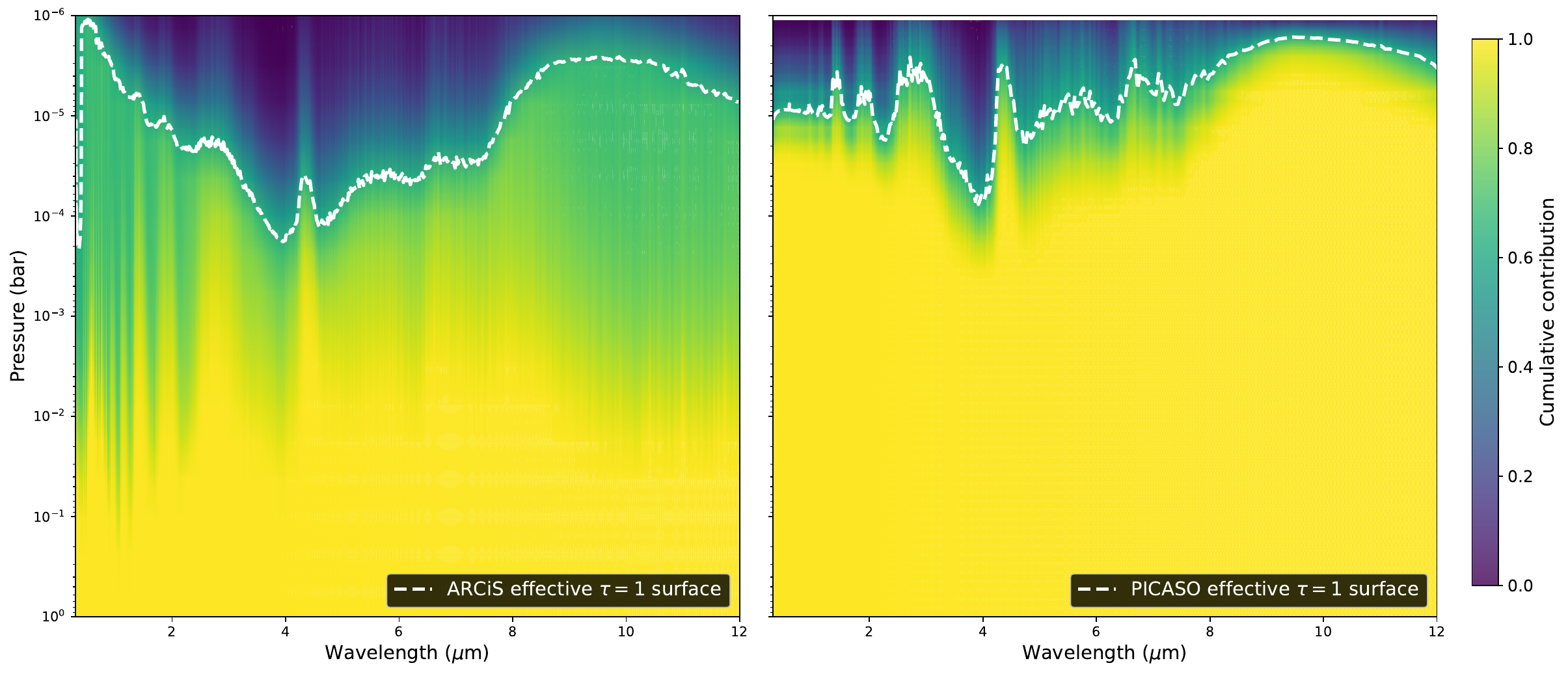}
\caption{The cumulative transmission contribution, as integrated from the top of the atmosphere towards deeper layers, for the combined \ce{Mg2SiO4} and \ce{MgSiO3} ARCiS retrieval (left) and \texttt{PICASO} grid-retrieval (right) on the full dataset. We overplot the $\tau$~=~1 surface at the level where the the cumulative contribution hits a value of 1~-~$e^{-1}$~$\approx$~0.63.}
\label{fig:trans_cf}
\end{figure*}


\bibliography{bib_ref}{}

@ARTICLE{corrales2023,
       author = {{Corrales}, Lia and {Gavilan}, Lisseth and {Teal}, D.~J. and {Kempton}, Eliza M. -R.},
        title = "{Photochemical hazes can trace the C/O ratio in exoplanet atmospheres}",
      journal = {arXiv e-prints},
         year = 2023,
        month = jan,
          eid = {arXiv:2301.01093},
        pages = {arXiv:2301.01093},
          doi = {10.48550/arXiv.2301.01093},
archivePrefix = {arXiv},
       eprint = {2301.01093},
 primaryClass = {astro-ph.EP},
       adsurl = {https://ui.adsabs.harvard.edu/abs/2023arXiv230101093C}
}

@TECHREPORT{Bergeron2021jwst.rept.7761B,
       author = {{Bergeron}, Louis and {Regan}, Mike},
        title = "{MIRI Last Frame Effect}",
  institution = {STScI},
         year = 2021,
       number = {Technical Report JWST-STScI-007761},
 howpublished = {JWST STScI Technical Report, JWST-STScI-007761},
       adsurl = {https://ui.adsabs.harvard.edu/abs/2021jwst.rept.7761B}
}

@article{kurucz1993atlas9,
  title={ATLAS9 Stellar Atmosphere Programs and 2km/s grid},
  author={Kurucz, R-L\_},
  journal={Kurucz CD-Rom},
  volume={13},
  year={1993},
  publisher={Smithsonian Astrophysical Observatory}
}

@article{magic2015stagger,
  title={The Stagger-grid: A grid of 3D stellar atmosphere models-IV. Limb darkening coefficients},
  author={Magic, Zazralt and Chiavassa, Andrea and Collet, Remo and Asplund, Martin},
  journal={Astronomy \& Astrophysics},
  volume={573},
  pages={A90},
  year={2015},
  publisher={EDP Sciences}
}

@ARTICLE{emc32019JOSS....4.1864F,
       author = {{Foreman-Mackey}, Daniel and {Farr}, Will and {Sinha}, Manodeep and {Archibald}, Anne and {Hogg}, David and {Sanders}, Jeremy and {Zuntz}, Joe and {Williams}, Peter and {Nelson}, Andrew and {de Val-Borro}, Miguel and {Erhardt}, Tobias and {Pashchenko}, Ilya and {Pla}, Oriol},
        title = "{emcee v3: A Python ensemble sampling toolkit for affine-invariant MCMC}",
      journal = {The Journal of Open Source Software},
         year = 2019,
        month = nov,
       volume = {4},
       number = {43},
          eid = {1864},
        pages = {1864},
          doi = {10.21105/joss.01864},
archivePrefix = {arXiv},
       eprint = {1911.07688},
 primaryClass = {astro-ph.IM},
       adsurl = {https://ui.adsabs.harvard.edu/abs/2019JOSS....4.1864F}
}

@ARTICLE{Schlawin2020AJ,
       author = {{Schlawin}, Everett and {Leisenring}, Jarron and {Misselt}, Karl and {Greene}, Thomas P. and {McElwain}, Michael W. and {Beatty}, Thomas and {Rieke}, Marcia},
        title = "{JWST Noise Floor. I. Random Error Sources in JWST NIRCam Time Series}",
      journal = {\aj},
         year = 2020,
        month = nov,
       volume = {160},
       number = {5},
          eid = {231},
        pages = {231},
          doi = {10.3847/1538-3881/abb811},
archivePrefix = {arXiv},
       eprint = {2010.03564},
 primaryClass = {astro-ph.IM},
       adsurl = {https://ui.adsabs.harvard.edu/abs/2020AJ....160..231S}
}

@INPROCEEDINGS{Birkmann2022SPIE,
       author = {{Birkmann}, Stephan M. and {Giardino}, Giovanna and {Sirianni}, Marco and {Ferruit}, Pierre and {Rauscher}, Bernhard and {Alves de Oliveira}, Catarina and {B{\"o}ker}, Torsten and {Kumari}, Nimisha and {L{\"u}tzgendorf}, Nora and {Manjavacas}, Elena and {Proffitt}, Charles and {Rawle}, Timothy D. and {te Plate}, Maurice and {Zeidler}, Peter},
        title = "{The in-flight noise performance of the JWST/NIRSpec detector system}",
    booktitle = {Space Telescopes and Instrumentation 2022: Optical, Infrared, and Millimeter Wave},
         year = 2022,
       editor = {{Coyle}, Laura E. and {Matsuura}, Shuji and {Perrin}, Marshall D.},
       series = {Society of Photo-Optical Instrumentation Engineers (SPIE) Conference Series},
       volume = {12180},
        month = aug,
          eid = {121802P},
        pages = {121802P},
          doi = {10.1117/12.2629545},
archivePrefix = {arXiv},
       eprint = {2208.12686},
 primaryClass = {astro-ph.IM},
       adsurl = {https://ui.adsabs.harvard.edu/abs/2022SPIE12180E..2PB}
}

@ARTICLE{Barat2025AJ_V1298Taub,
       author = {{Barat}, Saugata and {D{\'e}sert}, Jean-Michel and {Mukherjee}, Sagnick and {Goyal}, Jayesh M. and {Xue}, Qiao and {Kawashima}, Yui and {Vazan}, Allona and {Misener}, William and {Schlichting}, Hilke E. and {Fortney}, Jonathan J. and {Bean}, Jacob L. and {Avarsekar}, Swaroop and {Henry}, Gregory W. and {Baeyens}, Robin and {Line}, Michael R. and {Livingston}, John H. and {David}, Trevor and {Petigura}, Erik A. and {Sikora}, James T. and {Shivkumar}, Hinna and {Feinstein}, Adina D. and {Oklop{\v{c}}i{\'c}}, Antonija},
        title = "{A Metal-poor Atmosphere with a Hot Interior for a Young Sub-Neptune Progenitor: JWST/NIRSpec Transmission Spectrum of V1298 Tau b}",
      journal = {\aj},
         year = 2025,
        month = sep,
       volume = {170},
       number = {3},
          eid = {165},
        pages = {165},
          doi = {10.3847/1538-3881/adec89},
archivePrefix = {arXiv},
       eprint = {2507.08837},
 primaryClass = {astro-ph.EP},
       adsurl = {https://ui.adsabs.harvard.edu/abs/2025AJ....170..165B}
}

@Article{harris2020array,
 title         = {Array programming with {NumPy}},
 author        = {Charles R. Harris and K. Jarrod Millman and St{\'{e}}fan J.
                 van der Walt and Ralf Gommers and Pauli Virtanen and David
                 Cournapeau and Eric Wieser and Julian Taylor and Sebastian
                 Berg and Nathaniel J. Smith and Robert Kern and Matti Picus
                 and Stephan Hoyer and Marten H. van Kerkwijk and Matthew
                 Brett and Allan Haldane and Jaime Fern{\'{a}}ndez del
                 R{\'{i}}o and Mark Wiebe and Pearu Peterson and Pierre
                 G{\'{e}}rard-Marchant and Kevin Sheppard and Tyler Reddy and
                 Warren Weckesser and Hameer Abbasi and Christoph Gohlke and
                 Travis E. Oliphant},
 year          = {2020},
 month         = sep,
 journal       = {Nature},
 volume        = {585},
 number        = {7825},
 pages         = {357--362},
 doi           = {10.1038/s41586-020-2649-2},
 publisher     = {Springer Science and Business Media {LLC}},
 url           = {https://doi.org/10.1038/s41586-020-2649-2}
}

@ARTICLE{2020SciPy-NMeth,
  author  = {Virtanen, Pauli and Gommers, Ralf and Oliphant, Travis E. and
            Haberland, Matt and Reddy, Tyler and Cournapeau, David and
            Burovski, Evgeni and Peterson, Pearu and Weckesser, Warren and
            Bright, Jonathan and {van der Walt}, St{\'e}fan J. and
            Brett, Matthew and Wilson, Joshua and Millman, K. Jarrod and
            Mayorov, Nikolay and Nelson, Andrew R. J. and Jones, Eric and
            Kern, Robert and Larson, Eric and Carey, C J and
            Polat, {\.I}lhan and Feng, Yu and Moore, Eric W. and
            {VanderPlas}, Jake and Laxalde, Denis and Perktold, Josef and
            Cimrman, Robert and Henriksen, Ian and Quintero, E. A. and
            Harris, Charles R. and Archibald, Anne M. and
            Ribeiro, Ant{\^o}nio H. and Pedregosa, Fabian and
            {van Mulbregt}, Paul and {SciPy 1.0 Contributors}},
  title   = {{{SciPy} 1.0: Fundamental Algorithms for Scientific
            Computing in Python}},
  journal = {Nature Methods},
  year    = {2020},
  volume  = {17},
  pages   = {261--272},
  adsurl  = {https://rdcu.be/b08Wh},
  doi     = {10.1038/s41592-019-0686-2},
}

@Article{Hunter:2007,
  Author    = {Hunter, J. D.},
  Title     = {Matplotlib: A 2D graphics environment},
  Journal   = {Computing in Science \& Engineering},
  Volume    = {9},
  Number    = {3},
  Pages     = {90--95},
  publisher = {IEEE COMPUTER SOC},
  doi       = {10.1109/MCSE.2007.55},
  year      = 2007
}

@ARTICLE{2018AJ....156..123A,
       author = {{Astropy Collaboration} and {Price-Whelan}, A.~M. and {Sip{\H{o}}cz}, B.~M. and {G{\"u}nther}, H.~M. and {Lim}, P.~L. and {Crawford}, S.~M. and {Conseil}, S. and {Shupe}, D.~L. and {Craig}, M.~W. and {Dencheva}, N. and {Ginsburg}, A. and {VanderPlas}, J.~T. and {Bradley}, L.~D. and {P{\'e}rez-Su{\'a}rez}, D. and {de Val-Borro}, M. and {Aldcroft}, T.~L. and {Cruz}, K.~L. and {Robitaille}, T.~P. and {Tollerud}, E.~J. and {Ardelean}, C. and {Babej}, T. and {Bach}, Y.~P. and {Bachetti}, M. and {Bakanov}, A.~V. and {Bamford}, S.~P. and {Barentsen}, G. and {Barmby}, P. and {Baumbach}, A. and {Berry}, K.~L. and {Biscani}, F. and {Boquien}, M. and {Bostroem}, K.~A. and {Bouma}, L.~G. and {Brammer}, G.~B. and {Bray}, E.~M. and {Breytenbach}, H. and {Buddelmeijer}, H. and {Burke}, D.~J. and {Calderone}, G. and {Cano Rodr{\'\i}guez}, J.~L. and {Cara}, M. and {Cardoso}, J.~V.~M. and {Cheedella}, S. and {Copin}, Y. and {Corrales}, L. and {Crichton}, D. and {D'Avella}, D. and {Deil}, C. and {Depagne}, {\'E}. and {Dietrich}, J.~P. and {Donath}, A. and {Droettboom}, M. and {Earl}, N. and {Erben}, T. and {Fabbro}, S. and {Ferreira}, L.~A. and {Finethy}, T. and {Fox}, R.~T. and {Garrison}, L.~H. and {Gibbons}, S.~L.~J. and {Goldstein}, D.~A. and {Gommers}, R. and {Greco}, J.~P. and {Greenfield}, P. and {Groener}, A.~M. and {Grollier}, F. and {Hagen}, A. and {Hirst}, P. and {Homeier}, D. and {Horton}, A.~J. and {Hosseinzadeh}, G. and {Hu}, L. and {Hunkeler}, J.~S. and {Ivezi{\'c}}, {\v{Z}}. and {Jain}, A. and {Jenness}, T. and {Kanarek}, G. and {Kendrew}, S. and {Kern}, N.~S. and {Kerzendorf}, W.~E. and {Khvalko}, A. and {King}, J. and {Kirkby}, D. and {Kulkarni}, A.~M. and {Kumar}, A. and {Lee}, A. and {Lenz}, D. and {Littlefair}, S.~P. and {Ma}, Z. and {Macleod}, D.~M. and {Mastropietro}, M. and {McCully}, C. and {Montagnac}, S. and {Morris}, B.~M. and {Mueller}, M. and {Mumford}, S.~J. and {Muna}, D. and {Murphy}, N.~A. and {Nelson}, S. and {Nguyen}, G.~H. and {Ninan}, J.~P. and {N{\"o}the}, M. and {Ogaz}, S. and {Oh}, S. and {Parejko}, J.~K. and {Parley}, N. and {Pascual}, S. and {Patil}, R. and {Patil}, A.~A. and {Plunkett}, A.~L. and {Prochaska}, J.~X. and {Rastogi}, T. and {Reddy Janga}, V. and {Sabater}, J. and {Sakurikar}, P. and {Seifert}, M. and {Sherbert}, L.~E. and {Sherwood-Taylor}, H. and {Shih}, A.~Y. and {Sick}, J. and {Silbiger}, M.~T. and {Singanamalla}, S. and {Singer}, L.~P. and {Sladen}, P.~H. and {Sooley}, K.~A. and {Sornarajah}, S. and {Streicher}, O. and {Teuben}, P. and {Thomas}, S.~W. and {Tremblay}, G.~R. and {Turner}, J.~E.~H. and {Terr{\'o}n}, V. and {van Kerkwijk}, M.~H. and {de la Vega}, A. and {Watkins}, L.~L. and {Weaver}, B.~A. and {Whitmore}, J.~B. and {Woillez}, J. and {Zabalza}, V. and {Astropy Contributors}},
        title = "{The Astropy Project: Building an Open-science Project and Status of the v2.0 Core Package}",
      journal = {\aj},
         year = 2018,
        month = sep,
       volume = {156},
       number = {3},
          eid = {123},
        pages = {123},
          doi = {10.3847/1538-3881/aabc4f},
archivePrefix = {arXiv},
       eprint = {1801.02634},
 primaryClass = {astro-ph.IM},
       adsurl = {https://ui.adsabs.harvard.edu/abs/2018AJ....156..123A}
}

@ARTICLE{2013A&A...558A..33A,
       author = {{Astropy Collaboration} and {Robitaille}, Thomas P. and
         {Tollerud}, Erik J. and {Greenfield}, Perry and {Droettboom}, Michael and
         {Bray}, Erik and {Aldcroft}, Tom and {Davis}, Matt and
         {Ginsburg}, Adam and {Price-Whelan}, Adrian M. and
         {Kerzendorf}, Wolfgang E. and {Conley}, Alexander and {Crighton}, Neil and
         {Barbary}, Kyle and {Muna}, Demitri and {Ferguson}, Henry and
         {Grollier}, Fr{\'e}d{\'e}ric and {Parikh}, Madhura M. and
         {Nair}, Prasanth H. and {Unther}, Hans M. and {Deil}, Christoph and
         {Woillez}, Julien and {Conseil}, Simon and {Kramer}, Roban and
         {Turner}, James E.~H. and {Singer}, Leo and {Fox}, Ryan and
         {Weaver}, Benjamin A. and {Zabalza}, Victor and {Edwards}, Zachary I. and
         {Azalee Bostroem}, K. and {Burke}, D.~J. and {Casey}, Andrew R. and
         {Crawford}, Steven M. and {Dencheva}, Nadia and {Ely}, Justin and
         {Jenness}, Tim and {Labrie}, Kathleen and {Lim}, Pey Lian and
         {Pierfederici}, Francesco and {Pontzen}, Andrew and {Ptak}, Andy and
         {Refsdal}, Brian and {Servillat}, Mathieu and {Streicher}, Ole},
        title = "{Astropy: A community Python package for astronomy}",
      journal = {\aap},
         year = "2013",
        month = "Oct",
       volume = {558},
          eid = {A33},
        pages = {A33},
          doi = {10.1051/0004-6361/201322068},
archivePrefix = {arXiv},
       eprint = {1307.6212},
 primaryClass = {astro-ph.IM},
       adsurl = {https://ui.adsabs.harvard.edu/abs/2013A&A...558A..33A}
}

@article{price2022astropy,
  title={The Astropy Project: Sustaining and Growing a Community-oriented Open-source Project and the Latest Major Release (v5. 0) of the Core Package},
  author={Price-Whelan, Adrian M and Lim, Pey Lian and Earl, Nicholas and Starkman, Nathaniel and Bradley, Larry and Shupe, David L and Patil, Aarya A and Corrales, Lia and Brasseur, CE and N{\"o}the, Maximilian and others},
  journal={The Astrophysical Journal},
  volume={935},
  number={2},
  pages={167},
  year={2022},
  publisher={IOP Publishing}
}

@article{hoyer2017xarray,
  title     = {xarray: {N-D} labeled arrays and datasets in {Python}},
  author    = {Hoyer, S. and J. Hamman},
  journal   = {Journal of Open Research Software},
  volume    = {5},
  number    = {10},
  year      = {2017},
  publisher = {Ubiquity Press},
  doi       = {10.5334/jors.148},
  url       = {https://doi.org/10.5334/jors.148}
}

@software{hoyer_stephan_2022_6323468,
  author       = {Hoyer, Stephan and
                  Roos, Maximilian and
                  Joseph, Hamman and
                  Magin, Justus and
                  Cherian, Deepak and
                  Fitzgerald, Clark and
                  Hauser, Mathias and
                  Fujii, Keisuke and
                  Maussion, Fabien and
                  Imperiale, Guido and
                  Clark, Spencer and
                  Kleeman, Alex and
                  Nicholas, Thomas and
                  Kluyver, Thomas and
                  Westling, Jimmy and
                  Munroe, James and
                  Amici, Alessandro and
                  Barghini, Aureliana and
                  Banihirwe, Anderson and
                  Bell, Ray and
                  Hatfield-Dodds, Zac and
                  Abernathey, Ryan and
                  Bovy, Benoît and
                  Omotani, John and
                  Mühlbauer, Kai and
                  Roszko, Maximilian K. and
                  Wolfram, Phillip J.},
  title        = {xarray},
  month        = mar,
  year         = 2022,
  note         = {If you use this software, please cite it as below.},
  publisher    = {Zenodo},
  version      = {v2022.03.0},
  doi          = {10.5281/zenodo.6323468},
  url          = {https://doi.org/10.5281/zenodo.6323468}
}

@ARTICLE{Kreidberg2015_batman,
       author = {{Kreidberg}, Laura},
        title = "{batman: BAsic Transit Model cAlculatioN in Python}",
      journal = {\pasp},
         year = 2015,
        month = nov,
       volume = {127},
       number = {957},
        pages = {1161},
          doi = {10.1086/683602},
archivePrefix = {arXiv},
       eprint = {1507.08285},
 primaryClass = {astro-ph.EP},
       adsurl = {https://ui.adsabs.harvard.edu/abs/2015PASP..127.1161K}
}

@ARTICLE{Stevenson2025,
       author = {{Stevenson}, Kevin B. and {Lustig-Yaeger}, Jacob and {May}, E.~M. and {Kopparapu}, Ravi K. and {Fauchez}, Thomas J. and {Haqq-Misra}, Jacob and {Limbach}, Mary Anne and {Schwieterman}, Edward W. and {Sotzen}, Kristin S. and {Tsai}, Shang-Min},
        title = "{K2-18b Does Not Meet the Standards of Evidence for Life}",
      journal = {\aj},
         year = 2025,
        month = nov,
       volume = {170},
       number = {5},
          eid = {257},
        pages = {257},
          doi = {10.3847/1538-3881/ae0338},
archivePrefix = {arXiv},
       eprint = {2508.05961},
 primaryClass = {astro-ph.EP},
       adsurl = {https://ui.adsabs.harvard.edu/abs/2025AJ....170..257S}
}

@ARTICLE{Chen2025WeatherReport,
       author = {{Chen}, Xueqing and {Biller}, Beth A. and {Tan}, Xianyu and {Vos}, Johanna M. and {Zhou}, Yifan and {Su{\'a}rez}, Genaro and {McCarthy}, Allison M. and {Morley}, Caroline V. and {Whiteford}, Niall and {Dupuy}, Trent J. and {Faherty}, Jacqueline and {Sutlieff}, Ben J. and {Oliveros-Gomez}, Natalia and {Manjavacas}, Elena and {Limbach}, Mary Anne and {Lee}, Elspeth K.~H. and {Karalidi}, Theodora and {Crossfield}, Ian J.~M. and {Liu}, Pengyu and {Molliere}, Paul and {Muirhead}, Philip S. and {Henning}, Thomas and {Mace}, Gregory and {Crouzet}, Nicolas and {Kataria}, Tiffany},
        title = "{The JWST weather report from the nearest brown dwarfs II: consistent variability mechanisms over 7 months revealed by 1{\textendash}14 {\ensuremath{\mu}}m NIRSpec + MIRI monitoring of WISE 1049AB}",
      journal = {\mnras},
         year = 2025,
        month = jun,
       volume = {539},
       number = {4},
        pages = {3758-3777},
          doi = {10.1093/mnras/staf737},
archivePrefix = {arXiv},
       eprint = {2505.00794},
 primaryClass = {astro-ph.SR},
       adsurl = {https://ui.adsabs.harvard.edu/abs/2025MNRAS.539.3758C}
}

@ARTICLE{Biller2024WeatherReport,
       author = {{Biller}, Beth A. and {Vos}, Johanna M. and {Zhou}, Yifan and {McCarthy}, Allison M. and {Tan}, Xianyu and {Crossfield}, Ian J.~M. and {Whiteford}, Niall and {Suarez}, Genaro and {Faherty}, Jacqueline and {Manjavacas}, Elena and {Chen}, Xueqing and {Liu}, Pengyu and {Sutlieff}, Ben J. and {Limbach}, Mary Anne and {Molliere}, Paul and {Dupuy}, Trent J. and {Oliveros-Gomez}, Natalia and {Muirhead}, Philip S. and {Henning}, Thomas and {Mace}, Gregory and {Crouzet}, Nicolas and {Karalidi}, Theodora and {Morley}, Caroline V. and {Tremblin}, Pascal and {Kataria}, Tiffany},
        title = "{The JWST weather report from the nearest brown dwarfs I: multiperiod JWST NIRSpec + MIRI monitoring of the benchmark binary brown dwarf WISE 1049AB}",
      journal = {\mnras},
         year = 2024,
        month = aug,
       volume = {532},
       number = {2},
        pages = {2207-2233},
          doi = {10.1093/mnras/stae1602},
archivePrefix = {arXiv},
       eprint = {2407.09194},
 primaryClass = {astro-ph.SR},
       adsurl = {https://ui.adsabs.harvard.edu/abs/2024MNRAS.532.2207B}
}

@ARTICLE{Murphy2025,
       author = {{Murphy}, Matthew M. and {Beatty}, Thomas G. and {Schlawin}, Everett and {Bell}, Taylor J. and {Radica}, Michael and {Kennedy}, Thomas D. and {Mehta}, Nishil and {Welbanks}, Luis and {Line}, Michael R. and {Parmentier}, Vivien and {Greene}, Thomas P. and {Mukherjee}, Sagnick and {Fortney}, Jonathan J. and {Ohno}, Kazumasa and {Wiser}, Lindsey and {Arnold}, Kenneth and {Rauscher}, Emily and {Edelman}, Isaac R. and {Rieke}, Marcia J.},
        title = "{A Panchromatic Characterization of the Evening and Morning Atmosphere of WASP-107 b: Composition and Cloud Variations, and Insight into the Effect of Stellar Contamination}",
      journal = {\aj},
         year = 2025,
        month = jul,
       volume = {170},
       number = {1},
          eid = {61},
        pages = {61},
          doi = {10.3847/1538-3881/addf38},
archivePrefix = {arXiv},
       eprint = {2505.13602},
 primaryClass = {astro-ph.EP},
       adsurl = {https://ui.adsabs.harvard.edu/abs/2025AJ....170...61M}
}

@ARTICLE{sing2024,
       author = {{Sing}, David K. and {Rustamkulov}, Zafar and {Thorngren}, Daniel P. and {Barstow}, Joanna K. and {Tremblin}, Pascal and {Alves de Oliveira}, Catarina and {Beck}, Tracy L. and {Birkmann}, Stephan M. and {Challener}, Ryan C. and {Crouzet}, Nicolas and {Espinoza}, N{\'e}stor and {Ferruit}, Pierre and {Giardino}, Giovanna and {Gressier}, Am{\'e}lie and {Lee}, Elspeth K.~H. and {Lewis}, Nikole K. and {Maiolino}, Roberto and {Manjavacas}, Elena and {Rauscher}, Bernard J. and {Sirianni}, Marco and {Valenti}, Jeff A.},
        title = "{A warm Neptune's methane reveals core mass and vigorous atmospheric mixing}",
      journal = {\nat},
         year = 2024,
        month = jun,
       volume = {630},
       number = {8018},
        pages = {831-835},
          doi = {10.1038/s41586-024-07395-z},
archivePrefix = {arXiv},
       eprint = {2405.11027},
 primaryClass = {astro-ph.EP},
       adsurl = {https://ui.adsabs.harvard.edu/abs/2024Natur.630..831S}
}

@ARTICLE{Polanski2022,
       author = {{Polanski}, Alex S. and {Crossfield}, Ian J.~M. and {Howard}, Andrew W. and {Isaacson}, Howard and {Rice}, Malena},
        title = "{Chemical Abundances for 25 JWST Exoplanet Host Stars with KeckSpec}",
      journal = {Research Notes of the American Astronomical Society},
         year = 2022,
        month = aug,
       volume = {6},
       number = {8},
          eid = {155},
        pages = {155},
          doi = {10.3847/2515-5172/ac8676},
archivePrefix = {arXiv},
       eprint = {2207.13662},
 primaryClass = {astro-ph.EP},
       adsurl = {https://ui.adsabs.harvard.edu/abs/2022RNAAS...6..155P}
}

@ARTICLE{Calamari2024,
       author = {{Calamari}, Emily and {Faherty}, Jacqueline K. and {Visscher}, Channon and {Gemma}, Marina E. and {Burningham}, Ben and {Rothermich}, Austin},
        title = "{Predicting Cloud Conditions in Substellar Mass Objects Using Ultracool Dwarf Companions}",
      journal = {\apj},
         year = 2024,
        month = mar,
       volume = {963},
       number = {1},
          eid = {67},
        pages = {67},
          doi = {10.3847/1538-4357/ad1f6d},
archivePrefix = {arXiv},
       eprint = {2401.11038},
 primaryClass = {astro-ph.SR},
       adsurl = {https://ui.adsabs.harvard.edu/abs/2024ApJ...963...67C}
}

@ARTICLE{Moran2024,
       author = {{Moran}, Sarah E. and {Marley}, Mark S. and {Crossley}, Samuel D.},
        title = "{Neglected Silicon Dioxide Polymorphs as Clouds in Substellar Atmospheres}",
      journal = {\apjl},
         year = 2024,
        month = sep,
       volume = {973},
       number = {1},
          eid = {L3},
        pages = {L3},
          doi = {10.3847/2041-8213/ad72e7},
archivePrefix = {arXiv},
       eprint = {2408.00698},
 primaryClass = {astro-ph.EP},
       adsurl = {https://ui.adsabs.harvard.edu/abs/2024ApJ...973L...3M}
}

@dataset{lupu_2022_6600976,
  author       = {Lupu, Roxana and
                  Freedman, Richard and
                  Gharib-Nezhad, Ehsan and
                  Molliere, Paul},
  title        = {High resolution opacities for H2/He atmospheres},
  month        = jun,
  year         = 2022,
  publisher    = {Zenodo},
  doi          = {10.5281/zenodo.6600976},
  url          = {https://doi.org/10.5281/zenodo.6600976},
}

@ARTICLE{Boehm2025,
       author = {{Boehm}, V.~A. and {Lewis}, N.~K. and {Fairman}, C.~E. and {Moran}, S.~E. and {Gasc{\'o}n}, C. and {Wakeford}, H.~R. and {Alam}, M.~K. and {Alderson}, L. and {Barstow}, J. and {Batalha}, N.~E. and {Grant}, D. and {L{\'o}pez-Morales}, M. and {MacDonald}, R.~J. and {Marley}, Mark S. and {Ohno}, K.},
        title = "{The HUSTLE Program: The UV to Near-infrared HST WFC3/UVIS G280 Transmission Spectrum of WASP-127b}",
      journal = {\aj},
         year = 2025,
        month = jan,
       volume = {169},
       number = {1},
          eid = {23},
        pages = {23},
          doi = {10.3847/1538-3881/ad8dde},
archivePrefix = {arXiv},
       eprint = {2410.17368},
 primaryClass = {astro-ph.EP},
       adsurl = {https://ui.adsabs.harvard.edu/abs/2025AJ....169...23B}
}

@ARTICLE{hitran2012,
	author= {Rothman, L. S. and I. E. Gordon and Y. Babikov and A. Barbe and D. Chris Benner and P. F. Bernath and M. Birk and L. Bizzocchi and V. Boudon and L. R. Brown and A. Campargue and K. Chance and E. A. Cohen and L. H. Coudert and V. M. Devi and B. J. Drouin and A. Fayt and J.-M. Flaud and R. R. Gamache and J. J. Harrison and J.-M. Hartmann and C. Hill and J. T. Hodges and D. Jacquemart and A. Jolly and J. Lamouroux and R. J. Le Roy and G. Li and D. A. Long and O. M. Lyulin and C. J. Mackie and S. T. Massie and S. Mikhailenko and H. S. P. Müller and O. V. Naumenko and A. V. Nikitin and J. Orphal and V. Perevalov and A. Perrin and E. R. Polovtseva and C. Richard and M. A. H. Smith and E. Starikova and K. Sung and S. Tashkun and J. Tennyson and G. C. Toon and V. G. Tyuterev and  G. Wagner},
	title= {The HITRAN2012 molecular spectroscopic database},
	journal= {Journal of Quantitative Spectroscopy and Radiative Transfer},
	volume= {130},
	year= {2013},
	doi= {10.1016/j.jqsrt.2013.07.002},
}

@ARTICLE{yurchenko13vibrational,
       author = {{Yurchenko}, Sergei N. and {Tennyson}, Jonathan and {Barber}, Robert J. and
         {Thiel}, Walter},
        title = "{Vibrational transition moments of CH$_{4}$ from first principles}",
      journal = {Journal of Molecular Spectroscopy},
         year = 2013,
        month = sep,
       volume = {291},
        pages = {69-76},
          doi = {10.1016/j.jms.2013.05.014},
archivePrefix = {arXiv},
       eprint = {1302.1720},
 primaryClass = {astro-ph.SR},
       adsurl = {https://ui.adsabs.harvard.edu/abs/2013JMoSp.291...69Y}
}

@article{yurchenko_2014,
    author = {Yurchenko, Sergei N. and Tennyson, Jonathan},
    title = "{ExoMol line lists – IV. The rotation–vibration spectrum of methane up to 1500 K}",
    journal = {Monthly Notices of the Royal Astronomical Society},
    volume = {440},
    number = {2},
    pages = {1649-1661},
    year = {2014},
    month = {03},
    issn = {0035-8711},
    doi = {10.1093/mnras/stu326},
    url = {https://doi.org/10.1093/mnras/stu326},
    eprint = {https://academic.oup.com/mnras/article-pdf/440/2/1649/18753919/stu326.pdf},
}

@ARTICLE{HITEMP2010,
	author= {Rothman, L. S. and I. E. Gordon and R. J. Barber and H. Dothe and R. R. Gamache and A. Goldman and V. I. Perevalov and S. A. Tashkun and  J. Tennyson},
	title= {HITEMP, the high-temperature molecular spectroscopic database},
	journal= {Journal of Quantitative Spectroscopy and Radiative Transfer},
	volume= {111},
	year= {2010},
	doi= {10.1016/j.jqsrt.2010.05.001},
}

@INPROCEEDINGS{HITRAN2016,
       author = {{Gordon}, Iouli E. and {Rothman}, Laurence S. and {Tan}, Yan and {Kochanov}, Roman V. and {Hill}, Christian},
        title = "{HITRAN2016: Part I. Line lists for H\_2O, CO\_2, O\_3, N\_2O, CO, CH\_4, and O\_2}",
    booktitle = {72nd International Symposium on Molecular Spectroscopy},
         year = 2017,
        month = jun,
          eid = {TJ08},
        pages = {TJ08},
          doi = {10.15278/isms.2017.TJ08},
       adsurl = {https://ui.adsabs.harvard.edu/abs/2017isms.confETJ08G}
}

@ARTICLE{Mizus2017H3p,
       author = {{Mizus}, Irina I. and {Alijah}, Alexander and {Zobov}, Nikolai F. and {Lodi}, Lorenzo and {Kyuberis}, Aleksandra A. and {Yurchenko}, Sergei N. and {Tennyson}, Jonathan and {Polyansky}, Oleg L.},
        title = {ExoMol molecular line lists - XX. A comprehensive line list for H3+},
      journal = {\mnras},
         year = {2017},
        month = {jun},
       volume = {468},
       number = {2},
        pages = {1717-1725},
          doi = {10.1093/mnras/stx502},
archivePrefix = {arXiv},
       eprint = {1704.04096},
 primaryClass = {astro-ph.GA},
       adsurl = {https://ui.adsabs.harvard.edu/abs/2017MNRAS.468.1717M}
}

@ARTICLE{Polyansky2018H2O,
       author = {{Polyansky}, Oleg L. and {Kyuberis}, Aleksandra A. and {Zobov}, Nikolai F. and {Tennyson}, Jonathan and {Yurchenko}, Sergei N. and {Lodi}, Lorenzo},
        title = "{ExoMol molecular line lists XXX: a complete high-accuracy line list for water}",
      journal = {\mnras},
         year = 2018,
        month = oct,
       volume = {480},
       number = {2},
        pages = {2597-2608},
          doi = {10.1093/mnras/sty1877},
archivePrefix = {arXiv},
       eprint = {1807.04529},
 primaryClass = {astro-ph.EP},
       adsurl = {https://ui.adsabs.harvard.edu/abs/2018MNRAS.480.2597P}
}

@article{azzam16exomol,
    author = {Azzam, Ala'a A. A. and Tennyson, Jonathan and Yurchenko, Sergei N. and Naumenko, Olga V.},
    title = "{ExoMol molecular line lists – XVI. The rotation–vibration spectrum of hot H2S}",
    journal = {Monthly Notices of the Royal Astronomical Society},
    volume = {460},
    number = {4},
    pages = {4063-4074},
    year = {2016},
    month = {05},
    issn = {0035-8711},
    doi = {10.1093/mnras/stw1133},
    url = {https://doi.org/10.1093/mnras/stw1133},
    eprint = {https://academic.oup.com/mnras/article-pdf/460/4/4063/13773124/stw1133.pdf},
}

@ARTICLE{Harris2006hcn,
       author = {{Harris}, G.~J. and {Tennyson}, J. and {Kaminsky}, B.~M. and {Pavlenko}, Ya. V. and {Jones}, H.~R.~A.},
        title = "{Improved HCN/HNC linelist, model atmospheres and synthetic spectra for WZ Cas}",
      journal = {\mnras},
         year = 2006,
        month = mar,
       volume = {367},
       number = {1},
        pages = {400-406},
          doi = {10.1111/j.1365-2966.2005.09960.x},
archivePrefix = {arXiv},
       eprint = {astro-ph/0512363},
 primaryClass = {astro-ph},
       adsurl = {https://ui.adsabs.harvard.edu/abs/2006MNRAS.367..400H}
}

@ARTICLE{Barber2014HCN,
       author = {{Barber}, R.~J. and {Strange}, J.~K. and {Hill}, C. and {Polyansky}, O.~L. and {Mellau}, G. Ch. and {Yurchenko}, S.~N. and {Tennyson}, Jonathan},
        title = "{ExoMol line lists - III. An improved hot rotation-vibration line list for HCN and HNC}",
      journal = {\mnras},
         year = 2014,
        month = jan,
       volume = {437},
       number = {2},
        pages = {1828-1835},
          doi = {10.1093/mnras/stt2011},
archivePrefix = {arXiv},
       eprint = {1311.1328},
 primaryClass = {astro-ph.SR},
       adsurl = {https://ui.adsabs.harvard.edu/abs/2014MNRAS.437.1828B}
}

@ARTICLE{hitran2020,
       author = {{Gordon}, I.~E. and {Rothman}, L.~S. and {Hargreaves}, R.~J. and {Hashemi}, R. and {Karlovets}, E.~V. and {Skinner}, F.~M. and {Conway}, E.~K. and {Hill}, C. and {Kochanov}, R.~V. and {Tan}, Y. and {Wcis{\l}o}, P. and {Finenko}, A.~A. and {Nelson}, K. and {Bernath}, P.~F. and {Birk}, M. and {Boudon}, V. and {Campargue}, A. and {Chance}, K.~V. and {Coustenis}, A. and {Drouin}, B.~J. and {Flaud}, J. -M. and {Gamache}, R.~R. and {Hodges}, J.~T. and {Jacquemart}, D. and {Mlawer}, E.~J. and {Nikitin}, A.~V. and {Perevalov}, V.~I. and {Rotger}, M. and {Tennyson}, J. and {Toon}, G.~C. and {Tran}, H. and {Tyuterev}, V.~G. and {Adkins}, E.~M. and {Baker}, A. and {Barbe}, A. and {Can{\`e}}, E. and {Cs{\'a}sz{\'a}r}, A.~G. and {Dudaryonok}, A. and {Egorov}, O. and {Fleisher}, A.~J. and {Fleurbaey}, H. and {Foltynowicz}, A. and {Furtenbacher}, T. and {Harrison}, J.~J. and {Hartmann}, J. -M. and {Horneman}, V. -M. and {Huang}, X. and {Karman}, T. and {Karns}, J. and {Kassi}, S. and {Kleiner}, I. and {Kofman}, V. and {Kwabia-Tchana}, F. and {Lavrentieva}, N.~N. and {Lee}, T.~J. and {Long}, D.~A. and {Lukashevskaya}, A.~A. and {Lyulin}, O.~M. and {Makhnev}, V. Yu. and {Matt}, W. and {Massie}, S.~T. and {Melosso}, M. and {Mikhailenko}, S.~N. and {Mondelain}, D. and {M{\"u}ller}, H.~S.~P. and {Naumenko}, O.~V. and {Perrin}, A. and {Polyansky}, O.~L. and {Raddaoui}, E. and {Raston}, P.~L. and {Reed}, Z.~D. and {Rey}, M. and {Richard}, C. and {T{\'o}bi{\'a}s}, R. and {Sadiek}, I. and {Schwenke}, D.~W. and {Starikova}, E. and {Sung}, K. and {Tamassia}, F. and {Tashkun}, S.~A. and {Vander Auwera}, J. and {Vasilenko}, I.~A. and {Vigasin}, A.~A. and {Villanueva}, G.~L. and {Vispoel}, B. and {Wagner}, G. and {Yachmenev}, A. and {Yurchenko}, S.~N.},
        title = "{The HITRAN2020 molecular spectroscopic database}",
      journal = {\jqsrt},
         year = 2022,
        month = jan,
       volume = {277},
          eid = {107949},
        pages = {107949},
          doi = {10.1016/j.jqsrt.2021.107949},
       adsurl = {https://ui.adsabs.harvard.edu/abs/2022JQSRT.27707949G}
}

@ARTICLE{Bittner2018Lis,
       author = {{Bittner}, D.~M. and {Bernath}, P.~F.},
        title = "{VizieR Online Data Catalog: Line lists for X$^{1}$\{Sigma\}$^{+}$ LiF and LiCl (}",
      journal = {VizieR Online Data Catalog},
         year = 2018,
        month = aug,
          eid = {J/ApJS/235/8},
        pages = {J/ApJS/235/8},
       adsurl = {https://ui.adsabs.harvard.edu/abs/2018yCat..22350008B}
}

@ARTICLE{Coppola2011LiH,
       author = {{Coppola}, C.~M. and {Lodi}, L. and {Tennyson}, J.},
        title = "{Radiative cooling functions for primordial molecules}",
      journal = {\mnras},
         year = 2011,
        month = jul,
       volume = {415},
       number = {1},
        pages = {487-493},
          doi = {10.1111/j.1365-2966.2011.18723.x},
archivePrefix = {arXiv},
       eprint = {1103.2957},
 primaryClass = {astro-ph.CO},
       adsurl = {https://ui.adsabs.harvard.edu/abs/2011MNRAS.415..487C}
}

@ARTICLE{Yadin2012MgH,
       author = {{Yadin}, Benjamin and {Veness}, Thomas and {Conti}, Pierandrea and {Hill}, Christian and {Yurchenko}, Sergei N. and {Tennyson}, Jonathan},
        title = "{ExoMol line lists - I. The rovibrational spectrum of BeH, MgH and CaH in the X $^{2}${\ensuremath{\Sigma}}$^{+}$ state}",
      journal = {\mnras},
         year = 2012,
        month = sep,
       volume = {425},
       number = {1},
        pages = {34-43},
          doi = {10.1111/j.1365-2966.2012.21367.x},
archivePrefix = {arXiv},
       eprint = {1204.0137},
 primaryClass = {astro-ph.SR},
       adsurl = {https://ui.adsabs.harvard.edu/abs/2012MNRAS.425...34Y}
}

@ARTICLE{GharibNezhad2013MgH,
       author = {{GharibNezhad}, Ehsan and {Shayesteh}, Alireza and {Bernath}, Peter F.},
        title = "{Einstein A coefficients for rovibronic lines of the A $^{2}$ {\ensuremath{\Pi}} {\textrightarrow} X $^{2}${\ensuremath{\Sigma}}$^{+}$ and B' $^{2}${\ensuremath{\Sigma}}$^{+}${\textrightarrow} X $^{2}$ {\ensuremath{\Sigma}}$^{+}$ transitions of MgH}",
      journal = {\mnras},
         year = 2013,
        month = jul,
       volume = {432},
       number = {3},
        pages = {2043-2047},
          doi = {10.1093/mnras/stt510},
       adsurl = {https://ui.adsabs.harvard.edu/abs/2013MNRAS.432.2043G}
}

@ARTICLE{GharibNezhad2021,
       author = {{Gharib-Nezhad}, Ehsan and {Iyer}, Aishwarya R. and {Line}, Michael R. and {Freedman}, Richard S. and {Marley}, Mark S. and {Batalha}, Natasha E.},
        title = "{EXOPLINES: Molecular Absorption Cross-section Database for Brown Dwarf and Giant Exoplanet Atmospheres}",
      journal = {\apjs},
         year = 2021,
        month = jun,
       volume = {254},
       number = {2},
          eid = {34},
        pages = {34},
          doi = {10.3847/1538-4365/abf504},
archivePrefix = {arXiv},
       eprint = {2104.00264},
 primaryClass = {astro-ph.EP},
       adsurl = {https://ui.adsabs.harvard.edu/abs/2021ApJS..254...34G}
}

@ARTICLE{li15rovibrational,
       author = {{Li}, Gang and {Gordon}, Iouli E. and {Rothman}, Laurence S. and
         {Tan}, Yan and {Hu}, Shui-Ming and {Kassi}, Samir and
         {Campargue}, Alain and {Medvedev}, Emile S.},
        title = "{Rovibrational Line Lists for Nine Isotopologues of the CO Molecule in the X $^{1}${\ensuremath{\Sigma}}$^{+}$ Ground Electronic State}",
      journal = {\apjs},
         year = 2015,
        month = jan,
       volume = {216},
       number = {1},
          eid = {15},
        pages = {15},
          doi = {10.1088/0067-0049/216/1/15},
       adsurl = {https://ui.adsabs.harvard.edu/abs/2015ApJS..216...15L}
}

@article{HUANG2014reliable,
title = "Reliable infrared line lists for 13 CO2 isotopologues up to E′=18,000cm−1 and 1500K, with line shape parameters",
journal = "Journal of Quantitative Spectroscopy and Radiative Transfer",
volume = "147",
pages = "134 - 144",
year = "2014",
issn = "0022-4073",
doi = "https://doi.org/10.1016/j.jqsrt.2014.05.015",
url = "http://www.sciencedirect.com/science/article/pii/S0022407314002246",
author = "Xinchuan {Huang} and Robert R. Gamache and Richard S. Freedman and David W. Schwenke and Timothy J. Lee"
}

@ARTICLE{Burrows02_CrH,
	author= {Burrows, A. and R. S. Ram and P. Bernath and C. M. Sharp and  J. A. Milsom},
	title= {New CrH Opacities for the Study of L and Brown Dwarf Atmospheres},
	journal= {The Astrophysical Journal},
	volume= {577},
	year= {2002},
	doi= {10.1086/342242},
}

@article{Ryabchikova2015,
	doi = {10.1088/0031-8949/90/5/054005},
	url = {https://doi.org/10.1088/0031-8949/90/5/054005},
	year = 2015,
	month = {apr},
	publisher = {{IOP} Publishing},
	volume = {90},
	number = {5},
	pages = {054005},
	author = {T Ryabchikova and N Piskunov and R L Kurucz and H C Stempels and U Heiter and Yu Pakhomov and P S Barklem},
	title = {A major upgrade of the {VALD} database},
	journal = {Physica Scripta}
}

@ARTICLE{oBrian1991Fe,
       author = {{O'Brian}, T.~R. and {Wickliffe}, M.~E. and {Lawler}, J.~E. and {Whaling}, W. and {Brault}, James W.},
        title = "{Lifetimes, transition probabilities, and level energies in Fe i}",
      journal = {Journal of the Optical Society of America B Optical Physics},
         year = 1991,
        month = jan,
       volume = {8},
       number = {6},
        pages = {1185-1201},
          doi = {10.1364/JOSAB.8.001185},
       adsurl = {https://ui.adsabs.harvard.edu/abs/1991JOSAB...8.1185O}
}

@ARTICLE{Fuhr1988Fe,
       author = {{Fuhr}, J.~R. and {Martin}, G.~A. and {Wiese}, W.~L.},
        title = "{Atomic transition probabilities. Iron through Nickel}",
      journal = {Journal of Physical and Chemical Reference Data},
         year = 1988,
        month = jan,
       volume = {17},
       adsurl = {https://ui.adsabs.harvard.edu/abs/1988JPCRD..17S....F}
}

@ARTICLE{Bard1991Fe,
       author = {{Bard}, A. and {Kock}, A. and {Kock}, M.},
        title = "{Fe I oscillator strengths of lines of astrophysical interest.}",
      journal = {\aap},
         year = 1991,
        month = aug,
       volume = {248},
        pages = {315-322},
       adsurl = {https://ui.adsabs.harvard.edu/abs/1991A&A...248..315B}
}

@ARTICLE{Dulick2003FeH,
       author = {{Dulick}, M. and {Bauschlicher}, C.~W., Jr. and {Burrows}, Adam and {Sharp}, C.~M. and {Ram}, R.~S. and {Bernath}, Peter},
        title = "{Line Intensities and Molecular Opacities of the FeH F $^{4}${\ensuremath{\Delta}}$_{i}$-X $^{4}${\ensuremath{\Delta}}$_{i}$ Transition}",
      journal = {\apj},
         year = 2003,
        month = sep,
       volume = {594},
       number = {1},
        pages = {651-663},
          doi = {10.1086/376791},
archivePrefix = {arXiv},
       eprint = {astro-ph/0305162},
 primaryClass = {astro-ph},
       adsurl = {https://ui.adsabs.harvard.edu/abs/2003ApJ...594..651D}
}

@ARTICLE{Hargreaves2010FeH,
       author = {{Hargreaves}, Robert J. and {Hinkle}, Kenneth H. and {Bauschlicher}, Charles W., Jr. and {Wende}, Sebastian and {Seifahrt}, Andreas and {Bernath}, Peter F.},
        title = "{High-resolution 1.6 {\ensuremath{\mu}}m Spectra of FeH in M and L Dwarfs}",
      journal = {\aj},
         year = 2010,
        month = oct,
       volume = {140},
       number = {4},
        pages = {919-924},
          doi = {10.1088/0004-6256/140/4/919},
       adsurl = {https://ui.adsabs.harvard.edu/abs/2010AJ....140..919H}
}

@ARTICLE{Bard1994Fe,
       author = {{Bard}, A. and {Kock}, M.},
        title = "{Fe I oscillator strengths for lines with excitation energies between 3 and 7eV.}",
      journal = {\aap},
         year = 1994,
        month = feb,
       volume = {282},
        pages = {1014-1020},
       adsurl = {https://ui.adsabs.harvard.edu/abs/1994A&A...282.1014B}
}

@article{yurchenko11vibrationally,
    author = {Yurchenko, S. N. and Barber, R. J. and Tennyson, J.},
    title = "{A variationally computed line list for hot NH3}",
    journal = {Monthly Notices of the Royal Astronomical Society},
    volume = {413},
    number = {3},
    pages = {1828-1834},
    year = {2011},
    month = {05},
    issn = {0035-8711},
    doi = {10.1111/j.1365-2966.2011.18261.x},
    url = {https://doi.org/10.1111/j.1365-2966.2011.18261.x},
    eprint = {https://academic.oup.com/mnras/article-pdf/413/3/1828/2883744/mnras0413-1828.pdf},
}

@article{Wilzewski16,
title = {H2, He, and CO2 line-broadening coefficients, pressure shifts and temperature-dependence exponents for the HITRAN database. Part 1: SO2, NH3, HF, HCl, OCS and C2H2},
journal = {Journal of Quantitative Spectroscopy and Radiative Transfer},
volume = {168},
pages = {193-206},
year = {2016},
issn = {0022-4073},
doi = {https://doi.org/10.1016/j.jqsrt.2015.09.003},
url = {https://www.sciencedirect.com/science/article/pii/S0022407315002988},
author = {Jonas S. Wilzewski and Iouli E. Gordon and Roman V. Kochanov and Christian Hill and Laurence S. Rothman}
}

@article{sousa14exomol,
    author = {Sousa-Silva, Clara and Al-Refaie, Ahmed F. and Tennyson, Jonathan and Yurchenko, Sergei N.},
    title = "{ExoMol line lists – VII. The rotation–vibration spectrum of phosphine up to 1500 K}",
    journal = {Monthly Notices of the Royal Astronomical Society},
    volume = {446},
    number = {3},
    pages = {2337-2347},
    year = {2014},
    month = {11},
    issn = {0035-8711},
    doi = {10.1093/mnras/stu2246},
    url = {https://doi.org/10.1093/mnras/stu2246},
    eprint = {https://academic.oup.com/mnras/article-pdf/446/3/2337/13766614/stu2246.pdf},
}

@ARTICLE{Barton2013SiO,
       author = {{Barton}, Emma J. and {Yurchenko}, Sergei N. and {Tennyson}, Jonathan},
        title = "{ExoMol line lists - II. The ro-vibrational spectrum of SiO}",
      journal = {\mnras},
         year = 2013,
        month = sep,
       volume = {434},
       number = {2},
        pages = {1469-1475},
          doi = {10.1093/mnras/stt1105},
archivePrefix = {arXiv},
       eprint = {1307.2300},
 primaryClass = {astro-ph.SR},
       adsurl = {https://ui.adsabs.harvard.edu/abs/2013MNRAS.434.1469B}
}

@ARTICLE{McKemmish2019TiO,
       author = {{McKemmish}, Laura K. and {Masseron}, Thomas and {Hoeijmakers}, H. Jens and {P{\'e}rez-Mesa}, V{\'\i}ctor and {Grimm}, Simon L. and {Yurchenko}, Sergei N. and {Tennyson}, Jonathan},
        title = "{ExoMol molecular line lists - XXXIII. The spectrum of Titanium Oxide}",
      journal = {\mnras},
         year = 2019,
        month = sep,
       volume = {488},
       number = {2},
        pages = {2836-2854},
          doi = {10.1093/mnras/stz1818},
archivePrefix = {arXiv},
       eprint = {1905.04587},
 primaryClass = {astro-ph.SR},
       adsurl = {https://ui.adsabs.harvard.edu/abs/2019MNRAS.488.2836M}
}

@ARTICLE{McKemmish16,
       author = {{McKemmish}, Laura K. and {Yurchenko}, Sergei N. and
         {Tennyson}, Jonathan},
        title = "{ExoMol line lists - XVIII. The high-temperature spectrum of VO}",
      journal = {\mnras},
         year = 2016,
        month = nov,
       volume = {463},
       number = {1},
        pages = {771-793},
          doi = {10.1093/mnras/stw1969},
archivePrefix = {arXiv},
       eprint = {1609.06120},
 primaryClass = {astro-ph.SR},
       adsurl = {https://ui.adsabs.harvard.edu/abs/2016MNRAS.463..771M}
}

@ARTICLE{Allard2016,
       author = {{Allard}, N.~F. and {Spiegelman}, F. and {Kielkopf}, J.~F.},
        title = "{K-H$_{2}$ line shapes for the spectra of cool brown dwarfs}",
      journal = {\aap},
         year = 2016,
        month = may,
       volume = {589},
          eid = {A21},
        pages = {A21},
          doi = {10.1051/0004-6361/201628270},
       adsurl = {https://ui.adsabs.harvard.edu/abs/2016A&A...589A..21A}
}

@ARTICLE{Allard2007EPJD,
       author = {{Allard}, N.~F. and {Kielkopf}, J.~F. and {Allard}, F.},
        title = "{Impact broadening of alkali lines in brown dwarfs}",
      journal = {European Physical Journal D},
         year = 2007,
        month = sep,
       volume = {44},
       number = {3},
        pages = {507-514},
          doi = {10.1140/epjd/e2007-00230-6},
       adsurl = {https://ui.adsabs.harvard.edu/abs/2007EPJD...44..507A}
}

@ARTICLE{Allard2007AA,
       author = {{Allard}, F. and {Allard}, N.~F. and {Homeier}, D. and {Kielkopf}, J. and {McCaughrean}, M.~J. and {Spiegelman}, F.},
        title = "{K-H$_{2}$ quasi-molecular absorption detected in the T-dwarf \textbackslashvarepsilon Indi Ba}",
      journal = {\aap},
         year = 2007,
        month = nov,
       volume = {474},
       number = {2},
        pages = {L21-L24},
          doi = {10.1051/0004-6361:20078362},
archivePrefix = {arXiv},
       eprint = {0709.1192},
 primaryClass = {astro-ph},
       adsurl = {https://ui.adsabs.harvard.edu/abs/2007A&A...474L..21A}
}

@ARTICLE{Allard2019,
       author = {{Allard}, N.~F. and {Spiegelman}, F. and {Leininger}, T. and {Molliere}, P.},
        title = "{New study of the line profiles of sodium perturbed by H$_{2}$}",
      journal = {\aap},
         year = 2019,
        month = aug,
       volume = {628},
          eid = {A120},
        pages = {A120},
          doi = {10.1051/0004-6361/201935593},
archivePrefix = {arXiv},
       eprint = {1908.01989},
 primaryClass = {astro-ph.SR},
       adsurl = {https://ui.adsabs.harvard.edu/abs/2019A&A...628A.120A}
}

@ARTICLE{suarez2022,
       author = {{Su{\'a}rez}, Genaro and {Metchev}, Stanimir},
        title = "{Ultracool dwarfs observed with the Spitzer infrared spectrograph - II. Emergence and sedimentation of silicate clouds in L dwarfs, and analysis of the full M5-T9 field dwarf spectroscopic sample}",
      journal = {\mnras},
         year = 2022,
        month = jul,
       volume = {513},
       number = {4},
        pages = {5701-5726},
          doi = {10.1093/mnras/stac1205},
archivePrefix = {arXiv},
       eprint = {2205.00168},
 primaryClass = {astro-ph.SR},
       adsurl = {https://ui.adsabs.harvard.edu/abs/2022MNRAS.513.5701S}
}

@ARTICLE{buchner2021,
       author = {{Buchner}, Johannes},
        title = "{UltraNest - a robust, general purpose Bayesian inference engine}",
      journal = {The Journal of Open Source Software},
         year = 2021,
        month = apr,
       volume = {6},
       number = {60},
          eid = {3001},
        pages = {3001},
          doi = {10.21105/joss.03001},
archivePrefix = {arXiv},
       eprint = {2101.09604},
 primaryClass = {stat.CO},
       adsurl = {https://ui.adsabs.harvard.edu/abs/2021JOSS....6.3001B}
}

@ARTICLE{Kiefer2024,
       author = {{Kiefer}, S. and {Lecoq-Molinos}, H. and {Helling}, Ch. and {Bangera}, N. and {Decin}, L.},
        title = "{Fully time-dependent cloud formation from a non-equilibrium gas-phase in exoplanetary atmospheres}",
      journal = {\aap},
         year = 2024,
        month = feb,
       volume = {682},
          eid = {A150},
        pages = {A150},
          doi = {10.1051/0004-6361/202347441},
archivePrefix = {arXiv},
       eprint = {2311.03244},
 primaryClass = {astro-ph.EP},
       adsurl = {https://ui.adsabs.harvard.edu/abs/2024A&A...682A.150K}
}

@ARTICLE{Marley2021,
       author = {{Marley}, Mark S. and {Saumon}, Didier and {Visscher}, Channon and {Lupu}, Roxana and {Freedman}, Richard and {Morley}, Caroline and {Fortney}, Jonathan J. and {Seay}, Christopher and {Smith}, Adam J.~R.~W. and {Teal}, D.~J. and {Wang}, Ruoyan},
        title = "{The Sonora Brown Dwarf Atmosphere and Evolution Models. I. Model Description and Application to Cloudless Atmospheres in Rainout Chemical Equilibrium}",
      journal = {\apj},
         year = 2021,
        month = oct,
       volume = {920},
       number = {2},
          eid = {85},
        pages = {85},
          doi = {10.3847/1538-4357/ac141d},
archivePrefix = {arXiv},
       eprint = {2107.07434},
 primaryClass = {astro-ph.SR},
       adsurl = {https://ui.adsabs.harvard.edu/abs/2021ApJ...920...85M}
}

@ARTICLE{bell2022,
       author = {{Bell}, Taylor and {Ahrer}, Eva-Maria and {Brande}, Jonathan and {Carter}, Aarynn and {Feinstein}, Adina and {Caloca}, Giannina and {Mansfield}, Megan and {Zieba}, Sebastian and {Piaulet}, Caroline and {Benneke}, Bj{\"o}rn and {Filippazzo}, Joseph and {May}, Erin and {Roy}, Pierre-Alexis and {Kreidberg}, Laura and {Stevenson}, Kevin},
        title = "{Eureka!: An End-to-End Pipeline for JWST Time-Series Observations}",
      journal = {The Journal of Open Source Software},
         year = 2022,
        month = nov,
       volume = {7},
       number = {79},
          eid = {4503},
        pages = {4503},
          doi = {10.21105/joss.04503},
archivePrefix = {arXiv},
       eprint = {2207.03585},
 primaryClass = {astro-ph.IM},
       adsurl = {https://ui.adsabs.harvard.edu/abs/2022JOSS....7.4503B}
}

@ARTICLE{Molliere2025SiO,
       author = {{Molli{\`e}re}, P. and {K{\"u}hnle}, H. and {Matthews}, E.~C. and {Henning}, Th. and {Min}, M. and {Patapis}, P. and {Lagage}, P. -O. and {Waters}, L.~B.~F.~M. and {G{\"u}del}, M. and {J{\"a}ger}, Cornelia and {Zhang}, Z. and {Decin}, L. and {Biller}, B.~A. and {Absil}, O. and {Argyriou}, I. and {Barrado}, D. and {Cossou}, C. and {Glasse}, A. and {Olofsson}, G. and {Pye}, J.~P. and {Rouan}, D. and {Samland}, M. and {Scheithauer}, S. and {Tremblin}, P. and {Whiteford}, N. and {van Dishoeck}, E.~F. and {{\"O}stlin}, G. and {Ray}, T.},
        title = "{Evidence for SiO cloud nucleation in the rogue planet PSO J318}",
      journal = {arXiv e-prints},
         year = 2025,
        month = jul,
          eid = {arXiv:2507.18691},
        pages = {arXiv:2507.18691},
          doi = {10.48550/arXiv.2507.18691},
archivePrefix = {arXiv},
       eprint = {2507.18691},
 primaryClass = {astro-ph.EP},
       adsurl = {https://ui.adsabs.harvard.edu/abs/2025arXiv250718691M}
}

@ARTICLE{Hoch2025,
       author = {{Hoch}, K.~K.~W. and {Rowland}, M. and {Petrus}, S. and {Nasedkin}, E. and {Ingebretsen}, C. and {Kammerer}, J. and {Perrin}, M. and {D'Orazi}, V. and {Balmer}, W.~O. and {Barman}, T. and {Bonnefoy}, M. and {Chauvin}, G. and {Chen}, C. and {De Rosa}, R.~J. and {Girard}, J. and {Gonzales}, E. and {Kenworthy}, M. and {Konopacky}, Q.~M. and {Macintosh}, B. and {Moran}, S.~E. and {Morley}, C.~V. and {Palma-Bifani}, P. and {Pueyo}, L. and {Ren}, B. and {Rickman}, E. and {Ruffio}, J. -B. and {Theissen}, C.~A. and {Ward-Duong}, K. and {Zhang}, Y.},
        title = "{Silicate clouds and a circumplanetary disk in the YSES-1 exoplanet system}",
      journal = {\nat},
         year = 2025,
        month = jul,
       volume = {643},
       number = {8073},
        pages = {938-942},
          doi = {10.1038/s41586-025-09174-w},
archivePrefix = {arXiv},
       eprint = {2507.18861},
 primaryClass = {astro-ph.EP},
       adsurl = {https://ui.adsabs.harvard.edu/abs/2025Natur.643..938H}
}

@ARTICLE{Looper2008,
       author = {{Looper}, Dagny L. and {Gelino}, Christopher R. and {Burgasser}, Adam J. and {Kirkpatrick}, J. Davy},
        title = "{Discovery of a T Dwarf Binary with the Largest Known J-Band Flux Reversal}",
      journal = {\apj},
         year = 2008,
        month = oct,
       volume = {685},
       number = {2},
        pages = {1183-1192},
          doi = {10.1086/590382},
archivePrefix = {arXiv},
       eprint = {0803.0544},
 primaryClass = {astro-ph},
       adsurl = {https://ui.adsabs.harvard.edu/abs/2008ApJ...685.1183L}
}

@ARTICLE{Scott1996,
       author = {{Scott}, A. and {Duley}, W.~W.},
        title = "{Ultraviolet and Infrared Refractive Indices of Amorphous Silicates}",
      journal = {\apjs},
         year = 1996,
        month = aug,
       volume = {105},
        pages = {401},
          doi = {10.1086/192321},
       adsurl = {https://ui.adsabs.harvard.edu/abs/1996ApJS..105..401S}
}

@ARTICLE{lunamorley2021,
       author = {{Luna}, Jessica L. and {Morley}, Caroline V.},
        title = "{Empirically Determining Substellar Cloud Compositions in the Era of the James Webb Space Telescope}",
      journal = {\apj},
         year = 2021,
        month = oct,
       volume = {920},
       number = {2},
          eid = {146},
        pages = {146},
          doi = {10.3847/1538-4357/ac1865},
archivePrefix = {arXiv},
       eprint = {2108.03161},
 primaryClass = {astro-ph.SR},
       adsurl = {https://ui.adsabs.harvard.edu/abs/2021ApJ...920..146L}
}

@ARTICLE{jaeger1998,
       author = {{Jaeger}, C. and {Molster}, F.~J. and {Dorschner}, J. and {Henning}, Th. and {Mutschke}, H. and {Waters}, L.~B.~F.~M.},
        title = "{Steps toward interstellar silicate mineralogy. IV. The crystalline revolution}",
      journal = {\aap},
         year = 1998,
        month = nov,
       volume = {339},
        pages = {904-916},
       adsurl = {https://ui.adsabs.harvard.edu/abs/1998A&A...339..904J}
}

@ARTICLE{Mullens2024,
       author = {{Mullens}, Elijah and {Lewis}, Nikole K. and {MacDonald}, Ryan J.},
        title = "{Implementation of Aerosol Mie Scattering in POSEIDON with Application to the Hot Jupiter HD 189733 b's Transmission, Emission, and Reflected Light Spectrum}",
      journal = {\apj},
         year = 2024,
        month = dec,
       volume = {977},
       number = {1},
          eid = {105},
        pages = {105},
          doi = {10.3847/1538-4357/ad8575},
archivePrefix = {arXiv},
       eprint = {2410.19253},
 primaryClass = {astro-ph.EP},
       adsurl = {https://ui.adsabs.harvard.edu/abs/2024ApJ...977..105M}
}

@ARTICLE{Mukherjee2023,
       author = {{Mukherjee}, Sagnick and {Batalha}, Natasha E. and {Fortney}, Jonathan J. and {Marley}, Mark S.},
        title = "{PICASO 3.0: A One-dimensional Climate Model for Giant Planets and Brown Dwarfs}",
      journal = {\apj},
         year = 2023,
        month = jan,
       volume = {942},
       number = {2},
          eid = {71},
        pages = {71},
          doi = {10.3847/1538-4357/ac9f48},
archivePrefix = {arXiv},
       eprint = {2208.07836},
 primaryClass = {astro-ph.EP},
       adsurl = {https://ui.adsabs.harvard.edu/abs/2023ApJ...942...71M}
}

@ARTICLE{Batalha2025,
       author = {{Batalha}, Natasha E. and {Rooney}, Caoimhe M. and {Visscher}, Channon and {Moran}, Sarah E. and {Marley}, Mark S. and {Sengupta}, Aditya R. and {Kiefer}, Sven and {Lodge}, Matt G. and {Mang}, James and {Morley}, Caroline V. and {Mukherjee}, Sagnick and {Fortney}, Jonathan J. and {Gao}, Peter and {Lewis}, Nikole K. and {Mayorga}, L.~C. and {Pearce}, Logan A. and {Wakeford}, Hannah R.},
        title = "{Condensation Clouds in Substellar Atmospheres with Virga}",
      journal = {arXiv e-prints},
         year = 2025,
        month = aug,
          eid = {arXiv:2508.15102},
        pages = {arXiv:2508.15102},
          doi = {10.48550/arXiv.2508.15102},
archivePrefix = {arXiv},
       eprint = {2508.15102},
 primaryClass = {astro-ph.EP},
       adsurl = {https://ui.adsabs.harvard.edu/abs/2025arXiv250815102B}
}

@ARTICLE{Mullens2025,
       author = {{Mullens}, Elijah and {Lewis}, Nikole K.},
        title = "{Silicate Sundogs: Probing the Effects of Grain Directionality in Exoplanet Observations}",
      journal = {\apjl},
         year = 2025,
        month = aug,
       volume = {988},
       number = {2},
          eid = {L43},
        pages = {L43},
          doi = {10.3847/2041-8213/ade885},
archivePrefix = {arXiv},
       eprint = {2508.01839},
 primaryClass = {astro-ph.EP},
       adsurl = {https://ui.adsabs.harvard.edu/abs/2025ApJ...988L..43M}
}

@ARTICLE{wetzel2013,
       author = {{Wetzel}, S. and {Klevenz}, M. and {Gail}, H. -P. and {Pucci}, A. and {Trieloff}, M.},
        title = "{Laboratory measurement of optical constants of solid SiO and application to circumstellar dust}",
      journal = {\aap},
         year = 2013,
        month = may,
       volume = {553},
          eid = {A92},
        pages = {A92},
          doi = {10.1051/0004-6361/201220803},
archivePrefix = {arXiv},
       eprint = {1210.4430},
 primaryClass = {astro-ph.SR},
       adsurl = {https://ui.adsabs.harvard.edu/abs/2013A&A...553A..92W}
}

@ARTICLE{Helling2006,
       author = {{Helling}, Ch. and {Thi}, W. -F. and {Woitke}, P. and {Fridlund}, M.},
        title = "{Detectability of dirty dust grains in brown dwarf atmospheres}",
      journal = {\aap},
         year = 2006,
        month = may,
       volume = {451},
       number = {2},
        pages = {L9-L12},
          doi = {10.1051/0004-6361:20064944},
archivePrefix = {arXiv},
       eprint = {astro-ph/0603341},
 primaryClass = {astro-ph},
       adsurl = {https://ui.adsabs.harvard.edu/abs/2006A&A...451L...9H}
}

@ARTICLE{Burningham2021,
       author = {{Burningham}, Ben and {Faherty}, Jacqueline K. and {Gonzales}, Eileen C. and {Marley}, Mark S. and {Visscher}, Channon and {Lupu}, Roxana and {Gaarn}, Josefine and {Fabienne Bieger}, Michelle and {Freedman}, Richard and {Saumon}, Didier},
        title = "{Cloud busting: enstatite and quartz clouds in the atmosphere of 2M2224-0158}",
      journal = {\mnras},
         year = 2021,
        month = sep,
       volume = {506},
       number = {2},
        pages = {1944-1961},
          doi = {10.1093/mnras/stab1361},
archivePrefix = {arXiv},
       eprint = {2105.04268},
 primaryClass = {astro-ph.SR},
       adsurl = {https://ui.adsabs.harvard.edu/abs/2021MNRAS.506.1944B}
}

@ARTICLE{Schlawin2024AJ,
       author = {{Schlawin}, Everett and {Mukherjee}, Sagnick and {Ohno}, Kazumasa and {Bell}, Taylor J. and {Beatty}, Thomas G. and {Greene}, Thomas P. and {Line}, Michael and {Challener}, Ryan C. and {Parmentier}, Vivien and {Fortney}, Jonathan J. and {Rauscher}, Emily and {Wiser}, Lindsey and {Welbanks}, Luis and {Murphy}, Matthew and {Edelman}, Isaac and {Batalha}, Natasha and {Moran}, Sarah E. and {Mehta}, Nishil and {Rieke}, Marcia},
        title = "{Multiple Clues for Dayside Aerosols and Temperature Gradients in WASP-69 b from a Panchromatic JWST Emission Spectrum}",
      journal = {\aj},
         year = 2024,
        month = sep,
       volume = {168},
       number = {3},
          eid = {104},
        pages = {104},
          doi = {10.3847/1538-3881/ad58e0},
archivePrefix = {arXiv},
       eprint = {2406.15543},
 primaryClass = {astro-ph.EP},
       adsurl = {https://ui.adsabs.harvard.edu/abs/2024AJ....168..104S}
}

@ARTICLE{Dyrek2024Nature,
       author = {{Dyrek}, Achr{\`e}ne and {Min}, Michiel and {Decin}, Leen and {Bouwman}, Jeroen and {Crouzet}, Nicolas and {Molli{\`e}re}, Paul and {Lagage}, Pierre-Olivier and {Konings}, Thomas and {Tremblin}, Pascal and {G{\"u}del}, Manuel and {Pye}, John and {Waters}, Rens and {Henning}, Thomas and {Vandenbussche}, Bart and {Ardevol Martinez}, Francisco and {Argyriou}, Ioannis and {Ducrot}, Elsa and {Heinke}, Linus and {van Looveren}, Gwenael and {Absil}, Olivier and {Barrado}, David and {Baudoz}, Pierre and {Boccaletti}, Anthony and {Cossou}, Christophe and {Coulais}, Alain and {Edwards}, Billy and {Gastaud}, Ren{\'e} and {Glasse}, Alistair and {Glauser}, Adrian and {Greene}, Thomas P. and {Kendrew}, Sarah and {Krause}, Oliver and {Lahuis}, Fred and {Mueller}, Michael and {Olofsson}, Goran and {Patapis}, Polychronis and {Rouan}, Daniel and {Royer}, Pierre and {Scheithauer}, Silvia and {Waldmann}, Ingo and {Whiteford}, Niall and {Colina}, Luis and {van Dishoeck}, Ewine F. and {{\"O}stlin}, G{\"o}ran and {Ray}, Tom P. and {Wright}, Gillian},
        title = "{SO$_{2}$, silicate clouds, but no CH$_{4}$ detected in a warm Neptune}",
      journal = {\nat},
         year = 2024,
        month = jan,
       volume = {625},
       number = {7993},
        pages = {51-54},
          doi = {10.1038/s41586-023-06849-0},
archivePrefix = {arXiv},
       eprint = {2311.12515},
 primaryClass = {astro-ph.EP},
       adsurl = {https://ui.adsabs.harvard.edu/abs/2024Natur.625...51D}
}

@ARTICLE{Inglis2024ApJ,
       author = {{Inglis}, Julie and {Batalha}, Natasha E. and {Lewis}, Nikole K. and {Kataria}, Tiffany and {Knutson}, Heather A. and {Kilpatrick}, Brian M. and {Gagnebin}, Anna and {Mukherjee}, Sagnick and {Pettyjohn}, Maria M. and {Crossfield}, Ian J.~M. and {Foote}, Trevor O. and {Grant}, David and {Henry}, Gregory W. and {Lally}, Maura and {McKemmish}, Laura K. and {Sing}, David K. and {Wakeford}, Hannah R. and {Zapata Trujillo}, Juan C. and {Zellem}, Robert T.},
        title = "{Quartz Clouds in the Dayside Atmosphere of the Quintessential Hot Jupiter HD 189733 b}",
      journal = {\apjl},
         year = 2024,
        month = oct,
       volume = {973},
       number = {2},
          eid = {L41},
        pages = {L41},
          doi = {10.3847/2041-8213/ad725e},
archivePrefix = {arXiv},
       eprint = {2409.11395},
 primaryClass = {astro-ph.EP},
       adsurl = {https://ui.adsabs.harvard.edu/abs/2024ApJ...973L..41I}
}

@ARTICLE{Xue2024ApJ,
       author = {{Xue}, Qiao and {Bean}, Jacob L. and {Zhang}, Michael and {Welbanks}, Luis and {Lunine}, Jonathan and {August}, Prune},
        title = "{JWST Transmission Spectroscopy of HD 209458b: A Supersolar Metallicity, a Very Low C/O, and No Evidence of CH$_{4}$, HCN, or C$_{2}$H$_{2}$}",
      journal = {\apjl},
         year = 2024,
        month = mar,
       volume = {963},
       number = {1},
          eid = {L5},
        pages = {L5},
          doi = {10.3847/2041-8213/ad2682},
archivePrefix = {arXiv},
       eprint = {2310.03245},
 primaryClass = {astro-ph.EP},
       adsurl = {https://ui.adsabs.harvard.edu/abs/2024ApJ...963L...5X}
}

@ARTICLE{Stassun2017,
       author = {{Stassun}, Keivan G. and {Collins}, Karen A. and {Gaudi}, B. Scott},
        title = "{Accurate Empirical Radii and Masses of Planets and Their Host Stars with Gaia Parallaxes}",
      journal = {\aj},
         year = 2017,
        month = mar,
       volume = {153},
       number = {3},
          eid = {136},
        pages = {136},
          doi = {10.3847/1538-3881/aa5df3},
archivePrefix = {arXiv},
       eprint = {1609.04389},
 primaryClass = {astro-ph.EP},
       adsurl = {https://ui.adsabs.harvard.edu/abs/2017AJ....153..136S}
}

@software{grant_david_2023_8211207,
  author       = {Grant, David and
                  Valentine, Daniel and
                  Wakeford, Hannah R.},
  title        = {Exo-TiC/ExoTiC-MIRI: ExoTiC-MIRI v1.0.0},
  month        = aug,
  year         = 2023,
  publisher    = {Zenodo},
  version      = {v1.0.0},
  doi          = {10.5281/zenodo.8211207},
  url          = {https://doi.org/10.5281/zenodo.8211207}
}

@ARTICLE{bouwman2023,
       author = {{Bouwman}, Jeroen and {Kendrew}, Sarah and {Greene}, Thomas P. and {Bell}, Taylor J. and {Lagage}, Pierre-Olivier and {Schreiber}, J{\"u}rgen and {Dicken}, Daniel and {Sloan}, G.~C. and {Espinoza}, N{\'e}stor and {Scheithauer}, Silvia and {Coulais}, Alain and {Fox}, Ori D. and {Gastaud}, Ren{\'e} and {Glauser}, Adrian M. and {Jones}, Olivia C. and {Labiano}, Alvaro and {Lahuis}, Fred and {Morrison}, Jane E. and {Murray}, Katherine and {Mueller}, Michael and {Nayak}, Omnarayani and {Wright}, Gillian S. and {Glasse}, Alistair and {Rieke}, George},
        title = "{Spectroscopic Time Series Performance of the Mid-infrared Instrument on the JWST}",
      journal = {\pasp},
         year = 2023,
        month = mar,
       volume = {135},
       number = {1045},
          eid = {038002},
        pages = {038002},
          doi = {10.1088/1538-3873/acbc49},
archivePrefix = {arXiv},
       eprint = {2211.16123},
 primaryClass = {astro-ph.IM},
       adsurl = {https://ui.adsabs.harvard.edu/abs/2023PASP..135c8002B}
}

@ARTICLE{kendrew2015,
       author = {{Kendrew}, Sarah and {Scheithauer}, Silvia and {Bouchet}, Patrice and {Amiaux}, Jerome and {Azzollini}, Ruym{\'a}n and {Bouwman}, Jeroen and {Chen}, C.~H. and {Dubreuil}, D. and {Fischer}, Sebastian and {Glasse}, Alistair and {Greene}, T.~P. and {Lagage}, P. -O. and {Lahuis}, Fred and {Ronayette}, Samuel and {Wright}, David and {Wright}, G.~S.},
        title = "{The Mid-Infrared Instrument for the James Webb Space Telescope, IV: The Low-Resolution Spectrometer}",
      journal = {\pasp},
         year = 2015,
        month = jul,
       volume = {127},
       number = {953},
        pages = {623},
          doi = {10.1086/682255},
archivePrefix = {arXiv},
       eprint = {1512.03000},
 primaryClass = {astro-ph.IM},
       adsurl = {https://ui.adsabs.harvard.edu/abs/2015PASP..127..623K}
}

@article{Lodge2024,
	adsurl = {https://ui.adsabs.harvard.edu/abs/2024MNRAS.527.11113},
	archiveprefix = {arXiv},
	author = {{Lodge}, M.~G. and {Wakeford}, H.~R. and {Leinhardt}, Z.~M.},
	doi = {10.1093/mnras/stad3743},
	eprint = {2312.02301},
	journal = {\mnras},
	month = feb,
	number = {4},
	pages = {11113-11137},
	primaryclass = {astro-ph.EP},
	title = {{Aerosols are not spherical cows: using discrete dipole approximation to model the properties of fractal particles}},
	volume = {527},
	year = 2024}

@article{Grant2023quartz,
	author = {{Grant}, David and {Lewis}, Nikole K. and {Wakeford}, Hannah R. and {Batalha}, Natasha E. and {Glidden}, Ana and {Goyal}, Jayesh and {Mullens}, Elijah and {MacDonald}, Ryan J. and {May}, Erin M. and {Seager}, Sara and {Stevenson}, Kevin B. and {Valenti}, Jeff A. and {Visscher}, Channon and {Alderson}, Lili and {Allen}, Natalie H. and {Ca{\~n}as}, Caleb I. and {Col{\'o}n}, Knicole and {Clampin}, Mark and {Espinoza}, N{\'e}stor and {Gressier}, Am{\'e}lie and {Huang}, Jingcheng and {Lin}, Zifan and {Long}, Douglas and {Louie}, Dana R. and {Pe{\~n}a-Guerrero}, Maria and {Ranjan}, Sukrit and {Sotzen}, Kristin S. and {Valentine}, Daniel and {Anderson}, Jay and {Balmer}, William O. and {Bellini}, Andrea and {Hoch}, Kielan K.~W. and {Kammerer}, Jens and {Libralato}, Mattia and {Mountain}, C. Matt and {Perrin}, Marshall D. and {Pueyo}, Laurent and {Rickman}, Emily and {Rebollido}, Isabel and {Sohn}, Sangmo Tony and {van der Marel}, Roeland P. and {Watkins}, Laura L.},
	doi = {10.3847/2041-8213/acfc3b},
	journal = {The Astrophysical Journal Letters},
	pages = {L32},
    volume = {956},
    number = {2},
	title = {{JWST-TST DREAMS: Quartz Clouds in the Atmosphere of WASP-17b}},
	year = 2023
}

@article{Alderson2023_ERS,
	adsurl = {https://ui.adsabs.harvard.edu/abs/2023Natur.614..664A},
	archiveprefix = {arXiv},
	author = {{Alderson}, Lili and {Wakeford}, Hannah R. and {Alam}, Munazza K. and {Batalha}, Natasha E. and {Lothringer}, Joshua D. and {Adams Redai}, Jea and {Barat}, Saugata and {Brande}, Jonathan and {Damiano}, Mario and {Daylan}, Tansu and {Espinoza}, N{\'e}stor and {Flagg}, Laura and {Goyal}, Jayesh M. and {Grant}, David and {Hu}, Renyu and {Inglis}, Julie and {Lee}, Elspeth K.~H. and {Mikal-Evans}, Thomas and {Ramos-Rosado}, Lakeisha and {Roy}, Pierre-Alexis and {Wallack}, Nicole L. and {Batalha}, Natalie M. and {Bean}, Jacob L. and {Benneke}, Bj{\"o}rn and {Berta-Thompson}, Zachory K. and {Carter}, Aarynn L. and {Changeat}, Quentin and {Col{\'o}n}, Knicole D. and {Crossfield}, Ian J.~M. and {D{\'e}sert}, Jean-Michel and {Foreman-Mackey}, Daniel and {Gibson}, Neale P. and {Kreidberg}, Laura and {Line}, Michael R. and {L{\'o}pez-Morales}, Mercedes and {Molaverdikhani}, Karan and {Moran}, Sarah E. and {Morello}, Giuseppe and {Moses}, Julianne I. and {Mukherjee}, Sagnick and {Schlawin}, Everett and {Sing}, David K. and {Stevenson}, Kevin B. and {Taylor}, Jake and {Aggarwal}, Keshav and {Ahrer}, Eva-Maria and {Allen}, Natalie H. and {Barstow}, Joanna K. and {Bell}, Taylor J. and {Blecic}, Jasmina and {Casewell}, Sarah L. and {Chubb}, Katy L. and {Crouzet}, Nicolas and {Cubillos}, Patricio E. and {Decin}, Leen and {Feinstein}, Adina D. and {Fortney}, Joanthan J. and {Harrington}, Joseph and {Heng}, Kevin and {Iro}, Nicolas and {Kempton}, Eliza M. -R. and {Kirk}, James and {Knutson}, Heather A. and {Krick}, Jessica and {Leconte}, J{\'e}r{\'e}my and {Lendl}, Monika and {MacDonald}, Ryan J. and {Mancini}, Luigi and {Mansfield}, Megan and {May}, Erin M. and {Mayne}, Nathan J. and {Miguel}, Yamila and {Nikolov}, Nikolay K. and {Ohno}, Kazumasa and {Palle}, Enric and {Parmentier}, Vivien and {Petit dit de la Roche}, Dominique J.~M. and {Piaulet}, Caroline and {Powell}, Diana and {Rackham}, Benjamin V. and {Redfield}, Seth and {Rogers}, Laura K. and {Rustamkulov}, Zafar and {Tan}, Xianyu and {Tremblin}, P. and {Tsai}, Shang-Min and {Turner}, Jake D. and {de Val-Borro}, Miguel and {Venot}, Olivia and {Welbanks}, Luis and {Wheatley}, Peter J. and {Zhang}, Xi},
	doi = {10.1038/s41586-022-05591-3},
	eprint = {2211.10488},
	journal = {\nat},
	month = feb,
	number = {7949},
	pages = {664-669},
	primaryclass = {astro-ph.EP},
	title = {{Early Release Science of the exoplanet WASP-39b with JWST NIRSpec G395H}},
	volume = {614},
	year = 2023}

@article{miles2023ERS,
	adsurl = {https://ui.adsabs.harvard.edu/abs/2022arXiv220900620M},
	archiveprefix = {arXiv},
	author = {{Miles}, Brittany E. and {Biller}, Beth A. and {Patapis}, Polychronis and {Worthen}, Kadin and {Rickman}, Emily and {Hoch}, Kielan K.~W. and {Skemer}, Andrew and {Perrin}, Marshall D. and {Chen}, Christine H. and {Mukherjee}, Sagnick and {Morley}, Caroline V. and {Moran}, Sarah E. and {Bonnefoy}, Mickael and {Petrus}, Simon and {Carter}, Aarynn L. and {Choquet}, Elodie and {Hinkley}, Sasha and {Ward-Duong}, Kimberly and {Leisenring}, Jarron M. and {Millar-Blanchaer}, Maxwell A. and {Pueyo}, Laurent and {Ray}, Shrishmoy and {Stapelfeldt}, Karl R. and {Stone}, Jordan M. and {Wang}, Jason J. and {Absil}, Olivier and {Balmer}, William O. and {Boccaletti}, Anthony and {Bonavita}, Mariangela and {Booth}, Mark and {Bowler}, Brendan P. and {Chauvin}, Gael and {Christiaens}, Valentin and {Currie}, Thayne and {Danielski}, Camilla and {Fortney}, Jonathan J. and {Girard}, Julien H. and {Greenbaum}, Alexandra Z. and {Henning}, Thomas and {Hines}, Dean C. and {Janson}, Markus and {Kalas}, Paul and {Kammerer}, Jens and {Kenworthy}, Matthew A. and {Kervella}, Pierre and {Lagage}, Pierre-Olivier and {Lew}, Ben W.~P. and {Liu}, Michael C. and {Macintosh}, Bruce and {Marino}, Sebastian and {Marley}, Mark S. and {Marois}, Christian and {Matthews}, Elisabeth C. and {Matthews}, Brenda C. and {Mawet}, Dimitri and {McElwain}, Michael W. and {Metchev}, Stanimir and {Meyer}, Michael R. and {Molliere}, Paul and {Pantin}, Eric and {Rebollido}, Andreas Quirrenbachm Isabel and {Ren}, Bin B. and {Vasist}, Malavika and {Wyatt}, Mark C. and {Zhou}, Yifan and {Briesemeister}, Zackery W. and {Bryan}, Marta L. and {Calissendorff}, Per and {Catalloube}, Faustine and {Cugno}, Gabriele and {De Furio}, Matthew and {Dupuy}, Trent J. and {Factor}, Samuel M. and {Faherty}, Jacqueline K. and {Fitzgerald}, Michael P. and {Franson}, Kyle and {Gonzales}, Eileen C. and {Hood}, Callie E. and {Howe}, Alex R. and {Kraus}, Adam L. and {Kuzuhara}, Masayuki and {Lawson}, Kellen and {Lazzoni}, Cecilia and {Liu}, Pengyu and {Llop-Sayson}, Jorge and {Lloyd}, James P. and {Martinez}, Raquel A. and {Mazoyer}, Johan and {Quanz}, Sascha P. and {Adams Redai}, Jea and {Samland}, Matthias and {Schlieder}, Joshua E. and {Tamura}, Motohide and {Tan}, Xianyu and {Uyama}, Taichi and {Vigan}, Arthur and {Vos}, Johanna M. and {Wagner}, Kevin and {Wolff}, Schuyler G. and {Ygouf}, Marie and {Zhang}, Keming and {Zhang}, Zhoujian},
	eid = {arXiv:2209.00620},
	eprint = {2209.00620},
	journal = {arXiv e-prints},
	month = sep,
	pages = {arXiv:2209.00620},
	primaryclass = {astro-ph.EP},
	title = {{The JWST Early Release Science Program for Direct Observations of Exoplanetary Systems II: A 1 to 20 Micron Spectrum of the Planetary-Mass Companion VHS 1256-1257 b}},
	year = 2022}

@article{Rustamkulov2022,
	adsurl = {https://ui.adsabs.harvard.edu/abs/2022arXiv220304173R},
	archiveprefix = {arXiv},
	author = {{Rustamkulov}, Zafar and {Sing}, David and {Liu}, Rongrong and {Wang}, Ashley},
	eid = {arXiv:2203.04173},
	eprint = {2203.04173},
	journal = {arXiv e-prints},
	month = mar,
	pages = {arXiv:2203.04173},
	primaryclass = {astro-ph.EP},
	title = {{Analysis of a JWST NIRSpec Lab Time Series: Characterizing Systematics, Recovering Exoplanet Transit Spectroscopy, and Constraining a Noise Floor}},
	year = 2022}

@article{may2020,
	adsurl = {https://ui.adsabs.harvard.edu/abs/2020AJ....160..140M},
	archiveprefix = {arXiv},
	author = {{May}, E.~M. and {Stevenson}, K.~B.},
	doi = {10.3847/1538-3881/aba833},
	eid = {140},
	eprint = {2007.06618},
	journal = {\aj},
	month = sep,
	number = {3},
	pages = {140},
	primaryclass = {astro-ph.EP},
	title = {{Introducing a New Spitzer Master BLISS Map to Remove the Instrument Systematic Phase-curve-parameter Degeneracy, as Demonstrated by a Reanalysis of the 4.5 {\ensuremath{\mu}}m WASP-43b Phase Curve}},
	volume = {160},
	year = 2020}

@article{he2020,
	adsurl = {https://ui.adsabs.harvard.edu/abs/2020NatAs.tmp...70H},
	archiveprefix = {arXiv},
	author = {{He}, Chao and {H{\"o}rst}, Sarah M. and {Lewis}, Nikole K. and {Yu}, Xinting and {Moses}, Julianne I. and {McGuiggan}, Patricia and {Marley}, Mark S. and {Kempton}, Eliza M. -R. and {Moran}, Sarah E. and {Morley}, Caroline V. and {Vuitton}, V{\'e}ronique},
	doi = {10.1038/s41550-020-1072-9},
	eprint = {2004.02728},
	journal = {Nature Astronomy},
	month = apr,
	primaryclass = {astro-ph.EP},
	title = {{Sulfur-driven haze formation in warm CO$_{2}$-rich exoplanet atmospheres}},
	year = 2020}

@article{Hiranaka2016,
	adsurl = {https://ui.adsabs.harvard.edu/abs/2016ApJ...830...96H},
	archiveprefix = {arXiv},
	author = {{Hiranaka}, Kay and {Cruz}, Kelle L. and {Douglas}, Stephanie T. and {Marley}, Mark S. and {Baldassare}, Vivienne F.},
	doi = {10.3847/0004-637X/830/2/96},
	eid = {96},
	eprint = {1606.09485},
	journal = {\apj},
	month = oct,
	number = {2},
	pages = {96},
	primaryclass = {astro-ph.SR},
	title = {{Exploring the Role of Sub-micron-sized Dust Grains in the Atmospheres of Red L0-L6 Dwarfs}},
	volume = {830},
	year = 2016}

@article{Pinhas2019,
	adsurl = {https://ui.adsabs.harvard.edu/abs/2019MNRAS.482.1485P},
	archiveprefix = {arXiv},
	author = {{Pinhas}, Arazi and {Madhusudhan}, Nikku and {Gandhi}, Siddharth and {MacDonald}, Ryan},
	doi = {10.1093/mnras/sty2544},
	eprint = {1811.00011},
	journal = {\mnras},
	month = jan,
	number = {2},
	pages = {1485-1498},
	primaryclass = {astro-ph.EP},
	title = {{H$_{2}$O abundances and cloud properties in ten hot giant exoplanets}},
	volume = {482},
	year = 2019}

@article{Powell2018,
	adsurl = {https://ui.adsabs.harvard.edu/abs/2018ApJ...860...18P},
	archiveprefix = {arXiv},
	author = {{Powell}, Diana and {Zhang}, Xi and {Gao}, Peter and {Parmentier}, Vivien},
	doi = {10.3847/1538-4357/aac215},
	eid = {18},
	eprint = {1805.01468},
	journal = {\apj},
	month = jun,
	number = {1},
	pages = {18},
	primaryclass = {astro-ph.EP},
	title = {{Formation of Silicate and Titanium Clouds on Hot Jupiters}},
	volume = {860},
	year = 2018}

@article{macdonald2017,
	adsurl = {https://ui.adsabs.harvard.edu/abs/2017MNRAS.469.1979M},
	archiveprefix = {arXiv},
	author = {{MacDonald}, Ryan J. and {Madhusudhan}, Nikku},
	doi = {10.1093/mnras/stx804},
	eprint = {1701.01113},
	journal = {\mnras},
	month = aug,
	number = {2},
	pages = {1979-1996},
	primaryclass = {astro-ph.EP},
	title = {{HD 209458b in new light: evidence of nitrogen chemistry, patchy clouds and sub-solar water}},
	volume = {469},
	year = 2017}

@article{Gao2020,
	adsurl = {https://ui.adsabs.harvard.edu/abs/2020NatAs.tmp..114G},
	archiveprefix = {arXiv},
	author = {{Gao}, Peter and {Thorngren}, Daniel P. and {Lee}, Graham K.~H. and {Fortney}, Jonathan J. and {Morley}, Caroline V. and {Wakeford}, Hannah R. and {Powell}, Diana K. and {Stevenson}, Kevin B. and {Zhang}, Xi},
	doi = {10.1038/s41550-020-1114-3},
	eprint = {2005.11939},
	journal = {Nature Astronomy},
	month = may,
	primaryclass = {astro-ph.EP},
	title = {{Aerosol composition of hot giant exoplanets dominated by silicates and hydrocarbon hazes}},
	year = 2020}

@article{thorngren2019b,
	adsurl = {https://ui.adsabs.harvard.edu/abs/2019ApJ...884L...6T},
	archiveprefix = {arXiv},
	author = {{Thorngren}, Daniel and {Gao}, Peter and {Fortney}, Jonathan J.},
	doi = {10.3847/2041-8213/ab43d0},
	eid = {L6},
	eprint = {1907.07777},
	journal = {\apjl},
	month = oct,
	number = {1},
	pages = {L6},
	primaryclass = {astro-ph.EP},
	title = {{The Intrinsic Temperature and Radiative-Convective Boundary Depth in the Atmospheres of Hot Jupiters}},
	volume = {884},
	year = 2019}

@article{Barstow2020,
	adsurl = {https://ui.adsabs.harvard.edu/abs/2020arXiv200202945B},
	archiveprefix = {arXiv},
	author = {{Barstow}, Joanna K.},
	eid = {arXiv:2002.02945},
	eprint = {2002.02945},
	journal = {arXiv e-prints},
	month = {Feb},
	pages = {arXiv:2002.02945},
	primaryclass = {astro-ph.EP},
	title = {{Unveiling cloudy exoplanets: the influence of cloud model choices on retrieval solutions}},
	year = {2020}}

@article{Horne1986,
	adsurl = {https://ui.adsabs.harvard.edu/abs/1986PASP...98..609H},
	author = {{Horne}, K.},
	doi = {10.1086/131801},
	journal = {\pasp},
	month = {Jun},
	pages = {609-617},
	title = {{An optimal extraction algorithm for CCD spectroscopy.}},
	volume = {98},
	year = {1986}}

@article{wakeford2018_w39,
	adsurl = {https://ui.adsabs.harvard.edu/abs/2018AJ....155...29W},
	archiveprefix = {arXiv},
	author = {{Wakeford}, H.~R. and {Sing}, D.~K. and {Deming}, D. and {Lewis}, N.~K. and {Goyal}, J. and {Wilson}, T.~J. and {Barstow}, J. and {Kataria}, T. and {Drummond}, B. and {Evans}, T.~M. and {Carter}, A.~L. and {Nikolov}, N. and {Knutson}, H.~A. and {Ballester}, G.~E. and {Mand ell}, A.~M.},
	doi = {10.3847/1538-3881/aa9e4e},
	eid = {29},
	eprint = {1711.10529},
	journal = {\aj},
	month = {Jan},
	number = {1},
	pages = {29},
	primaryclass = {astro-ph.EP},
	title = {{The Complete Transmission Spectrum of WASP-39b with a Precise Water Constraint}},
	volume = {155},
	year = {2018}}

@article{batalhaNE2019a,
	adsurl = {https://ui.adsabs.harvard.edu/abs/2019ApJ...878...70B},
	archiveprefix = {arXiv},
	author = {{Batalha}, Natasha E. and {Marley}, Mark S. and {Lewis}, Nikole K. and {Fortney}, Jonathan J.},
	doi = {10.3847/1538-4357/ab1b51},
	eid = {70},
	eprint = {1904.09355},
	journal = {\apj},
	month = {Jun},
	number = {1},
	pages = {70},
	primaryclass = {astro-ph.EP},
	title = {{Exoplanet Reflected-light Spectroscopy with PICASO}},
	volume = {878},
	year = {2019}}

@article{iyer2016,
	adsurl = {https://ui.adsabs.harvard.edu/abs/2016ApJ...823..109I},
	archiveprefix = {arXiv},
	author = {{Iyer}, Aishwarya R. and {Swain}, Mark R. and {Zellem}, Robert T. and {Line}, Michael R. and {Roudier}, Gael and {Rocha}, Gra{\c{c}}a and {Livingston}, John H.},
	doi = {10.3847/0004-637X/823/2/109},
	eid = {109},
	eprint = {1512.00151},
	journal = {\apj},
	month = {Jun},
	number = {2},
	pages = {109},
	primaryclass = {astro-ph.EP},
	title = {{A Characteristic Transmission Spectrum Dominated by H$_{2}$O Applies to the Majority of HST/WFC3 Exoplanet Observations}},
	volume = {823},
	year = {2016}}

@article{Fu2017,
	adsurl = {https://ui.adsabs.harvard.edu/abs/2017ApJ...847L..22F},
	archiveprefix = {arXiv},
	author = {{Fu}, Guangwei and {Deming}, Drake and {Knutson}, Heather and {Madhusudhan}, Nikku and {Mandell}, Avi and {Fraine}, Jonathan},
	doi = {10.3847/2041-8213/aa8e40},
	eid = {L22},
	eprint = {1709.07385},
	journal = {\apj},
	month = {Oct},
	number = {2},
	pages = {L22},
	primaryclass = {astro-ph.EP},
	title = {{Statistical Analysis of Hubble/WFC3 Transit Spectroscopy of Extrasolar Planets}},
	volume = {847},
	year = {2017}}

@article{SO2,
	author = {Vandaele, Ann Carine and Simon, Paul C and Guilmot, Jean Michel and Carleer, Michel and Colin, R{\'e}ginald},
	journal = {Journal of Geophysical Research: Atmospheres},
	number = {D12},
	pages = {25599--25605},
	publisher = {Wiley Online Library},
	title = {SO2 absorption cross section measurement in the UV using a Fourier transform spectrometer},
	volume = {99},
	year = {1994}}

@article{Barstow2017,
	adsurl = {https://ui.adsabs.harvard.edu/\#abs/2017ApJ...834...50B},
	archiveprefix = {arXiv},
	author = {{Barstow}, J.~K. and {Aigrain}, S. and {Irwin}, P.~G.~J. and {Sing}, D.~K.},
	doi = {10.3847/1538-4357/834/1/50},
	eid = {50},
	eprint = {1610.01841},
	journal = {\apj},
	month = {Jan},
	pages = {50},
	primaryclass = {astro-ph.EP},
	title = {{A Consistent Retrieval Analysis of 10 Hot Jupiters Observed in Transmission}},
	volume = {834},
	year = {2017}}

@article{Wakeford2019RNAAS,
	adsurl = {https://ui.adsabs.harvard.edu/\#abs/2019RNAAS...3a...7W},
	author = {{Wakeford}, H.~R. and {Wilson}, T.~J. and {Stevenson}, K.~B. and {Lewis}, N.~K.},
	doi = {10.3847/2515-5172/aafc63},
	eid = {7},
	journal = {Research Notes of the American Astronomical Society},
	month = {Jan},
	pages = {7},
	title = {{Exoplanet Atmosphere Forecast: Observers Should Expect Spectroscopic Transmission Features to be Muted to 33\%}},
	volume = {3},
	year = {2019}}

@article{kitzmann2018,
	adsurl = {http://adsabs.harvard.edu/abs/2018MNRAS.475...94K},
	archiveprefix = {arXiv},
	author = {{Kitzmann}, D. and {Heng}, K.},
	doi = {10.1093/mnras/stx3141},
	eprint = {1710.04946},
	journal = {\mnras},
	month = mar,
	pages = {94-107},
	primaryclass = {astro-ph.EP},
	title = {{Optical properties of potential condensates in exoplanetary atmospheres}},
	volume = 475,
	year = 2018}

@article{Atreya2003,
	author = {S.K Atreya and P.R Mahaffy and H.B Niemann and M.H Wong and T.C Owen},
	doi = {https://doi.org/10.1016/S0032-0633(02)00144-7},
	issn = {0032-0633},
	journal = {Planetary and Space Science},
	note = {Recent Advances on the Atmosphere of Outer Planets and Titan},
	number = {2},
	pages = {105 - 112},
	title = {Composition and origin of the atmosphere of Jupiter---an update, and implications for the extrasolar giant planets},
	url = {http://www.sciencedirect.com/science/article/pii/S0032063302001447},
	volume = {51},
	year = {2003}}

@article{lee2016,
	adsurl = {http://adsabs.harvard.edu/abs/2016A%26A...594A..48L},
	archiveprefix = {arXiv},
	author = {{Lee}, E.K.H. and {Dobbs-Dixon}, I. and {Helling}, C. and {Bognar}, K. and {Woitke}, P.},
	doi = {10.1051/0004-6361/201628606},
	eid = {A48},
	eprint = {1603.09098},
	journal = {\aap},
	month = oct,
	pages = {A48},
	primaryclass = {astro-ph.EP},
	title = {{Dynamic mineral clouds on HD 189733b. I. 3D RHD with kinetic, non-equilibrium cloud formation}},
	volume = 594,
	year = 2016}

@article{BatalhaN2017b,
	adsurl = {http://adsabs.harvard.edu/abs/2017arXiv170201820B},
	archiveprefix = {arXiv},
	author = {{Batalha}, N.~E. and {Mandell}, A. and {Pontoppidan}, K. and {Stevenson}, K.~B. and {Lewis}, N.~K. and {Kalirai}, J. and {Greene}, T. and {Albert}, L. and {Nielsen}, L.~D. and {Earl}, N.},
	eprint = {1702.01820},
	journal = {ArXiv e-prints},
	month = feb,
	primaryclass = {astro-ph.IM},
	title = {{PandExo: A Community Tool for Transiting Exoplanet Science with JWST {\amp} HST}},
	year = 2017}

@article{helling2016,
	adsurl = {http://adsabs.harvard.edu/abs/2016arXiv161201863H},
	archiveprefix = {arXiv},
	author = {{Helling}, C. and {Tootill}, D. and {Woitke}, P. and {Lee}, G.},
	eprint = {1612.01863},
	journal = {ArXiv e-prints},
	month = dec,
	primaryclass = {astro-ph.SR},
	title = {{Dust in brown dwarfs and extra-solar planets V. Cloud formation in carbon- and oxygen-rich environment}},
	year = 2016}

@article{stevenson2012,
	adsurl = {http://adsabs.harvard.edu/abs/2012ApJ...754..136S},
	archiveprefix = {arXiv},
	author = {{Stevenson}, K.~B. and {Harrington}, J. and {Fortney}, J.~J. and {Loredo}, T.~J. and {Hardy}, R.~A. and {Nymeyer}, S. and {Bowman}, W.~C. and {Cubillos}, P. and {Bowman}, M.~O. and {Hardin}, M.},
	doi = {10.1088/0004-637X/754/2/136},
	eid = {136},
	eprint = {1108.2057},
	journal = {\apj},
	month = aug,
	pages = {136},
	primaryclass = {astro-ph.EP},
	title = {{Transit and Eclipse Analyses of the Exoplanet HD 149026b Using BLISS Mapping}},
	volume = 754,
	year = 2012}

@article{kataria2016,
	adsurl = {http://adsabs.harvard.edu/abs/2016ApJ...821....9K},
	archiveprefix = {arXiv},
	author = {{Kataria}, T. and {Sing}, D.~K. and {Lewis}, N.~K. and {Visscher}, C. and {Showman}, A.~P. and {Fortney}, J.~J. and {Marley}, M.~S.},
	doi = {10.3847/0004-637X/821/1/9},
	eid = {9},
	eprint = {1602.06733},
	journal = {\apj},
	month = apr,
	pages = {9},
	primaryclass = {astro-ph.EP},
	title = {{The Atmospheric Circulation of a Nine-hot-Jupiter Sample: Probing Circulation and Chemistry over a Wide Phase Space}},
	volume = 821,
	year = 2016}

@article{wakeford2016,
    author = {{Wakeford}, H.~R. and {Sing}, D.~K. and {Evans}, T. and {Deming}, D. and {Mandell}, A.},
        title = "{Marginalizing Instrument Systematics in HST WFC3 Transit Light Curves}",
      journal = {\apj},
         year = 2016,
        month = mar,
       volume = {819},
       number = {1},
          eid = {10},
        pages = {10},
          doi = {10.3847/0004-637X/819/1/10},
archivePrefix = {arXiv},
       eprint = {1601.02587},
 primaryClass = {astro-ph.EP},
       adsurl = {https://ui.adsabs.harvard.edu/abs/2016ApJ...819...10W}
}

@article{Sing2016,
	author = {Sing, David K. and Fortney, Jonathan J. and Nikolov, Nikolay and Wakeford, Hannah R. and Kataria, Tiffany and Evans, Thomas M. and Aigrain, Suzanne and Ballester, Gilda E. and Burrows, Adam S. and Deming, Drake and D{\'e}sert, Jean-Michel and Gibson, Neale P. and Henry, Gregory W. and Huitson, Catherine M. and Knutson, Heather A. and Etangs, Alain Lecavelier des and Pont, Frederic and Showman, Adam P. and Vidal-Madjar, Alfred and Williamson, Michael H. and Wilson, Paul A.},
	date = {2016/01/07/print},
	day = {07},
	isbn = {0028-0836},
	journal = {Nature},
	l3 = {10.1038/nature16068},
	m3 = {Letter},
	month = {01},
	number = {7584},
	pages = {59--62},
	publisher = {Nature Publishing Group, a division of Macmillan Publishers Limited. All Rights Reserved.},
	title = {A continuum from clear to cloudy hot-Jupiter exoplanets without primordial water depletion},
	ty = {JOUR},
	url = {http://dx.doi.org/10.1038/nature16068},
	volume = {529},
	year = {2016}}

@article{visscher2010,
	adsurl = {http://adsabs.harvard.edu/abs/2010ApJ...716.1060V},
	archiveprefix = {arXiv},
	author = {{Visscher}, C. and {Lodders}, K. and {Fegley}, Jr., B.},
	doi = {10.1088/0004-637X/716/2/1060},
	eid = {1060},
	eprint = {1001.3639},
	journal = {\apj},
	month = jun,
	pages = {1060-1075},
	primaryclass = {astro-ph.EP},
	title = {{Atmospheric Chemistry in Giant Planets, Brown Dwarfs, and Low-mass Dwarf Stars. III. Iron, Magnesium, and Silicon}},
	volume = 716,
	year = 2010}

@article{wakeford2015,
	adsurl = {http://adsabs.harvard.edu/abs/2015A%26A...573A.122W},
	archiveprefix = {arXiv},
	author = {{Wakeford}, H.~R. and {Sing}, D.~K.},
	doi = {10.1051/0004-6361/201424207},
	eid = {A122},
	eprint = {1409.7594},
	journal = {\aap},
	month = jan,
	pages = {A122},
	primaryclass = {astro-ph.EP},
	title = {{Transmission spectral properties of clouds for hot Jupiter exoplanets}},
	volume = 573,
	year = 2015}

@article{henry2000,
	adsurl = {http://adsabs.harvard.edu/abs/2000ApJ...529L..41H},
	author = {{Henry}, G.~W. and {Marcy}, G.~W. and {Butler}, R.~P. and {Vogt}, S.~S.},
	doi = {10.1086/312458},
	journal = {\apjl},
	month = jan,
	pages = {L41-L44},
	title = {{A Transiting ``51 Peg-like'' Planet}},
	volume = 529,
	year = 2000}

@article{charbonneau2000,
	author = {Charbonneau, David and Brown, Timothy M and Latham, David W and Mayor, Michel},
	journal = {\apjl},
	number = {1},
	pages = {L45},
	publisher = {IOP Publishing},
	title = {Detection of planetary transits across a sun-like star},
	volume = {529},
	year = {2000}}

@article{kreidberg2014a,
	adsurl = {http://adsabs.harvard.edu/abs/2014Natur.505...69K},
	archiveprefix = {arXiv},
	author = {{Kreidberg}, L. and {Bean}, J.~L. and {D{\'e}sert}, J.-M. and {Benneke}, B. and {Deming}, D. and {Stevenson}, K.~B. and {Seager}, S. and {Berta-Thompson}, Z. and {Seifahrt}, A. and {Homeier}, D.},
	doi = {10.1038/nature12888},
	eprint = {1401.0022},
	journal = {\nat},
	month = jan,
	pages = {69-72},
	primaryclass = {astro-ph.EP},
	title = {{Clouds in the atmosphere of the super-Earth exoplanet GJ1214b}},
	volume = 505,
	year = 2014 # a}

@article{vidal2013,
	adsurl = {http://adsabs.harvard.edu/abs/2013A%26A...560A..54V},
	archiveprefix = {arXiv},
	author = {{Vidal-Madjar}, A. and {Huitson}, C.~M. and {Bourrier}, V. and {D{\'e}sert}, J.-M. and {Ballester}, G. and {Lecavelier des Etangs}, A. and {Sing}, D.~K. and {Ehrenreich}, D. and {Ferlet}, R. and {H{\'e}brard}, G. and {McConnell}, J.~C.},
	doi = {10.1051/0004-6361/201322234},
	eid = {A54},
	eprint = {1310.8104},
	journal = {\aap},
	month = dec,
	pages = {A54},
	primaryclass = {astro-ph.EP},
	title = {{Magnesium in the atmosphere of the planet HD 209458 b: observations of the thermosphere-exosphere transition region}},
	volume = 560,
	year = 2013}

@article{ackerman2001,
	author = {Andrew S. Ackerman and Mark S. Marley},
	journal = {\apj},
	number = {2},
	pages = {872},
	title = {Precipitating Condensation Clouds in Substellar Atmospheres},
	url = {http://stacks.iop.org/0004-637X/556/i=2/a=872},
	volume = {556},
	year = {2001}}

@article{cushing2006,
	adsurl = {http://adsabs.harvard.edu/abs/2006ApJ...648..614C},
	author = {{Cushing}, M.~C. and {Roellig}, T.~L. and {Marley}, M.~S. and {Saumon}, D. and {Leggett}, S.~K. and {Kirkpatrick}, J.~D. and {Wilson}, J.~C. and {Sloan}, G.~C. and {Mainzer}, A.~K. and {Van Cleve}, J.~E. and {Houck}, J.~R.},
	doi = {10.1086/505637},
	eprint = {astro-ph/0605639},
	journal = {\apj},
	month = sep,
	pages = {614-628},
	title = {{A Spitzer Infrared Spectrograph Spectral Sequence of M, L, and T Dwarfs}},
	volume = 648,
	year = 2006}

@article{dorschner1995,
	adsurl = {http://adsabs.harvard.edu/abs/1995A%26A...300..503D},
	author = {{Dorschner}, J. and {Begemann}, B. and {Henning}, T. and {Jaeger}, C. and {Mutschke}, H.},
	journal = {\aap},
	month = aug,
	pages = {503},
	title = {{Steps toward interstellar silicate mineralogy. II. Study of Mg-Fe-silicate glasses of variable composition.}},
	volume = 300,
	year = 1995}

@article{jager2003,
	author = {J{\"a}ger, Cornelia and B Il'in, Vladimir and Henning, Thomas and Mutschke, Harald and Fabian, Dirk and A Semenov, Dmitry and V Voshchinnikov, Nikolai},
	journal = {Journal of Quantitative Spectroscopy and Radiative Transfer},
	pages = {765--774},
	publisher = {Elsevier},
	title = {A database of optical constants of cosmic dust analogs},
	volume = {79},
	year = {2003}}

@article{visscher2006,
	adsurl = {http://adsabs.harvard.edu/abs/2006ApJ...648.1181V},
	author = {{Visscher}, C. and {Lodders}, K. and {Fegley}, Jr., B.},
	doi = {10.1086/506245},
	eprint = {arXiv:astro-ph/0511136},
	journal = {\apj},
	month = sep,
	pages = {1181-1195},
	title = {{Atmospheric Chemistry in Giant Planets, Brown Dwarfs, and Low-Mass Dwarf Stars. II. Sulfur and Phosphorus}},
	volume = 648,
	year = 2006}

@article{pont2006,
	adsurl = {http://adsabs.harvard.edu/abs/2006MNRAS.373..231P},
	author = {{Pont}, F. and {Zucker}, S. and {Queloz}, D.},
	doi = {10.1111/j.1365-2966.2006.11012.x},
	eprint = {arXiv:astro-ph/0608597},
	journal = {\mnras},
	month = nov,
	pages = {231-242},
	title = {{The effect of red noise on planetary transit detection}},
	volume = 373,
	year = 2006}

@article{sing2011b,
	adsurl = {http://adsabs.harvard.edu/abs/2011MNRAS.416.1443S},
	archiveprefix = {arXiv},
	author = {{Sing}, D.~K. and {Pont}, F. and {Aigrain}, S. and {Charbonneau}, D. and {D{\'e}sert}, J.-M. and {Gibson}, N. and {Gilliland}, R. and {Hayek}, W. and {Henry}, G. and {Knutson}, H. and {Lecavelier Des Etangs}, A. and {Mazeh}, T. and {Shporer}, A.},
	doi = {10.1111/j.1365-2966.2011.19142.x},
	eprint = {1103.0026},
	journal = {\mnras},
	month = sep,
	pages = {1443-1455},
	primaryclass = {astro-ph.EP},
	title = {{Hubble Space Telescope transmission spectroscopy of the exoplanet HD 189733b: high-altitude atmospheric haze in the optical and near-ultraviolet with STIS}},
	volume = 416,
	year = 2011 # b}

@article{charbonneau2002,
	author = {David Charbonneau and Timothy M. Brown and Robert W. Noyes and Ronald L. Gilliland},
	journal = {\apj},
	number = {1},
	pages = {377},
	title = {Detection of an Extrasolar Planet Atmosphere},
	url = {http://stacks.iop.org/0004-637X/568/i=1/a=377},
	volume = {568},
	year = {2002}}

@article{18OrMi,
	author = {Ormel, Chris W. and Min, Michiel},
	title = {{ARCiS} framework for exoplanet atmospheres - The cloud transport model},
	DOI= {10.1051/0004-6361/201833678},
	journal = {A\&A},
	year = 2019,
	volume = {622},
	pages = {A121}
}

@ARTICLE{Min2020,
       author = {{Min}, Michiel and {Ormel}, Chris W. and {Chubb}, Katy and {Helling}, Christiane and {Kawashima}, Yui},
        title = "{The ARCiS framework for exoplanet atmospheres. Modelling philosophy and retrieval}",
      journal = {\aap},
         year = 2020,
        month = oct,
       volume = {642},
          eid = {A28},
        pages = {A28},
          doi = {10.1051/0004-6361/201937377}
}

@ARTICLE{Feroz2009,
       author = {{Feroz}, F. and {Hobson}, M.~P. and {Bridges}, M.},
        title = "{MULTINEST: an efficient and robust Bayesian inference tool for cosmology and particle physics}",
      journal = {Mon. Not. Roy. Astron. Soc.},
         year = "2009",
        month = {10},
       volume = {398},
       number = {4},
        pages = {1601-1614},
          doi = {10.1111/j.1365-2966.2009.14548.x}
}

@ARTICLE{Feroz2019,
       author = {{Feroz}, Farhan and {Hobson}, Michael P. and {Cameron}, Ewan and
         {Pettitt}, Anthony N.},
        title = "{Importance Nested Sampling and the MultiNest Algorithm}",
      journal = {The Open Journal of Astrophysics},
         year = 2019,
        month = nov,
       volume = {2},
       number = {1},
          eid = {10},
        pages = {10},
          doi = {10.21105/astro.1306.2144}
}

@article{20ChRoAl.exo,
author = {K. L. Chubb and M. Rocchetto and A. F. Al-Refaie and I. Waldmann and M. Min and J. Barstow and P. Molli{\`e}re and M. W. Phillips and J. Tennyson and S. N. Yurchenko},
title = {The {ExoMolOP Database: C}ross-sections and k-tables for Molecules of Interest in High-Temperature Exoplanet Atmospheres},
journal = AA,
year  = {2021},
volume={646},
number={A21},
doi = {10.1051/0004-6361/202038350}
}

@Article{ExoMol_H2O,
author = {O. L. Polyansky and A. A. Kyuberis   and N. F. Zobov and J. Tennyson and  S. N. Yurchenko and  L. Lodi},
title = {{ExoMol molecular line lists XXX: a complete high-accuracy  line list for water}},
journal = {Mon. Not. R. Astron. Soc.},
doi={10.1093/mnras/sty1877},
Volume={480},
Pages={2597-2608},
year = {2018}}

@Article{ExoMol_NH3,
author = {P. A. Coles and and Sergei N. Yurchenko and Jonathan Tennyson},
title = {{ExoMol molecular line lists XXXV: a rotation-vibration line list for hot ammonia}},
journal = {Mon. Not. R. Astron. Soc.},
volume={490},
issue = {4},
Pages={4638--4647},
doi={10.1093/mnras/stz2778},
year = {2019}}

@article{ 15LiGoRo.CO,
Author = {Li, Gang and Gordon, Iouli E. and Rothman, Laurence S. and Tan, Yan and
   Hu, Shui-Ming and Kassi, Samir and Campargue, Alain and Medvedev, Emile
   S.},
Title = {Rovibrational line lists for nine isotopologues of the CO molecule in
the  X {$^1\Sigma^+$} ground electronic state},
Journal = {Astrophys. J. Suppl.},
Year = {{2015}},
Volume = {{216}},
DOI = {10.1088/0067-0049/216/1/15},
pages = {15}}

@article{ExoMol_CO2,
author = {Sergei N. Yurchenko and Thomas M. Mellor and Richard S. Freedman and Jonathan Tennyson},
title = {{ExoMol line lists – XXXIX. Ro-vibrational molecular line list for CO$_2$}},
journal = {Mon. Not. R. Astron. Soc.},
    volume = {496},
    number = {4},
    pages = {5282-5291},
year = {2020},
    doi = {10.1093/mnras/staa1874}
}

@ARTICLE{ExoMol_SO2,
  author = {D. S. Underwood and J. Tennyson and S. N. Yurchenko and Huang, Xinchuan and Schwenke, David W. and Lee, Timothy J. and S. Clausen and A. Fateev},
title = {{ExoMol line lists XIV: A line list for hot SO$_2$}},
journal = {Mon. Not. R. Astron. Soc.},
year  = {2016},
volume  = {459},
pages = {3890-3899},
doi= {10.1093/mnras/stw849}}

@Article{ExoMol_C2H2,
author = {K. L. Chubb and J. Tennyson and S. N. Yurchenko}, 
title = {{ExoMol Molecular linelists -- XXXVII: spectra of acetylene}},
journal = {Mon. Not. R. Astron. Soc.},
doi = {10.1093/mnras/staa229},
volume = {493},
number = {2},
pages = {1531-1545},
year={2020}
}

@Article{ExoMol_H2S, 
author = {A. A. A. Azzam  and S. N. Yurchenko and J. Tennyson and O. V. Naumenko},
title = {{ExoMol line lists XVI: A Hot Line List for H$_{2}$S}},
journal = {Mon. Not. R. Astron. Soc.},
volume = {460},
year  = {2016},
doi= {10.1093/mnras/stw1133},
pages = {4063-4074}}

@article{ExoMol_HCN,
author = {R J Barber and J K Strange and C Hill and O L  Polyansky and G Ch Mellau and S N Yurchenko and Jonathan Tennyson},
title = {{ExoMol line lists -- III. An improved hot rotation-vibration line list for HCN and HNC}},
journal = {Mon. Not. R. Astron. Soc.},
volume = {437}, 
pages = {1828-1835},
year = {2014},
doi = {10.1093/mnras/stt2011}
}

@Article{ExoMol_CH4,
author = {S. N. Yurchenko and D. S. Amundsen AND J. Tennyson  and I P Waldmann},
title = {{A hybrid line list for CH$_4$ and hot methane continuum}},
journal = {A\&A},
volume = {605},
pages = {A95},
doi = {10.1051/0004-6361/201731026},
year={2017}
}

@article{16AlSpKi.broad,
        author = {Allard, N. F. and Spiegelman, F. and Kielkopf, J. F.},
        title = {{K-H$_2$} line shapes for the spectra of cool brown dwarfs},
        DOI= {10.1051/0004-6361/201628270},
        journal = {A\&A},
        year = 2016,
        volume = 589,
        pages = {A21}
}

@article{19AlSpLe.broad,
        author = {Allard, N. F. and Spiegelman, F. and Leininger, T. and Molli\`ere, P.},
        title = {New study of the line profiles of sodium perturbed by {H$_2$}},
        DOI= {10.1051/0004-6361/201935593},
        journal = {A\&A},
        year = 2019,
        volume = 628,
        pages = {A120}
}

@misc{KURonline,
    author  = "R.L. Kurucz and B. Bell",
    title   = "Kurucz atomic database",
    note    = "http://kurucz.harvard.edu/atoms.html",
    year    = "1995"
}

@article{10Guillot.exo,
        author = {Guillot, T.},
        title = {On the radiative equilibrium of irradiated planetary atmospheres},
        DOI= {10.1051/0004-6361/200913396},
        journal = {A\&A},
        year = 2010,
        volume = 520,
        pages = {A27}
}

@article{ 16McYuTe,
Author = {McKemmish, Laura K. and Yurchenko, Sergei N. and Tennyson, Jonathan},
Title = {{ExoMol line lists - XVIII. The high-temperature spectrum of VO}},
journal = {Mon. Not. R. Astron. Soc.},
Year = {2016},
Volume = {463},
Pages = {771-793},
DOI = {10.1093/mnras/stw1969}
}

@article{19McMaHo,
    author = {McKemmish, Laura K and Masseron, Thomas and Hoeijmakers, H Jens and P\'{e}rez-Mesa, V\'{i}ctor and Grimm, Simon L and Yurchenko, Sergei N and Tennyson, Jonathan},
    title = {{ExoMol molecular line lists - XXXIII. The spectrum of Titanium Oxide}},
journal = {Mon. Not. R. Astron. Soc.},
    volume = {488},
    pages = {2836-2854},
    year = {2019},
    doi = {10.1093/mnras/stz1818}
}

@article{21YuTeSy,
    author = {Yurchenko, Sergei N and Tennyson, Jonathan and Syme, Anna-Maree and Adam, Ahmad Y and Clark, Victoria H J and Cooper, Bridgette and Dobney, C Pria and Donnelly, Shaun T E and Gorman, Maire N and Lynas-Gray, Anthony E and Meltzer, Thomas and Owens, Alec and Qu, Qianwei and Semenov, Mikhail and Somogyi, Wilfrid and Upadhyay, Apoorva and Wright, Samuel and Zapata-Trujillo, Juan C},
    title = "{ExoMol line lists - XLIV. Infrared and ultraviolet line list for silicon monoxide ($^{28}$Si$_16$O)}",
    journal = {Monthly Notices of the Royal Astronomical Society},
    volume = {510},
    number = {1},
    pages = {903-919},
    year = {2021},
    issn = {0035-8711},
    doi = {10.1093/mnras/stab3267}
}

@article{95KoKaYa,
title = {{Extinction Spectra of Corundum in the Wavelengths from UV to FIR}},
journal = {Icarus},
volume = {114},
pages = {203-214},
year = {1995},
issn = {0019-1035},
doi = {10.1006/icar.1995.1055},
author = {Chiyoe Koike and Chihiro Kaito and Tetsuo Yamamoto and Hiroshi Shibai and Seiji Kimura and Hiroshi Suto}
}

@BOOK{12Palik,
       author = {Edward D. Palik},
        title = {Handbook of Optical Constants of Solids: Vol 2},
        publisher = {Academic Press},
         year = 2012
}

@article{03JaDoMu,
        author = {J\"ager, C. and Dorschner, J. and Mutschke, H. and Posch, Th. and Henning, Th.},
        title = {Steps toward interstellar silicate mineralogy - {VII. S}pectral properties and crystallization behaviour of magnesium silicates produced by the sol-gel method},
        DOI= "10.1051/0004-6361:20030916",
        journal = {A\&A},
        year = 2003,
        volume = 408,
        pages = "193-204"
}

@article{ 13ZePoMu,
	author = {Zeidler, S. and Posch, Th. and Mutschke, H.},
	title = {Optical constants of refractory oxides at high temperatures - Mid-infrared properties of corundum, spinel, and $\alpha$-quartz,  potential carriers of the 13micron feature},
	DOI= "10.1051/0004-6361/201220459",
	journal = {A\&A},
	year = 2013,
	volume = 553,
	pages = "A81"
	}

@article{24DyMiDe,
	author = {Dyrek, Achr{\`e}ne and Min, Michiel and Decin, Leen and Bouwman, Jeroen and others},
	doi = {10.1038/s41586-023-06849-0},
	journal = {Nature},
	pages = {51--54},
	title = {{SO$_2$}, silicate clouds, but no {CH$_4$} detected in a warm {Neptune}},
volume = {625},
	year = {2024}
}

@ARTICLE{97HeMu,
       author = {{Henning}, T. and {Mutschke}, H.},
        title = "{Low-temperature infrared properties of cosmic dust analogues.}",
      journal = {\aap},
         year = {1997},
       volume = {327},
        pages = {743-754}
}

@ARTICLE{95DoBeHe,
       author = {Dorschner, J. and Begemann, B. and {Henning}, T. and {Jaeger}, C. and {Mutschke}, H.},
        title = {Steps toward interstellar silicate mineralogy. {II. S}tudy of {Mg-Fe}-silicate glasses of variable composition.},
      journal = {A\&A},
         year = 1995,
       volume = {300},
        pages = {503}
}

@ARTICLE{25BaKrMo,
       author = {Bachmann, N. and Kreidberg, L. and {Molli{\`e}re}, P. and Deming, D. and Tsai,S.-M.},
        title = {Osiris revisited: Confirming a solar metallicity and low {C/O} in {HD 209458b}},
      journal = {A\&A},
         year = 2025,
       volume = {},
        pages = {}, 
     doi = {10.1051/0004-6361/202555577},
note = {Forthcoming article}
}

@article{24FaWaMa,
doi = {10.3847/1538-3881/ad3454},
year = {2024},
month = {apr},
publisher = {The American Astronomical Society},
volume = {167},
number = {5},
pages = {240},
author = {Fairman, Charlotte and Wakeford, Hannah R. and MacDonald, Ryan J.},
title = {The Importance of Optical Wavelength Data on Atmospheric Retrievals of Exoplanet Transmission Spectra},
journal = {The Astronomical Journal}
}

@article{16HeLeDo,
    author = {Helling, Ch. and Lee, E. and Dobbs-Dixon, I. and Mayne, N. and Amundsen, D. S. and Khaimova, J. and Unger, A. A. and Manners, J. and Acreman, D. and Smith, C.},
    title = {The mineral clouds on HD~209458b and HD~189733b},
    journal = {Monthly Notices of the Royal Astronomical Society},
    volume = {460},
    number = {1},
    pages = {855-883},
    year = {2016},
    month = {04},
    issn = {0035-8711},
    doi = {10.1093/mnras/stw662},
    url = {https://doi.org/10.1093/mnras/stw662}
}

@article{21GiBrGa,
	author = {Giacobbe, Paolo and Brogi, Matteo and Gandhi, Siddharth and others},
	doi = {10.1038/s41586-021-03381-x},
	journal = {Nature},
	number = {7853},
	pages = {205--208},
	title = {Five carbon- and nitrogen-bearing species in a hot giant planet's atmosphere},
	volume = {592},
	year = {2021}
}

@article{25VeGoAv,
doi = {10.3847/1538-3881/addc5c},
year = {2025},
month = {jul},
publisher = {The American Astronomical Society},
volume = {170},
number = {2},
pages = {69},
author = {Verma, Avinash and Goyal, Jayesh and Avarsekar, Swaroop and Shukla, Gaurav},
title = {A Detailed Investigation of {HD 209458 b HST} and {JWST} Transmission Spectra with {SANSAR}},
journal = {The Astronomical Journal}
}

@article{ressler2015mid,
  title={The mid-infrared instrument for the James Webb Space Telescope, VIII: the MIRI focal plane system},
  author={Ressler, ME and Sukhatme, KG and Franklin, BR and Mahoney, JC and Thelen, MP and Bouchet, P and Colbert, JW and Cracraft, Misty and Dicken, D and Gastaud, R and others},
  journal={Publications of the Astronomical Society of the Pacific},
  volume={127},
  number={953},
  pages={675},
  year={2015},
  publisher={IOP Publishing}
}

@article{wright2023mid,
  title={The Mid-infrared Instrument for JWST and Its In-flight Performance},
  author={Wright, Gillian S and Rieke, George H and Glasse, Alistair and Ressler, Michael and Mar{\'\i}n, Macarena Garc{\'\i}a and Aguilar, Jonathan and Alberts, Stacey and {\'A}lvarez-M{\'a}rquez, Javier and Argyriou, Ioannis and Banks, Kimberly and others},
  journal={Publications of the Astronomical Society of the Pacific},
  volume={135},
  number={1046},
  pages={048003},
  year={2023},
  publisher={IOP Publishing}
}

@article{argyriou2023brighter,
  title={The Brighter-Fatter Effect in the JWST MIRI Si: As IBC detectors I. Observations, impact on science, and modelling},
  author={Argyriou, Ioannis and Lage, Craig and Rieke, George H and Gasman, Danny and Bouwman, Jeroen and Morrison, Jane and Libralato, Mattia and Dicken, Daniel and Brandl, Bernhard R and {\'A}lvarez-M{\'a}rquez, Javier and others},
  journal={arXiv preprint arXiv:2303.13517},
  year={2023}
}

@article{bell2023first,
  title={A First Look at the JWST MIRI/LRS Phase Curve of WASP-43b},
  author={Bell, Taylor J and Kreidberg, Laura and Kendrew, Sarah and Bean, Jacob and Crouzet, Nicolas and Ducrot, Elsa and Dyrek, Achr{\`e}ne and Gao, Peter and Lagage, Pierre-Olivier and Moses, Julianne I},
  journal={arXiv preprint arXiv:2301.06350},
  year={2023}
}

@article{Grant2024,
  title = {ExoTiC-LD: thirty seconds to stellar limb-darkening coefficients},
  author = {David Grant and Hannah R. Wakeford},
  journal = {Journal of Open Source Software},
  publisher = {The Open Journal},
  year = {2024},
  volume = {9},
  number = {100},
  pages = {6816},
  doi = {10.21105/joss.06816},
  url = {https://doi.org/10.21105/joss.06816}
}

@article{22GaMiCh,
	author = {Gasman, Danny and Min, Michiel and Chubb, Katy L.},
	title = {Investigating the detectability of hydrocarbons in exoplanet atmospheres with JWST★},
	DOI= "10.1051/0004-6361/202141468",
	journal = {A\&A},
	year = 2022,
	volume = 659,
	pages = "A114"
}

@article{24WiChMa,
	author = {Wilkinson, C. and Charnay, B. and Mazevet, S. and Lagrange, A.-M. and Chomez, A. and Squicciarini, V. and Panek, E. and Mazoyer, J.},
	title = {Breaking degeneracies in exoplanetary parameters through self-consistent atmosphere–interior modelling},
	DOI= "10.1051/0004-6361/202348945",
	journal = {A\&A},
	year = 2024,
	volume = 692,
	pages = "A113"
}

@article{PowellZhang2024,
doi = {10.3847/1538-4357/ad3de4},
year = {2024},
publisher = {The American Astronomical Society},
volume = {969},
number = {1},
pages = {5},
author = {Powell, Diana and Zhang, Xi},
title = {Two-dimensional Models of Microphysical Clouds on {Hot Jupiters. I. C}loud Properties},
journal = {The Astrophysical Journal}
}

@article{Brogi_2017,
doi = {10.3847/2041-8213/aa6933},
year = {2017},
month = {apr},
publisher = {The American Astronomical Society},
volume = {839},
number = {1},
pages = {L2},
author = {Brogi, M. and Line, M. and Bean, J. and Désert, J.-M. and Schwarz, H.},
title = {A Framework to Combine Low- and High-resolution Spectroscopy for the Atmospheres of Transiting Exoplanets},
journal = {The Astrophysical Journal Letters}}

@article{35Bruggeman,
author = {Bruggeman, D. A. G.},
title = {Berechnung verschiedener physikalischer Konstanten von heterogenen Substanzen. I. Dielektrizit{\"a}tskonstanten und Leitf{\"a}higkeiten der Mischk{\"o}rper aus isotropen Substanzen},
journal = {Annalen der Physik},
volume = {416},
number = {7},
pages = {636-664},
doi = {10.1002/andp.19354160705},
year = {1935}
}

@ARTICLE{Woitke2018,
       author = {{Woitke}, P. and {Helling}, Ch. and {Hunter}, G.~H. and {Millard}, J.~D. and {Turner}, G.~E. and {Worters}, M. and {Blecic}, J. and {Stock}, J.~W.},
        title = "{Equilibrium chemistry down to 100 K. Impact of silicates and phyllosilicates on the carbon to oxygen ratio}",
      journal = {\aap},
         year = 2018,
        month = jun,
       volume = {614},
          eid = {A1},
        pages = {A1},
          doi = {10.1051/0004-6361/201732193},
archivePrefix = {arXiv},
       eprint = {1712.01010},
 primaryClass = {astro-ph.EP},
       adsurl = {https://ui.adsabs.harvard.edu/abs/2018A&A...614A...1W}
}

@article{thorngren2025bayesian,
  title={Bayesian Model Comparison and Significance: {Widespread} Errors and how to Correct Them},
  author={Thorngren, Daniel P and Sing, David K and Mukherjee, Sagnick},
  journal={arXiv preprint arXiv:2510.00169},
  year={2025}
}

@ARTICLE{Eckes2013,
       author = {M. Eckes and B. Gibert and D. De Sousa Meneses, M. Malki & P. Echegut}, 
        title = "{High-temperature infrared properties of forsterite}",
      journal = {Phys Chem Minerals},
         year = 2013,
        month = {},
       volume = {40},
        pages = {287--298},
          doi = {10.1007/s00269-013-0570-z}
}

@article{23ChStHe,
    author = {Chubb, Katy L and Stam, Daphne M and Helling, Christiane and Samra, Dominic and Carone, Ludmila},
    title = {Modelling reflected polarized light from close-in giant exoplanet {WASP-96b} using {PolHEx} (Polarization of hot exoplanets)},
    journal = {Monthly Notices of the Royal Astronomical Society},
    volume = {527},
    number = {3},
    pages = {4955-4982},
    year = {2023},
    month = {11},
    doi = {10.1093/mnras/stad3413}
}

@article{Moran_2025,
doi = {10.3847/1538-4357/ae0583},
year = {2025},
month = {nov},
publisher = {The American Astronomical Society},
volume = {994},
number = {1},
pages = {116},
author = {Moran, Sarah E. and Lodge, Matt G. and Batalha, Natasha E. and Ohno, Kazumasa and Vahidinia, Sanaz and Marley, Mark S. and Wakeford, Hannah R. and Leinhardt, Zo{\"e} M.},
title = {Fractal Aggregate Aerosols in the Virga Cloud Code. I. Model Description and Application to a Benchmark Cloudy Exoplanet},
journal = {The Astrophysical Journal}
}

@ARTICLE{Calamari2026,
       author = {{Calamari}, Emily and {Faherty}, Jacqueline K. and {Visscher}, Channon and {Gemma}, Marina E. and {Rothermich}, Austin and {Mart{\'\i}nez}, Francisco Ard{\'e}vol and {Merchan}, Sherelyn Alejandro and {Su{\'a}rez}, Genaro},
        title = "{Bridging the Gap: Using Brown Dwarfs to Examine Silicate Clouds in Giant Exoplanet Atmospheres}",
      journal = {\apjl},
         year = 2026,
        month = apr,
       volume = {1000},
       number = {2},
          eid = {L51},
        pages = {L51},
          doi = {10.3847/2041-8213/ae497d},
archivePrefix = {arXiv}
}

@Article{AlO_ExoMol,
author = {A. T. Patrascu and J. Tennyson and S. N. Yurchenko},
title = {{ExoMol molecular linelists - IX: The spectrum of AlO}},
journal = {MNRAS},
volume = {449},
pages = {3613-3619},
doi = {10.1093/mnras/stv507},
year={2015}}

@Article{BeH_ExoMol,
author = {D. Darby-Lewis and J. Tennyson and K. D. Lawson and S. N.
Yurchenko and M. F. Stamp and A. Shaw and S. Brezinsek and  {JET
Contributor}},
title = {{Synthetic spectra of BeH, BeD and BeT for emission modelling in JET plasmas}},
journal = {J. Phys. B: At. Mol. Opt. Phys.},
year  = {2018},
doi = {10.1088/1361-6455/aad6d0},
volume={51},
pages={185701}}

@article{14MaPlVa.CH,
Author = {Masseron, T. and Plez, B. and Van Eck, S. and Colin, R. and Daoutidis, I.
and Godefroid, M. and Coheur, P.-F. and Bernath, P. and Jorissen, A. and Christlieb, N.},
Title = {{CH in stellar atmospheres: an extensive linelist}},
Journal = AA,
Year = {2014},
Volume = {571},
pages = {A47},
DOI = {10.1051/0004-6361/201423956}
}

@Article{CS_ExoMol,
author = {G Paulose and E J Barton AND S. N. Yurchenko AND J. Tennyson},
title = {{ ExoMol Molecular linelists -- XII. Line lists for eight isotopologues of CS}},
journal = {MNRAS},
volume = {454},
pages = {1931-1939},
doi= {10.1093/mnras/stv1543},
year={2015}}

@Article{H2CO_ExoMol,
author = {A. F. Al-Refaie and S. N. Yurchenko and A. Yachmenev and J. Tennyson},
title = {{ExoMol line lists - VIII: A variationally computed line list for hot formaldehyde}},
journal = {MNRAS},
volume = {448},
pages = {1704-1714},
doi = {10.1093/mnras/stv091},
year = {2015}}

@article{ 13LiGoHa.HCl,
Author = {Li, Gang and Gordon, Iouli E. and Hajigeorgiou, Photos G. and Coxon,
   John A. and Rothman, Laurence S.},
Title = {{Reference spectroscopic data for hydrogen halides, Part II: The line lists}},
journal = {JQSRT},
Year = {2013},
Volume = {130},
Pages = {284-295},
DOI = {10.1016/j.jqsrt.2013.07.019}
}

@article{N2O_ExoMol,
    author = {Yurchenko, Sergei N and Mellor, Thomas M and Tennyson, Jonathan},
    title = {ExoMol line lists–LIX. High-temperature line list for N2O},
    journal = {Monthly Notices of the Royal Astronomical Society},
    volume = {534},
    number = {2},
    pages = {1364-1375},
    year = {2024},
    month = {10},
    issn = {0035-8711},
    doi = {10.1093/mnras/stae2201}
}

@Article{NO_ExoMol,
author = {Andy Wong and S. N. Yurchenko and Peter Bernath and Holger S. P. Mueller and Stephanie McConkey AND J. Tennyson},
title = {{ExoMol Line List XXI: Nitric Oxide (NO)}},
journal = {MNRAS},
volume = {470},
pages = {882-897},
DOI={10.1093/mnras/stx1211},
year={2017}}

@Article{NS_ExoMol,
author = {S. N. Yurchenko and W. Bond and M. N. Gorman and L. Lodi and  L. K. McKemmish and W. Nunn and R. Shah and J. Tennyson},
title = {{ExoMol Molecular linelists --  XXVI: spectra of SH and NS}},
journal = {MNRAS},
volume = {478},
pages = {270-282},
doi={10.1093/mnras/sty939},
year={2018}}

@article{16BrBeWe.OH,
   author = {James S. A. Brooke and Peter F. Bernath and Colin M. Western and Christopher Sneden
   and Melike Afsar and Gang Li and Iouli E. Gordon},
   title = {{Line strengths of rovibrational and rotational transitions in the $X~^{2}\Pi$ ground state of OH}},
   journal = JQSRT,
   year = {2016},
   volume = {168},
   pages = {142-157},
   doi = {10.1016/j.jqsrt.2015.07.021}
}

@Article{PS_ExoMol,
author = {L. Prajapat and P. Jagoda and L. Lodi and M. N. Gorman AND S. N. Yurchenko  and J. Tennyson},
title = {{ExoMol molecular line lists XXIII. Spectra of PO and PS}},
journal = {MNRAS},
  volume={472},
  pages={3648-3658},
doi={10.1093/mnras/stx2229},
year = {2017}}

@Article{SiH_ExoMol,
author = {S. N. Yurchenko and Frances Sinden and Lorenzo Lodi and Christian Hill and  Maire N. Gorman AND J. Tennyson},
title = {{ExoMol Molecular linelists --  XXIV: A new hot line list for silicon monohydride, SiH}},
journal = {MNRAS},
volume = {473},
pages = {5324-5333},
DOI={10.1093/mnras/stx2738},
year={2018}}

@Article{SiS_ExoMol,
author = {Apoorva Upadhyay and  Eamon K. Conway AND J. Tennyson and S. N. Yurchenko},
 title = {{ExoMol Molecular linelists --  XXV: A hot line list for silicon sulphide, SiS}},
journal = {MNRAS},
volume = {477},
pages = {1520-1527},
doi={10.1093/mnras/sty998},
year={2018}}

@ARTICLE{SO3_ExoMol,
  author = {D. S. Underwood and J. Tennyson and S. N. Yurchenko and  S. Clausen and A. Fateev},
title = {{ExoMol line lists XVII: A line list for hot SO$_3$}},
journal = {MNRAS},
year  = {2016},
volume = {462},
doi = {10.1093/mnras/stw1828},
pages = {4300-4313}}

@Article{SO_ExoMol,
author = {R. P. Brady and S. N. Yurchenko and J. Tennyson and G.-S. Kim},
title = {{ExoMol line lists -- LVI. The SO line list, MARVEL analysis of experimental transition data and  refinement of the  spectroscopic model}},
journal = {MNRAS},
volume  = {527},
pages = {6675-6690},
year = {2024},
doi = {10.1093/mnras/stad3508}}

@dataset{natasha_batalha_2025_14861730,
author = {Natasha Batalha and
Richard Freedman and
Ehsan Gharib-Nezhad and
Roxana Lupu},
title = {Resampled Opacity Database for PICASO},
month = feb,
year = 2025,
publisher = {Zenodo},
doi = {10.5281/zenodo.14861730},
url = {https://doi.org/10.5281/zenodo.14861730},
}

@dataset{HD209_zenodo,
author = {Katy L. Chubb and David Grant and Hannah R. Wakeford and Sarah E. Moran and others},
title = {Supplementary Information: Magnesium Silicate Clouds in the Atmosphere of HD 209458b from a Rule-Based Tree-Structured Data Reduction},
month = {},
year = 2026,
publisher = {Zenodo},
doi = {10.5281/zenodo.20089901},
url = {https://doi.org/10.5281/zenodo.20089901},
}

@article{Espinoza2024,
	author = {Espinoza, N{\'e}stor and Steinrueck, Maria E. and Kirk, James and MacDonald, Ryan J. and Savel, Arjun B. and Triantafillides, Anastasia and Kempton, Eliza M. -R. and Murphy, Matthew M. and Carone, Ludmila and Zamyatina, Maria and Lewis, David A. and Samra, Dominic and Kiefer, Sven and Rauscher, Emily and Christie, Duncan and Mayne, Nathan and Helling, Christiane and Rustamkulov, Zafar and Parmentier, Vivien and May, Erin M. and Carter, Aarynn L. and Zhang, Xi and L{\'o}pez-Morales, Mercedes and Allen, Natalie and Blecic, Jasmina and Decin, Leen and Mancini, Luigi and Molaverdikhani, Karan and Rackham, Benjamin V. and Palle, Enric and Tsai, Shang-Min and Ahrer, Eva-Maria and Bean, Jacob L. and Crossfield, Ian J. M. and Haegele, David and H{\'e}brard, Eric and Kreidberg, Laura and Powell, Diana and Schneider, Aaron D. and Welbanks, Luis and Wheatley, Peter and Brahm, Rafael and Crouzet, Nicolas},
	doi = {10.1038/s41586-024-07768-4},
	journal = {Nature},
	number = {8027},
	pages = {1017--1020},
	title = {Inhomogeneous terminators on the exoplanet {WASP-39 b}},
	url = {https://doi.org/10.1038/s41586-024-07768-4},
	volume = {632},
	year = {2024}
    }

@article{Fu2025,
doi = {10.3847/2041-8213/adf20f},
url = {https://doi.org/10.3847/2041-8213/adf20f},
year = {2025},
month = {aug},
publisher = {The American Astronomical Society},
volume = {989},
number = {1},
pages = {L17},
author = {Fu, Guangwei and Mukherjee, Sagnick and Stevenson, Kevin B. and Sing, David K. and Ashtari, Reza and Mayne, Nathan and Lothringer, Joshua D. and Zamyatina, Maria and Schmidt, Stephen P. and Gascón, Carlos and Allen, Natalie H. and Bennett, Katherine A. and López-Morales, Mercedes},
title = {Overcast Mornings and Clear Evenings in Hot Jupiter Exoplanet Atmospheres},
journal = {The Astrophysical Journal Letters}
}

@article{Mukherjee2025,
author = {Sagnick Mukherjee  and David K. Sing  and Guangwei Fu  and Kevin B. Stevenson  and Stephen P. Schmidt  and Harry Baskett  and Mei Ting Mak  and Patrick McCreery  and Natalie H. Allen  and Katherine A. Bennett  and Duncan A. Christie  and Carlos Gascón  and Jayesh Goyal  and Éric Hébrard  and Joshua D. Lothringer  and Mercedes López-Morales  and Jacob Lustig-Yaeger  and Erin M. May  and L. C. Mayorga  and Nathan Mayne  and Lakeisha M. Ramos Rosado  and Henrique Reggiani  and Zafar Rustamkulov  and Kevin C. Schlaufman  and Kristin S. Sotzen  and Daniel Thorngren  and Le-Chris Wang  and Maria Zamyatina },
title = {Cloudy mornings and clear evenings on a gas giant exoplanet},
journal = {Science},
volume = {392},
number = {6800},
pages = {858-862},
year = {2026},
doi = {10.1126/science.adx5903},
URL = {https://www.science.org/doi/abs/10.1126/science.adx5903}
}
\bibliographystyle{aasjournal}



\end{document}